\documentclass[preprint,notoc]{JHEP3} 
\usepackage{epsfig}

\def\eslt{\not\!\!{E_T}}
\def\eslt{E_T^{\rm miss}}

\def\to{\rightarrow}

\def\bi{\begin{itemize}}
\def\ei{\end{itemize}}
\def\te{\tilde e}

\def\tH{\tilde H}

\def\tu{\tilde u}
\def\tc{\tilde c}

\def\tb{\tilde b}
\def\tf{\tilde f}

\def\tH{\tilde H}
\def\tst{\tilde t}
\def\ttau{\tilde \tau}
\def\tmu{\tilde \mu}
\def\tg{\tilde g}
\def\tnu{\tilde\nu}
\def\tell{\tilde\ell}
\def\tq{\tilde q}

\def\tw{\widetilde W}
\def\tz{\widetilde Z}
\def\pbar{\bar{p}}
\def\alt{\lesssim}
\def\agt{\gtrsim}
\def\be{\begin{equation}}  
\def\ee{\end{equation}}  
\def\bea{\begin{eqnarray}}  
\def\eea{\end{eqnarray}}  

\title{Collider signals and neutralino dark matter detection in relic-density-consistent models without universality}
\author{Howard Baer$^a$, Azar Mustafayev$^b$, Eun-Kyung Park$^c$ and  Xerxes Tata$^d$\\
$^a$Department of Physics, Florida State University, Tallahassee, 
FL 32306, USA\\
$^b$Dept. of Physics and Astronomy,
University of Kansas, Lawrence, KS 66045, USA\\
$^c$Physikalisches Institut, Universit$\ddot{a}$t Bonn, Nussallee 12, D53115 Bonn, Germany\\
$^d$Dept. of Physics and Astronomy, University of Hawaii,
Honolulu, HI 96822, USA\\
E-mail: \email{baer@hep.fsu.edu},\email{amustaf@ku.edu},
\email{epark@th.physik.uni-bonn.de},\email{tata@phys.hawaii.edu}}

\preprint{\vbox{FSU-HEP-080214, UH-511-1118-08}}

\abstract{We present brief synopses of supersymmetric models
where either the neutralino composition 
or its mass is
adjusted so that thermal relic neutralinos from the Big Bang saturate
the measured abundance of cold dark matter in the universe.
We first review minimal supergravity (mSUGRA), and then examine its
various one-parameter extensions where we
relax the assumed universality of the soft supersymmetry breaking parameters. 
Our goal is to correlate relic-density-allowed parameter choices 
with expected phenomena in direct, indirect and collider 
dark matter search experiments.
For every non-universal model,  
we first provide plots to facilitate the selection of ``dark-matter allowed''
parameter space points, 
and then present salient features of each model with respect to
searches at Tevatron, LHC and ILC and also direct and indirect dark
matter searches. We present benchmark scenarios that allow one to compare
and contrast the non-universal models with one another and with the
paradigm mSUGRA framework. We show that many implications about
sparticle properties and collider signals drawn from the analysis
of the relic density constraint within mSUGRA do not carry over 
to simple one-parameter extensions of the mSUGRA framework.  
We find that in many relic-density-consistent models, 
there is one (or more) detectable
edge in the invariant mass distribution of same-flavour, opposite sign 
dileptons in SUSY cascade decay events at the LHC.
Finally,
we scan the parameter space of these various models, requiring
consistency with the LEP2 constraint on the chargino mass, and with the 
observed relic density, and examine prospects for direct and indirect
dark matter detection. We find that in a large number of cases the
mechanism that causes the early universe neutralino annihilation 
rate to be large (so as to produce the measured relic density) 
also enhances the direct detection rate, and often also the rates for
indirect detection of neutralino dark matter.}


\begin{document}
\tableofcontents
\section{Introduction and framework}
\label{sec:intro}

An abundance of evidence arising from a variety of cosmological
measurements shows that most of the matter in the Universe is not
baryonic, but rather composed of massive neutral stable (or at least
extremely long-lived), weakly (or super-weakly) interacting particles.
Since none of the particles of the Standard Model (SM) have these
properties, the existence of this so-called dark matter (DM) in the
universe provides unequivocal evidence for physics beyond the SM.

Cosmological measurements severely constrain the abundance of DM:
combining the results from the WMAP Collaboration with those from the
Sloan Digital Sky Survey gives~\cite{wmap}
\be
\Omega_{\rm DM}h^2 = 0.111^{+0.011}_{-0.015} \ \ (2\sigma)\;,
\label{eq:relic}
\ee
where $\Omega =\rho/\rho_c$ with $\rho_c$ the closure density of the
Universe, and $h$ is the scaled Hubble parameter, $h=0.73\pm 0.04$.
While the mass density of DM is rather precisely known, the identity of
the DM particle remains a mystery.  One class of candidates -- thermally
produced weakly interacting massive particles (WIMPS) -- are especially
appealing in that they naturally occur in a variety of well-motivated
models, and further, because their relic density today (which can be reliably
computed) is found to automatically have about the observed
magnitude, provided the WIMP mass is of order the weak scale: $m_{\rm
WIMP}\sim 100$~GeV. Of course, we should always bear in mind that, like
visible matter, DM may consist of several components, so that, in
standard Big Bang cosmology, (\ref{eq:relic}) really implies an upper
bound on the density of any single component.

Softly broken supersymmetry (SUSY), with a SUSY breaking scale below
1-2~TeV, is highly motivated for a variety of theoretical as well as
experimental reasons~\cite{wss,dbgrm}.  SUSY models with conserved $R$-parity
include a stable, massive weakly interacting particle -- the lightest
neutralino $\tz_1$ in many models -- which is perhaps the prototypical
thermal WIMP~\cite{haim}.  In any supersymmetric model with a stable
neutralino, the neutralino relic abundance can be reliably calculated
as a function of model parameter space~\cite{griest}. The
result depends inversely on the thermally averaged neutralino-neutralino
annihilation and co-annihilation cross sections, integrated over
time from freeze-out to the present day.  Once the parameters of the
model under study are known to match the measured relic abundance
(\ref{eq:relic})~\cite{wmapcon}, then these select parameter space
regions can be checked for phenomenological constraints from low energy
measurements and from non-observation of new physics signals in the LEP
and Fermilab Tevatron data.  Implications for the on-going Tevatron run
as well as for experiments soon-to-begin at the CERN LHC, and possibly
at a TeV linear electron-positron collider in the future can be
examined. Likewise, predictions can be made for rates of direct
detection of relic neutralinos via scattering on nuclear targets, or
rates for indirect neutralino detection, either via $\nu_\mu$ signals
from neutralino annihilation in the solar core, or via galactic halo
annihilations which can give rise to gamma ray or anti-matter
($\bar{p}$, $e^+$ or $\bar{D}$) signals.

Most analyses of neutralino dark matter have been carried out in the
context of the minimal supergravity model -- mSUGRA~\cite{msugra}
where SUSY breaking, which 
occurs in a hidden sector, is communicated to the observable sector
via gravitational interactions.  
The universality of soft SUSY breaking (SSB) parameters, renormalized at
a scale $Q \simeq M_{\rm GUT}-M_P$ is the hallmark
of this framework. Specifically, one assumes that the mediation mechanism 
induces a common mass parameter $m_0$ 
for all MSSM scalars,  a common gaugino mass $m_{1/2}$ for gauginos,
a common trilinear SSB parameter $A_0$ together with a bilinear Higgs
scalar mass $b$, 
in the effective MSSM Lagrangian, with parameters renormalized at
$Q=M_{\rm GUT}$. It is also assumed that the dimensionful SSB parameters all
have the magnitude of the weak scale.
The large top quark Yukawa coupling drives
the celebrated radiative electroweak symmetry breaking (REWSB)
mechanism, and
automatically leads to the $SU(3)_C\times U(1)_{\rm EM}$ symmetric
vacuum over a significant portion (but not all) of the model parameter space. 
The GUT scale SSB parameter $b$ can be traded for $\tan\beta$, the ratio
of Higgs field vevs, while the magnitude (but not the sign) of 
the superpotential mass parameter $\mu$ is fixed by the observed
value of $M_Z$.  
The mSUGRA model is thus completely specified  by the parameter set
\be
m_0,\ m_{1/2},\ A_0,\ \tan\beta,\ sign(\mu ) ,
\label{eq:msugra}
\ee
along with the value of the top quark mass $m_t$. Except where
explicitly 
mentioned, we fix
$m_t=171.4$~GeV, in accord with recent top mass measurements at the
Fermilab Tevatron~\cite{mtop}.

Unless sparticles are very light ($\sim 100$~GeV) the generic value of
the relic density in mSUGRA (as well as in many other SUSY models) tends
to be well in excess of the observed CDM relic density (\ref{eq:relic}). As a
result, only special regions of the mSUGRA parameter space where the
annihilation rate for neutralinos is enhanced are compatible with the
measured value of the relic density.\footnote{In our analysis, we are
assuming that thermally produced neutralinos in the standard Big Bang
cosmology make up the DM. While it is possible to get around these
assumptions, we feel that an examination of the conceptually simplest
scenario that does not invoke additional hypotheses warrants special
attention.}
In early work on the mSUGRA model, the low $m_0$, low $m_{1/2}$ region
(so-called ``bulk region''), where sparticles are indeed very light, was
favored~\cite{bulk} in that neutralino annihilation into leptons via
light $t-$channel slepton exchange occurred at large rates, leading to
relic densities $\Omega_{\tz_1}h^2\sim 0.3-1$. The rather lower measured
abundance in (\ref{eq:relic}), however, favors even lower values of
$m_0$ and $m_{1/2}$, resulting in considerable tension with the negative
search results from LEP2 for chargino and slepton pair
production. Within the mSUGRA framework, the remaining
relic-density-allowed regions consist of:
\bi
\item The stau-co-annihilation region at low $m_0$ and low-to-moderate
values of $m_{1/2}$ where $m_{\ttau_1}\sim m_{\tz_1}$, so that
neutralinos can co-annihilate with staus~\cite{stau} in the early
universe.

\item The hyperbolic branch/focus point (HB/FP) region~\cite{hb_fp} at
very large $m_0$ values where $|\mu |$ becomes small so that $\tz_1$
becomes a mixed bino-higgsino state. In this case, $\tz_1\tz_1$
annihilation to $WW$, $ZZ$ and $Zh$ via the $\tz_1$ higgsino 
component is enhanced in the early universe.

\item The Higgs funnel region  at large $\tan\beta\sim
50$~\cite{Afunnel}, where $2m_{\tz_1}\sim m_A$, so that neutralinos can
annihilate at an enhanced rate through the (wide at large $\tan\beta$)
$A$ (or $H$) resonance. An $h$-resonance annihilation strip can also
occur at low $m_{1/2}$, where $2m_{\tz_1}\simeq m_h$~\cite{drees_h}.
\item The top squark co-annihilation region at large negative $A_0$
values where $m_{\tz_1}\sim m_{\tst_1}$ so that $\tz_1$ can
co-annihilate with $\tst_1$ particles~\cite{stop}.  
\ei

The regions of mSUGRA parameter
space leading to a neutralino relic density in agreement with
(\ref{eq:relic}) are all near the edges of theoretically 
(or in the case of $h$ resonance annihilation, experimentally) allowed
parameter space, which can lead one to question whether the mSUGRA model
might be {\it disfavored} by the measured neutralino relic abundance. In
this vein, many authors have examined SUGRA-type models but with {\it
non-universal} soft term boundary conditions at $Q\sim M_{\rm GUT}$.  It
is appropriate to note here that unfettered non-universality of soft
terms generically leads to the occurrence of flavor changing processes
at levels far beyond experimental limits \cite{masiero}. With this in
mind, we work in a simplified parameter space wherein there exists
degeneracy or near degeneracy of first and second generation scalar
masses, leading to a suppression of FCNC processes via the super-GIM
mechanism.
At the same time, in order to maintain the obvious success of gauge
coupling unification, we must assume that the correct effective theory
between the weak and GUT scales is the MSSM, or the MSSM augmented by
gauge singlets, or the MSSM together with additional matter in
complete multiplets of $SU(5)$.

With these considerations in mind, we will assume that:
\begin{enumerate}
\item In the interests of minimality, while maintaining the successful
predictions of gauge coupling unification, that the MSSM is the correct
effective field theory between $M_{\rm weak}$ and $M_{\rm GUT}$.

\item The REWSB mechanism leads to an $SU(3)_C\times U(1)_{\rm EM}$
  symmetric ground state; {\it i.e.} 
  electric charge and color gauge symmetries are not spontaneously
  broken. 

\item $CP$ violating phases in the SSB parameters are
  suppressed so that supersymmetric contributions to $CP$ violating
  processes are sufficiently small \cite{susycp}.

\item There is a near-degeneracy of SSB of the first and second
  generation sfermions so that SUSY contributions to flavor-violating
  processes is automatically suppressed. We do allow some non-degeneracy
  between third generation scalars and first or second generation
  scalars.

\item $R$-parity is conserved so that the lightest supersymmetric
  particle (LSP) is stable.

\item The gravitino -- which in models with gravity-mediated SUSY breaking
  naturally has a mass of order $M_{\rm weak}$ -- is not the LSP, which we
  take instead to be the lightest neutralino. 
  For a discussion of the possibility
  that a gravitino LSP is the DM, see Ref.~\cite{superwimp}.
\end{enumerate}

In the spirit of our earlier discussion we need to relax the
theoretically least well-motivated universality assumption that
underlies the mSUGRA framework in a controlled manner (to avoid large
flavor-violating couplings) and 
explore non-minimal SUGRA
models with an expanded parameter space.  We could, for instance, consider
a non-minimal model where we independently vary,
\bea 
M_1,\ M_2,\ M_3 & \ \ \ \ ({\rm gaugino\ masses}), \label{nogaug}\\ 
m_0(1,2)& \ \ \ \ ({\rm common\ first/second\ generation\
SSB\ matter\ scalar\ masses}),\\ 
m_0(3) &\ \ \ \ ({\rm common\ third \ generation\ SSB\ matter\
scalar\ masses}),\\ m_{H_u}^2,\ m_{H_d}^2 &\ \ \ \ ({\rm non-universal\
SSB\ Higgs\ mass \ parameter}),\label{nohiggs}\\ A_t,\ A_b,\ A_\tau &\ \ \ \ ({\rm
non-universal\ third\ gen.}\ A\ {\rm terms}),\\ \tan\beta ,&\\ sign(\mu). & 
\eea
A different but equally reasonable option
may be to require common masses for matter scalars with
the same gauge quantum numbers but allow intra-generation splittings. In
this case, we would have common mass parameters $m_Q^2$, $m_U^2$,
$m_D^2$, $m_L^2$ and $m_E^2$ at $Q=M_{\rm GUT}$. 
In the extreme case, the matter
scalar masses can be further broken down into specific soft term masses
$m_{Q_i}^2,\ m_{U_i}^2,\ m_{D_i}^2,\ m_{L_i}^2,\ m_{E_i}^2$ where
$i=1-3$ for each generation.  
These  options have been explored elsewhere~\cite{models,dterms}, and
will not be discussed further in this paper.

One way to proceed is to perform scans over the much larger non-minimal
SUGRA parameter space and search for solutions which satisfy dark matter
(and also other) constraints.  While this approach has the virtue of
being unbiased in the scanning, it is practically difficult to
implement. Moreover, when a large number of free parameters are varied
simultaneously, it is frequently difficult to draw insights into the
associated dark matter and collider phenomenology that follow.
Instead, many groups~\cite{nmh,nuhm1,nuhm2,mwdm,bwca,m3dm,hm2dm,king1} have
examined the impact of relaxing the underlying universality of the
mSUGRA model in a controlled way, by allowing non-universal parameters
only in one sector of parameter space at a time. For instance, we may
consider the mSUGRA parameter space augmented to accommodate
non-universal gaugino mass parameters~\cite{nonunivgaug} as in
(\ref{nogaug}), or instead extended to allow Higgs boson SSB mass
parameters to be different from matter scalar mass parameters as in
(\ref{nohiggs}), but not both. Other directions in the parameter space
of the non-minimal SUGRA models can be similarly explored.  These
extensions generally require augmenting the mSUGRA space by just one
(sometimes, two) additional parameter that is adjusted to yield
agreement with the observed DM relic density. The phenomenological
implications of the extended model as a function of the remaining mSUGRA
parameters can be readily examined, and directly compared
with the paradigm mSUGRA framework.
This approach has led to new
insights and to exciting new possibilities for collider and dark matter
phenomenology that can be expected in models with non-universal soft
SUSY breaking terms.

Examination of these simple one-parameter extensions of mSUGRA leads to
another important pay-off. Since, as discussed above, analyses of
the relic density constraint in mSUGRA force parameters to be in the 
bulk region,
the stop or stau co-annihilation region, the Higgs funnel region or the
HB/FP region of parameter space, many groups have inferred that at least
one of the following must hold:
\begin{enumerate}
\item Sfermions have masses $\sim 100$~GeV (bulk region), and so must be
  accessible at the LHC.
\item There is at least one charged sparticle close in mass to the LSP,
  so that this should be accessible at the LHC, unless the LSP is so
  heavy that the hard-won gauge hierarchy is again destabilized
  (co-annihilation).
\item The additional Higgs scalars of the MSSM are relatively light with
  $m_A\sim 2m_{\tz_1}$ so that these can be searched for at the LHC,
  which requires large values of $\tan\beta$ where sparticles
  preferentially decay to third generation quarks and leptons 
  (Higgs resonance region).
\item The lightest neutralino has a significant higgsino component,
  which is possible only if $m_0$ is so large that squarks and sleptons
  are essentially inaccessible at the LHC (HB/FP region).
\end{enumerate}
It is imperative, of course, to check the
robustness of these ``predictions'' to minor variations of the
assumptions underlying mSUGRA before
drawing broad conclusions about what the relic density determination
implies for experiments at the LHC, as well as for direct and indirect
detection of DM. Our study of the various extensions of mSUGRA
naturally permits this. 

In this paper, we have two broad goals. The first, in
Sec.~\ref{sec:models}, is to present an overview of a number of
different models, wherein by tuning one additional parameter beyond
those of the mSUGRA model we can match the predicted neutralino relic
abundance with (\ref{eq:relic}).  Models wherein the {\it composition}
of the neutralino is adjusted to obtain the measured relic abundance are
referred to as ``well-tempered neutralino'' models (WTN) \cite{wtn}.  We
also examine several models wherein neutralino or other sparticle {\it
masses} are adjusted to obtain the correct relic abundance of dark
matter.

In each case, we
present {\it i}.) motivation, {\it ii}.) the parameter space, and
selected parameter values that allow the reader to generate spectra and
collider events for the particular model. We also comment on the salient
features of {\it iii}.) collider and {\it iv}.) dark matter search
phenomenology associated with each particular model. For our analysis,
we adopt the SUSY spectrum generator Isasugra, a part of the event
generator ISAJET 7.76~\cite{isajet}. For any given parameter set
satisfying the DM relic density constraint (\ref{eq:relic}), the
sparticle mass spectrum and associated neutralino relic density and
direct and indirect detection rates may be calculated, and associated
collider events may be generated for the Tevatron, LHC or ILC colliders.
To facilitate comparison with the paradigm mSUGRA case, we first
present an updated overview of allowed regions within mSUGRA. In
Sec. \ref{sec:bm}, we present some benchmark cases 
where the spectra and some results
from these various models are explicitly compared with the corresponding
situation in the mSUGRA case.

Our second goal, presented in Sec.~\ref{sec:scan}, is to extract several
general results from scans over the models examined in
Sec.~\ref{sec:models}, to gain an idea of some of the features relevant
to collider and dark matter searches that might be shared by many of
these models. For instance, it has already been pointed out 
that the subset of these models which resolve the dark
matter relic density problem via mixed gaugino/higgsino dark matter
({\it i.e.} models with a WTN)
collectively have neutralino-nucleon direct detection scattering rates
that asymptote around $\sim 10^{-8}$~pb\cite{wtn_dd}, 
putting them within reach of
direct dark matter search experiments currently being mounted, such as
SuperCDMS, LUX, Xenon-100, WARP and mini-CLEAN. Also, models with
non-universality where the composition is tempered to yield the observed
relic density, or where agreement with (\ref{eq:relic}) is obtained via
bino-wino co-annihilation, tend to have a neutralino mass gap
$m_{\tz_2}-m_{\tz_1}$ smaller than $M_Z$, so that three body decay modes
dominate the $\tz_2$ branching fraction. Unless the leptonic branching
fraction of the neutralino happens to be strongly suppressed
\cite{btwino}, this then yields an observable mass edge in the dilepton
mass spectrum at $m(\ell^+\ell^- )=m_{\tz_2}-m_{\tz_1}$, which serves as
a good starting point for sparticle mass reconstruction in gluino and
squark cascade decay events at the CERN LHC~\cite{mlledge}.

We conclude in Sec.~\ref{sec:conclude} with a summary of our results together
with some general comments.

\section{Brief synopses of SUSY models with neutralino dark matter}
\label{sec:models}

\subsection{The mSUGRA model}
\label{ssec:msugra}

We begin by presenting updated results on the allowed parameter space of
the minimal supergravity model.  The mSUGRA model is completely
specified by the parameter set (\ref{eq:msugra}).  To calculate the
sparticle mass spectrum, we use ISAJET 7.76~\cite{isajet}.  The relic
density is evaluated via the IsaReD program~\cite{isared}, which is part
of the IsaTools package.  IsaReD evaluates all $2\to 2$ tree-level
neutralino annihilation and co-annihilation processes and implements
relativistic thermal averaging in the relic density calculation.

For our first results, we show in Fig.~\ref{fig:msug} the $m_0\ vs.\
m_{1/2}$ plane for parameters $A_0=0$, $\tan\beta =10$ and $\mu >0$ with
{\it a}) $m_t=170$ GeV, {\it b}) $m_t=171.4$~GeV and {\it c})
$m_t=175$~GeV. The red-shaded regions on the left are excluded because
$\ttau_1$ becomes the LSP, while the red-shaded regions on the lower
right are excluded due to a failure to meet the EWSB minimization
conditions. The blue-shaded region is theoretically allowed, but is
experimentally excluded by LEP2 searches for chargino pair production
where we require $m_{\tw_1}>103.5$~GeV \cite{lepwino}. The negative
results of Higgs boson searches at LEP2~\cite{lep2_higgs} require that
the SM Higgs boson is heavier than 114.1~GeV.  This limit can be
translated to a lower limit on the MSSM Higgs boson mass. While $h
\simeq H_{\rm SM}$ if $m_A$ is large, in general the bound on $m_h$
depends on MSSM parameters, including $CP$ violating phases that we have
ignored in our analysis. The evaluation of $m_h$ is also uncertain to
about $\sim 3$~GeV  due to missing two-loop
corrections~\cite{Allanach:2004rh}.  For these reasons, we do not
include any bound on $m_h$ in the LEP2-excluded blue region, but only
show the boundary of the region $m_h \leq 110$~GeV by the magenta
contour (lower-left) in the figure.  The green regions have a neutralino relic
density in accord with (\ref{eq:relic}): $0.094<\Omega_{\tz_1}h^2
<0.129$. In the yellow regions, however, $\Omega_{\tz_1}h^2 <0.094$, so
that an additional component of dark matter particles is necessary to
saturate the observed DM relic density.  The remaining unshaded regions
all have $\Omega_{\tz_1}h^2 >0.129$, {\it i.e.}, they give rise to {\it
too much} dark matter: thus, these are excluded in standard Big Bang
cosmology.  We also show contours of gluino and first generation squark
masses; these contours hardly change under variation of $A_0$,
$\tan\beta$ and $sign(\mu )$, except at the level of one loop
corrections and $D$-term contributions to their masses.
\FIGURE[tbh]{
\epsfig{file=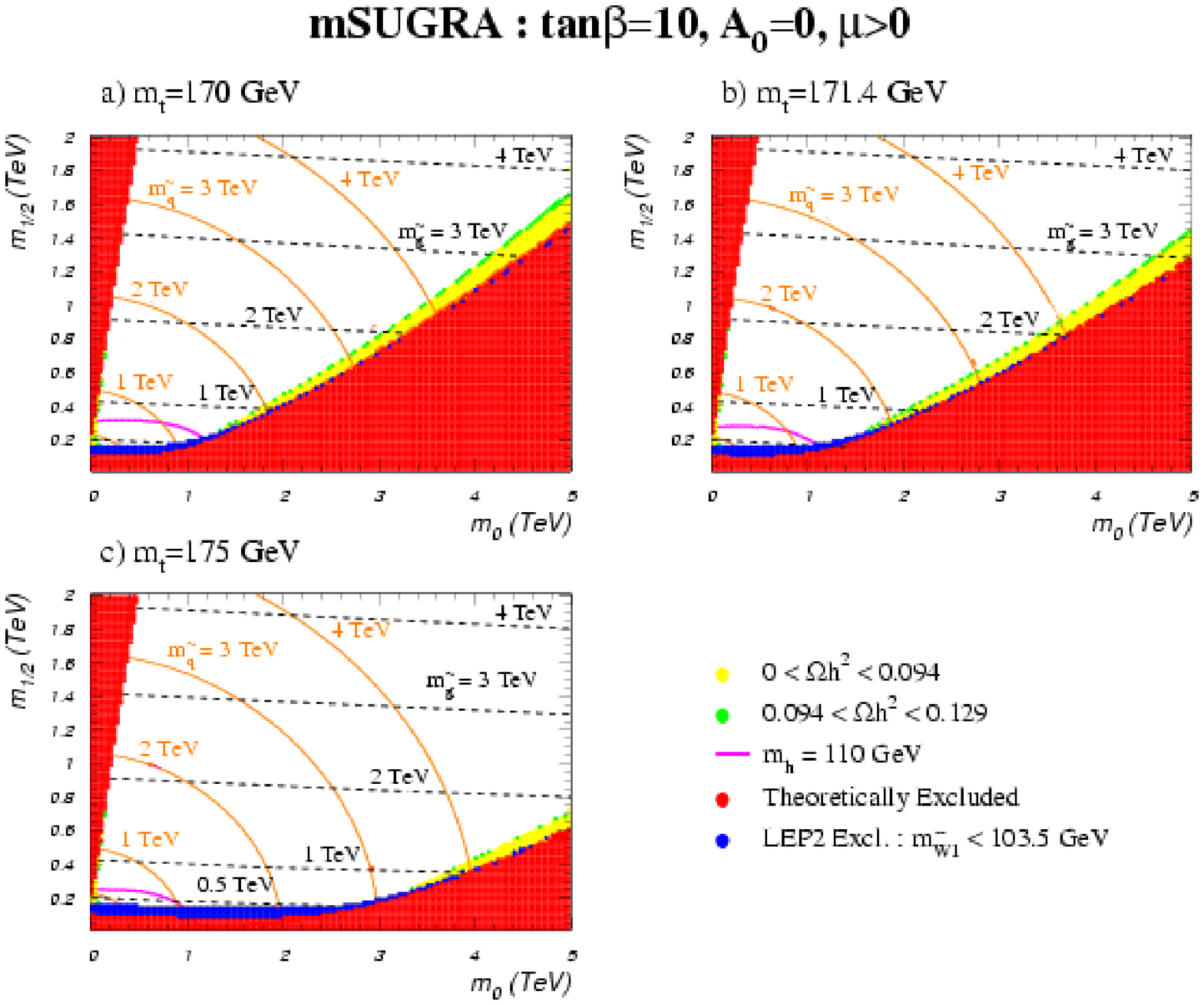,width=12cm,angle=0} 
\caption{\label{fig:msug} A plot of the $m_0\ vs.\ m_{1/2}$ plane in
mSUGRA for $A_0=0$, $\tan\beta =10$ with $\mu >0$ and {\it a})
$m_t=170$~GeV, {\it b}) $m_t=171.4$~GeV and {\it c}) $m_t=175$~GeV.  The
red-shaded regions are excluded because electroweak symmetry is not
correctly broken, or because the LSP is
charged. Blue regions are excluded by direct SUSY searches at LEP2.
Yellow and green shaded regions are WMAP-allowed, while white regions
are excluded owing to $\Omega_{\tz_1}h^2>0.129$.  Also shown are
gluino and first generation squark mass contours, as well as a magenta
contour below which $m_h\leq 110$~GeV.}}

The hard-to-see green/yellow region adjacent to the $\ttau_1$-$\tz_1$
region shows up as a very narrow sliver where the $\ttau_1-\tz_1$ mass
gap is small enough so that stau co-annihilation occurs at a large
rate. This region in fact appears jagged only due to the resolution of
our parameter space scans.  One can also see the HB/FP region -- where
$|\mu |$ becomes comparable to the $SU(2)$ and $U(1)$ gaugino masses and
the $\tz_1$ becomes mixed higgsino dark matter -- adjacent to the EWSB
forbidden region as the wider green/yellow shaded region at large $m_0$,
starting at $m_{1/2}\sim 300$~GeV, which corresponds to the turn-on
point for $\tz_1\tz_1\to W^+W^-$; for lower $m_{1/2}$ values, this
annihilation channel is closed, and the neutralino annihilation rate
(via $Z^*$ exchange) generally becomes too small to bring the relic
density into accord with (\ref{eq:relic}).

While these three frames for the different $m_t$ values are
qualitatively similar, the main effect of $m_t$ variation shows up in
the location of the EWSB excluded region, and hence the location of the
adjacent HB/FP region: on the low side of the allowed $m_t$ range, the
HB/FP region moves to $m_0$ values as low as 1.5~TeV, while at the high 
end of this range, the HB/FP region only starts when $m_0 \agt 3$~TeV~\cite{tadas_tev}.

In Fig.~\ref{fig:sugpmu}, we show the $m_0\ vs.\ m_{1/2}$ plane for
$A_0=0$, $\mu >0$, $m_t=171.4$~GeV, and for six different values of
$\tan\beta$. We see that for $\tan\beta =10$ only the stau
co-annihilation and HB/FP regions are DM-allowed. As $\tan\beta$ is increased,
more and more parameter space comes into accord with
(\ref{eq:relic}). Already for $\tan\beta =45$, a small DM-allowed
region appears at low $m_0$ and low $m_{1/2}$. The reason is that as
$\tan\beta$ grows, the $b$ and $\tau$ Yukawa couplings become large,
causing $m_A$ to drop. Thus, $\tz_1\tz_1\to A^*\to b\bar{b},\
\tau^+\tau^-$ becomes more and more important, even if one is not
``right on
the $A$ resonance''~\cite{drees_A}.  The low $m_0,\ m_{1/2}$ allowed
region grows even more at $\tan\beta =50$. At $\tan\beta =52$, the $A$-
funnel annihilation region has just come into view on the left edge of
parameter space. By $\tan\beta =54$, the $A$-funnel is extremely broad
due to the large $A$ width: $\Gamma_A\sim 10$~GeV (50~GeV) for low
(high) $m_{1/2}$. We notice at $\tan\beta =55$, a small red-shaded wedge
invades the plot at low $m_0$ and $m_{1/2}$. In this region, the value
of $m_h^2<0$, signaling collapse of EWSB. For somewhat higher
$\tan\beta$ values, the entire parameter space collapses.
\FIGURE[tbh]{
\epsfig{file=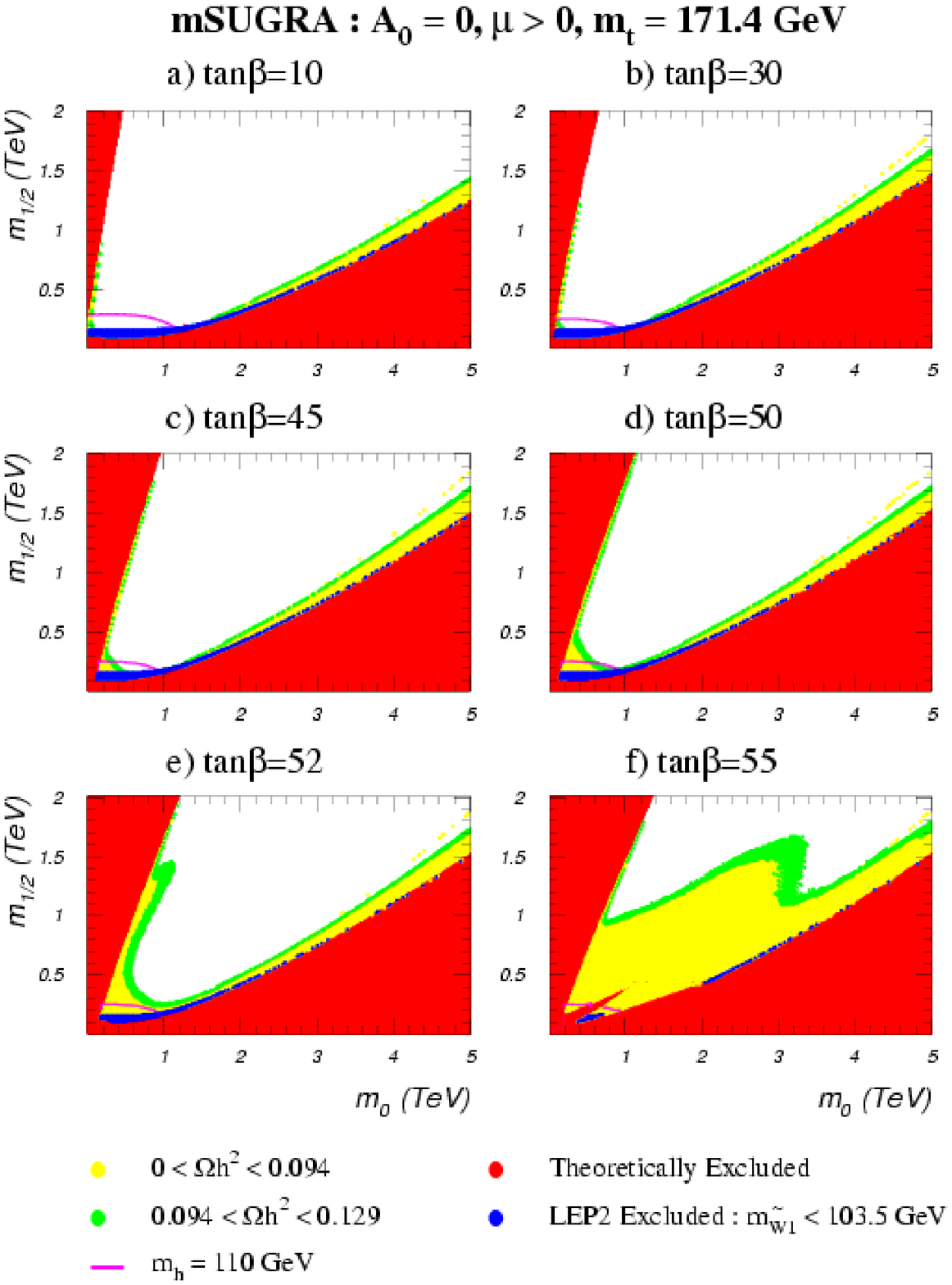,width=12cm,angle=0} 
\caption{\label{fig:sugpmu} 
A plot of the $m_0\ vs.\ m_{1/2}$ plane in mSUGRA for $A_0=0$
and various values of $\tan\beta$, with $\mu >0$ and $m_t=171.4$~GeV.
The red-shaded regions are excluded because electroweak symmetry is not
correctly broken, or because the 
LSP is charged. Blue regions are excluded by direct SUSY searches at LEP2.
Yellow and green shaded regions are WMAP-allowed, while white 
regions are excluded owing to $\Omega_{\tz_1}h^2>0.129$. Below the
magenta contour in each frame, $m_h< 110$~GeV. 
}}

In Fig.~\ref{fig:sugnmu}, we show the same $m_0\ vs.\ m_{1/2}$ planes as
in Fig.~\ref{fig:sugpmu} for multiple $\tan\beta$ values, but this time
for $\mu <0$.  This sign of $\mu$ is disfavored by the Muon $g-2$
Collaboration measurements of anomalous magnetic moment
$(g-2)_\mu$ at low $m_0$ and low $m_{1/2}$~\cite{pdb}. At high $m_0$ and
$m_{1/2}$ values, sparticle contributions to the muon QED vertex
decouple, and the deviation from SM predictions is tiny for either sign
of $\mu$.  We also see that the $A$-funnel arises in the $m_0\
vs.\ m_{1/2}$ plane at somewhat lower $\tan\beta$ values. The $A$-funnel
is actually narrower than in the $\mu >0$ case, in part because the $A$
width is narrower. We also see a bulge of incorrect  EWSB beginning
already at $\tan\beta =45$, and growing so as to engulf nearly all
parameter space by $\tan\beta =55$.
\FIGURE[tbh]{
\epsfig{file=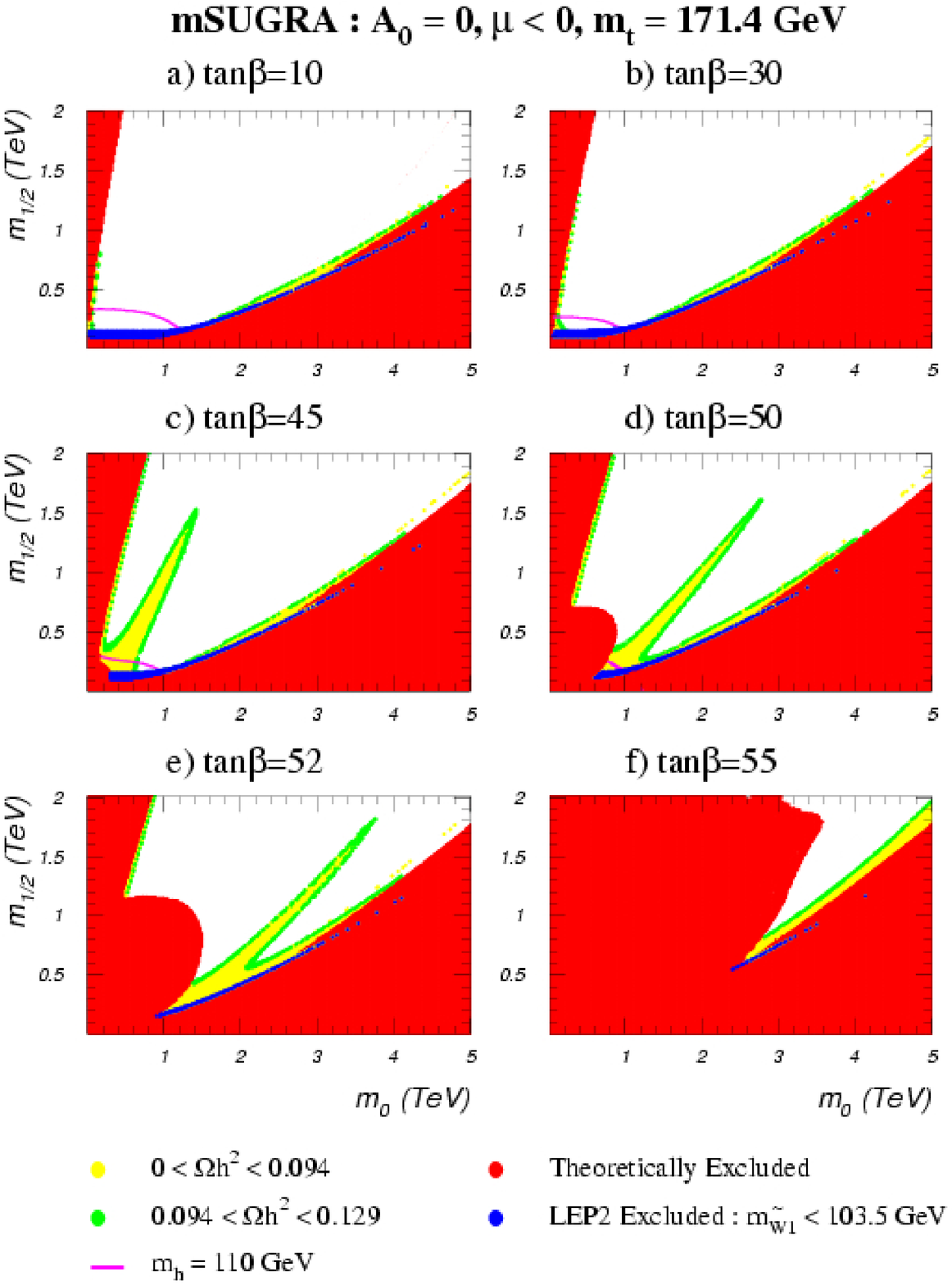,width=12cm,angle=0} 
\caption{\label{fig:sugnmu} 
A plot of the $m_0\ vs.\ m_{1/2}$ plane in mSUGRA for $A_0=0$
and various values of $\tan\beta$, with $\mu <0$ and $m_t=171.4$~GeV.
The red-shaded regions are excluded by lack of  correct EWSB or by the 
presence of
a charged LSP. Blue regions are excluded by direct SUSY searches at LEP2.
Yellow and green shaded regions are WMAP-allowed, while white 
regions are excluded owing to $\Omega_{\tz_1}h^2>0.129$. Below the
magenta contour in frames {\it a})-{\it d}) $m_h < 110$~GeV , while
$m_h> 110$~GeV all over the LEP2 allowed region in frames {\it e}) and
{\it f}).
}}

We see from Fig.~\ref{fig:sugnmu} that the Higgs funnel moves to larger
values of $m_0$ as we increase $\tan\beta$. To understand this, we take
a point $(m_0, m_{1/2})$ in the funnel where $m_A \simeq 2m_{\tz_1}$,
and examine what would happen if we increase $\tan\beta$ keeping the
other parameters fixed. We first remark that because $\tz_1$ is
essentially a bino, $m_{\tz_1} \simeq M_1$ remains essentially
unaltered. The behaviour of $m_A^2 \sim m_{H_d}^2 - m_{H_u}^2$ is
governed by how the evolution of the Higgs scalar SSB parameters is
altered by the increase in $\tan\beta$. 
In the case where the evolution of
$m_{H_u}^2$ ($m_{H_d}^2$) is dominated by the term $3f_t^2 X_t$ ($3f_b^2
X_b$)\footnote{Here, $X_t=m_{Q_3}^2+m_{\tst_R}^2+m_{H_u}^2+A_t^2$
and $X_b=m_{Q_3}^2+m_{\tb_R}^2+m_{H_d}^2+A_b^2$.} in their
one-loop RGE \cite{wss},\footnote{This will be the case
as long as squark masses and $A$-parameters are not simultaneously very
small, since for the large values of $\tan\beta$ where the Higgs funnel
occurs, the Yukawa couplings are typically larger than the electroweak
gauge couplings.}  we see that -- since $f_b$ increases with $\tan\beta$
while $f_t$ is left essentially unaltered -- the weak scale values of
$m_{H_d}^2$ and $m_{H_u}^2$ move closer to each other so that $m_A^2$ is
reduced. As a result, for a larger $\tan\beta$ value the point will move
out of the $A$-funnel region (modulo effects of the width of $A$)
because $m_A$ becomes smaller than $2m_{\tz_1}$. To return to the
$A$-funnel region, we must have a larger value of $X_t$ to also move
$m_{H_u}^2$ to more negative values compensating for the reduction of
$m_A$ with the increase in $\tan\beta$. For a fixed value of $A_0$, this
means increasing $m_0$, explaining why the Higgs funnel moves to the
right as we increase $\tan\beta$ in Fig.~\ref{fig:sugnmu}. A
qualitatively similar behaviour can also be seen in
Fig.~\ref{fig:sugpmu}, but just in the last two frames since for $\mu
>0$ the Higgs funnel does not appear for the other choices of
$\tan\beta$.

A similar analysis can also help us to understand how the location of
the Higgs funnel in the $m_0-m_{1/2}$ plane depends on the choice of
$m_t$. For larger values of $m_t$, and thus of $f_t$, $m_{H_u}^2$
evolves to more negative values while the evolution of $m_{H_d}^2$
remains essentially unaltered. Thus we move out of the $A$-funnel
because now $m_A$ becomes {\em larger} than $2m_{\tz_1}$, and to return
to the $A$-funnel we must now reduce $m_0$, so that the $A$-funnel (if
it occurs) moves to smaller values of $m_0$ as $m_t$ is increased. 
Although we do not show figures here, we
have verified that this is indeed the case for representative slices
of the $m_0-m_{1/2}$ plane.

Up to now in all our plots we have assumed $A_0=0$. By changing the $A_0$
parameter, one is altering the intra-generation  mixing between third generation
sfermions, especially the top squarks. This mixing also reduces $m_{\tst_1}$.
In Fig.~\ref{fig:sugstop}, we
show the $m_0\ vs.\ m_{1/2}$ plane for $\tan\beta =10$, $A_0=-2$~TeV and
$\mu >0$. In this case, a forbidden region appears at low
$m_0$ and $m_{1/2}$, where the $\tst_1$ becomes the LSP. Along the
edge of this region, a yellow/green band appears: the stop
co-annihilation region, where the $\tst_1-\tz_1$ mass gap is positive
but quite small, so that $\tz_1$ can co-annihilate against $\tst_1$ in
the early universe, thus giving a relic density  matching 
(\ref{eq:relic}).  We also see  an $h$-annihilation strip
at low $m_{1/2}$ and $m_0\sim 2.75$~TeV, where $2m_{\tz_1} \simeq m_h$
and neutralino annihilations into SM fermions are resonantly enhanced.

\FIGURE[tbh]{
\epsfig{file=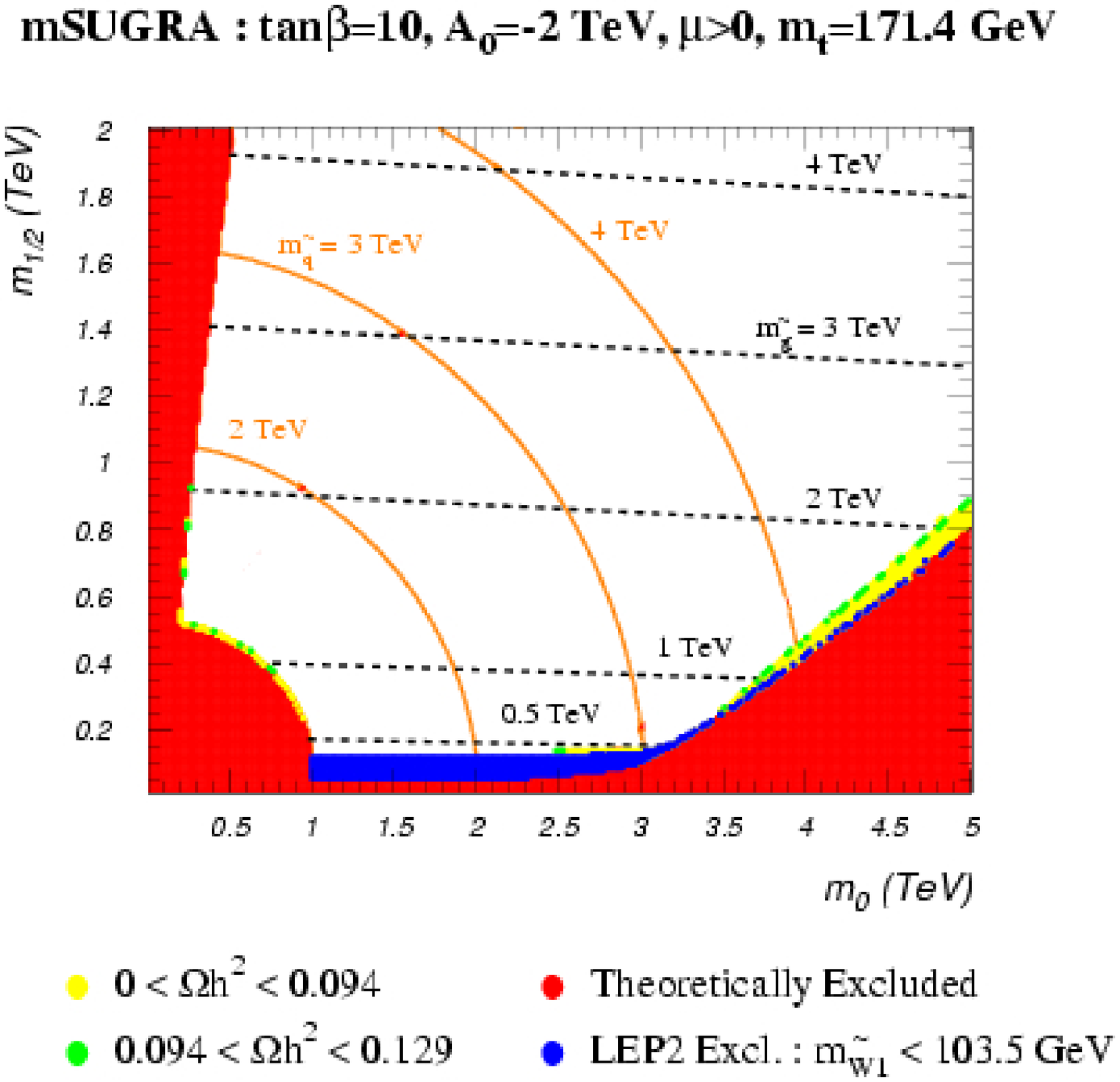,width=12cm,angle=0} 
\caption{\label{fig:sugstop} 
A plot of the $m_0\ vs.\ m_{1/2}$ plane in mSUGRA for $A_0=-2$~TeV,
$\tan\beta =10$ with $\mu >0$ and $m_t=171.4$~GeV.
The red-shaded regions are excluded by lack of correct EWSB or by presence of
a charged LSP. Blue regions are excluded by direct SUSY searches at LEP2.
Yellow and green shaded regions are WMAP-allowed, while white 
regions are excluded owing to $\Omega_{\tz_1}h^2>0.129$.
This plot includes a top-squark co-annihilation region adjacent to the 
excluded bulge at low $m_0$ and low $m_{1/2}$
as well as an $h-$annihilation strip at low $m_{1/2}$ and $m_0\sim 2.75$~TeV.
Throughout the LEP2 allowed region, $m_h > 114$~GeV.
}}

\subsection{Models with scalar mass non-universality}

\subsubsection{Generational non-universality: normal scalar mass hierarchy}
\label{ssec:nmh}

{\it Motivation:} The normal scalar mass hierarchy model (NMH) examines
the effect of generational non-universality in the SSB sfermion
mass parameters~\cite{auto,nmh,king}.  While constraints from $K-\bar{K}$ mass
difference restrict first and second generation scalar masses to be
nearly universal, the constraints arising from $B-\bar{B}$ mixing are
much less strict, and some non-universality of third generation matter
scalars compared to first/second generation matter scalars can be
allowed. In fact, it can be argued that the data actually favor such a
case: we know that the measured value of $BF(b\to s\gamma )$ is in
rather close accord with SM predictions, suggesting that the third
generation sparticles that enter the $b\to s\gamma$ loop diagrams are
rather heavy-- of order the TeV scale. Meanwhile, the $2-3\sigma$
discrepancy of the measured $(g-2)_\mu$ against the SM prediction seems
to favor rather light, sub-TeV scale smuon and muon sneutrino masses. A
normal scalar mass hierarchy at the GUT scale with $m_0(1,2)\ll m_0(3)$
can reconcile the apparent tension between $BF(b\to s\gamma )$ and
$(g-2)_\mu$ in SUSY models, and give a relic density in accord with
(\ref{eq:relic}).

{\it Parameter space:} The parameter space of the NMH model is given by
\be m_0(1,2),\ m_0,\ m_{1/2},\ A_0,\ \tan\beta ,\ sign(\mu ) \ee where
$m_0(1,2)$ is the common GUT scale matter scalar mass parameter for
first/second generation scalars at $M_{\rm GUT}$, while
$m_0(3)=m_{H_u}=m_{H_d}\equiv m_0$ defines the remaining scalar mass
parameter, again at $Q=M_{\rm GUT}$.  The Higgs scalar masses are taken here
to be degenerate with $m_0(3)$, but could be independent as well.

If we begin with a generic point in mSUGRA parameter space where
$\Omega_{\tz_1}h^2\gg 0.129$, and then dial $m_0(1,2)$ to successively
lower values, the first/second generation slepton masses fall until they
are low enough that bulk neutralino annihilation via light sleptons
and/or neutralino-slepton co-annihilation acts to reduce the relic
density to WMAP-allowed levels. As a result of lowering $m_0(1,2)$,
sleptons tend to be quite light. However, first/second generation
squark masses are typically pulled up via RG running into the several
hundred GeV to a TeV range.

In Fig.~\ref{fig:nmh}, we show the ratio $m_0(1,2)/m_0$ needed to reduce
the relic density to the WMAP allowed value versus $m_0$. We show
results for three choices of $m_{1/2}:\ 200,\ 300$ and 500~GeV. Solid
curves are for $\tan\beta =10$ while dashed curves are for $\tan\beta
=40$. We fix $A_0=0$ and take $\mu >0$.  At quite low $m_0$, we are
already in the stau co-annihilation region (bulk region for
$m_{1/2}=200$~GeV), so little or no reduction of $m_0(1,2)$ is
needed. The curves terminate at the left because for still smaller
values of $m_0$, we hit the stau LSP region. As $m_0$ increases, a large
reduction is needed to match the measured relic density, where ratios
$m_0(1,2)/m_0\sim 0.1$ are common. At very large $m_0$, no reduction in
$m_0(1,2)/m_0$ is again necessary as we enter the HB/FP region.
\FIGURE[tbh]{
\epsfig{file=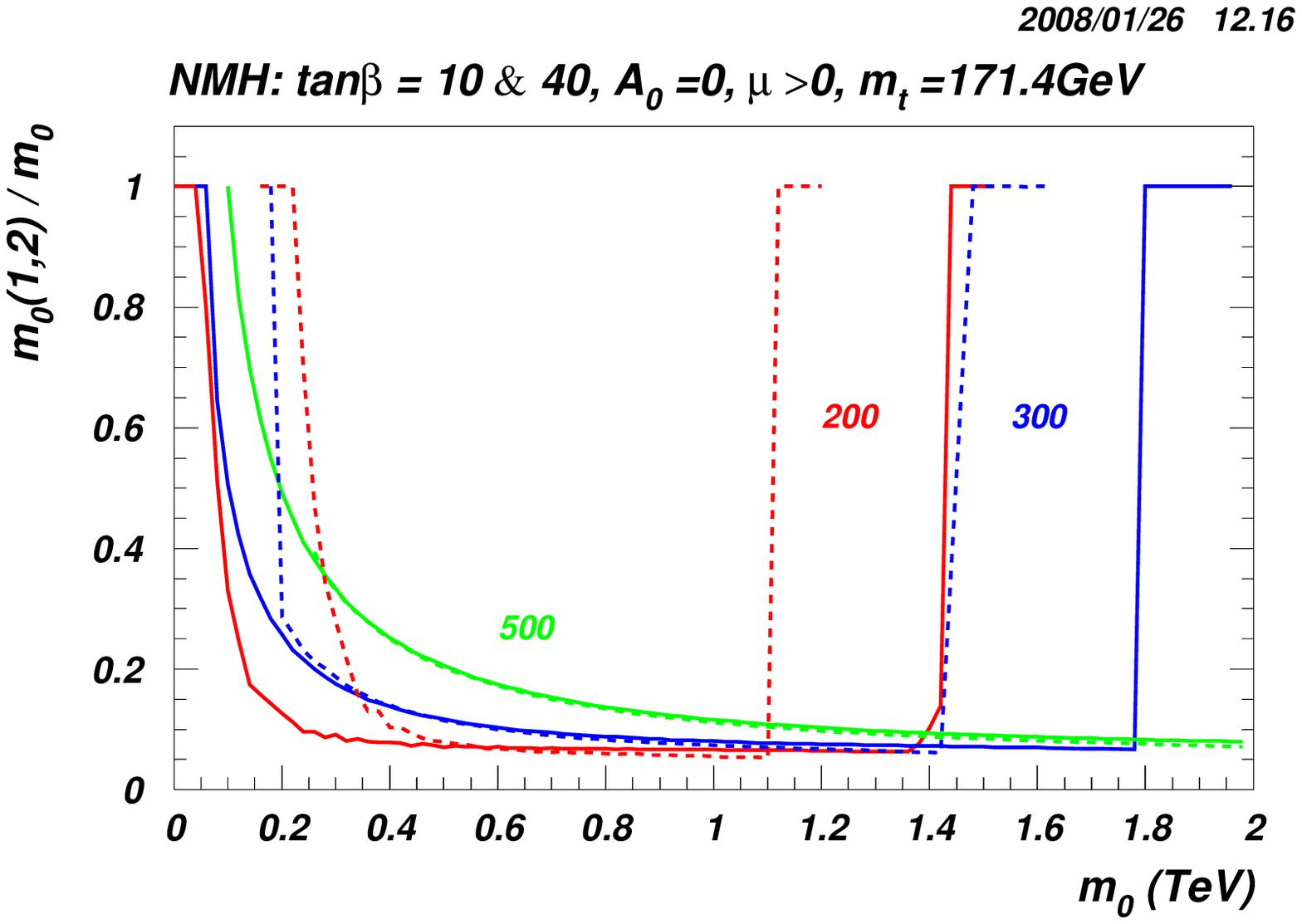,width=10cm,angle=0} 
\caption{\label{fig:nmh} The value of $m_0(1,2)/m_0$ needed to
bring various mSUGRA points into accord with the measured relic density
versus $m_0$, for $A_0=0$, $\mu>0$, and $\tan\beta =10$ (solid lines) and
40~(dashed lines) in the NMH scenario.  The three sets of curves correspond to
$m_{1/2}$ values of 200, 300 and 500~GeV.}}

{\it Implications for collider searches:} The very small first/second
generation slepton masses in the NMH model imply that these will
likely be directly accessible to LHC searches~\cite{sleptons}.  Even if
this is not the case, branching ratios for chargino and neutralino
decays to leptons via 2 or 3 body modes will be considerably enhanced,
leading to SUSY cascade decay events at the LHC that are much richer in
hard, isolated leptons than would be expected in the mSUGRA
model. Meanwhile, selectron and smuon pair production -- but not stau
pair production -- would likely be accessible to ILC searches.

{\it Implications for DM searches:} While neutralino annihilation in the
early universe is enhanced via light slepton exchange or slepton
co-annihilation, squarks remain relatively heavy, and the neutralino is
largely bino-like. Thus, both direct and indirect DM search predictions
will be qualitatively similar to those generated in the mSUGRA model
with $m_0(3)\sim m_0$. For the case that we study in
Table~\ref{tab:nusm} below, we have explicitly checked that even the
indirect detection signals at IceCube and  Pamela detectors remain small
despite the reduced masses of first/second generation sneutrinos and
charged sleptons.

\subsubsection{Non-universal Higgs mass: one extra parameter case}
\label{ssec:nuhm1}

{\it Motivation:} In supersymmetric grand unified theories based upon
the gauge group $SO(10)$, the matter superfields of a single generation
are contained in a 16-dimensional spinor representation of $SO(10)$,
$\psi_{16}$, which includes, in addition, a SM gauge singlet
right-handed neutrino superfield. The Higgs superfields can be most
simply accommodated in the fundamental 10-dimensional representation $\phi_{10}$.  
It is natural to expect that different multiplets would receive different soft
masses at the GUT scale.  Even if the soft masses 
for Higgs and matter scalars
were common at some scale near $M_{P}$, RG running effects in
the $SO(10)$ theory would split the soft terms at $M_{\rm
GUT}$ (see Ref.~\cite{models} for explicit examples of soft
term running in $SO(10)$ SUSY GUTs).

{\it Parameter space:} In the non-universal Higgs model with one
additional parameter (NUHM1)~\cite{nuhm1}, the matter scalars receive a
common squared mass parameter $m_0^2$ at $Q=M_{\rm GUT}$, while both
$SU(2)$ Higgs scalar doublets $H_u$ and $H_d$ acquire 
equal values for their SSB parameters, that are different from $m_0^2$. Note
that $m^2_{H_u}(=m^2_{H_d})$ is just a parameter, not a physical mass
squared,
so its value can be either positive or negative (as can
$m_0^2$~\cite{feng}). The parameter space is thus given by \be m_0,\
\delta_{\phi},\ m_{1/2},\ A_0,\ \tan\beta ,\ sign(\mu ) , \ee where
$m_{\phi}=m_0(1+\delta_\phi )$ and $m^2_{H_u}=m^2_{H_d} \equiv
sign(m_{\phi})\cdot |m_{\phi}|^2$ at the GUT scale.

Given any parameter space point in the mSUGRA model with typically too
high a relic density, one can always {\it increase} $m_\phi$ beyond its
mSUGRA value of $m_0$. A large value of $m_\phi> m_0$ implies via the
RGEs and EWSB minimization conditions a smaller weak scale value of
$|\mu |$,
and thus the possibility of {\it mixed higgsino dark matter} with a
WMAP-allowed relic density (as in the mSUGRA HB/FP region), even though
$m_0$ is not large. Alternatively, if $m_\phi$ is negative,
the value of $|\mu |$ increases, but $m_A$ decreases: 
thus, by dialing $m_\phi$ to a sufficiently negative value,
we can get $A$-funnel annihilation, {\em for any value of  $\tan\beta$}. 

In Fig.~\ref{fig:nuhm1}, we show curves which illustrate the value of
$\delta_\phi$ needed to move the mSUGRA relic density prediction into
accord with (\ref{eq:relic}), either by raising $\delta_\phi$ or
equivalently, $m_\phi$, as in frame {\it a}), or by lowering
$\delta_\phi$ (and hence $m_\phi$), to negative values as in frame
{\it b}).  We show curves versus $m_0$ for $m_{1/2}=200,\ 300$ and
500~GeV, and for $A_0=0$, $\tan\beta =10$ and 40 and $\mu >0$.  Curves
terminate at both ends where we reach an end of the scanned $m_0$ space
or hit a forbidden region.  At the low $m_0$ end, no dialing of
$\delta_\phi$ is needed, since we are in the narrow stau co-annihilation
region, or for $m_{1/2}=200$~GeV, in the bulk region.
When we move to larger $m_0$, we eventually leave that region
and large $|\delta_\phi|$ values become necessary to lower $|\mu |$ (or
$m_A$). As we continue to increase $m_0$ in the upper frame, smaller
values of $\delta_\phi$ are necessary because of another assisting effect--
the downward push of higgs mass-squared parameters from the top Yukawa
coupling -- is getting stronger. We smoothly reach the HB/FP region of
mSUGRA where no dialing is required. The behaviour in the lower frame,
where we adjust $\delta_\phi$ so as to hit the Higgs-funnel region is
qualitatively different at larger values of $m_0$: once we hit the HB/FP
region, there is no need to have $\delta_\phi$ different from zero. The
jump in the curves reflects the rapidity with which this region
is reached. 

\FIGURE[tbh]{
\epsfig{file=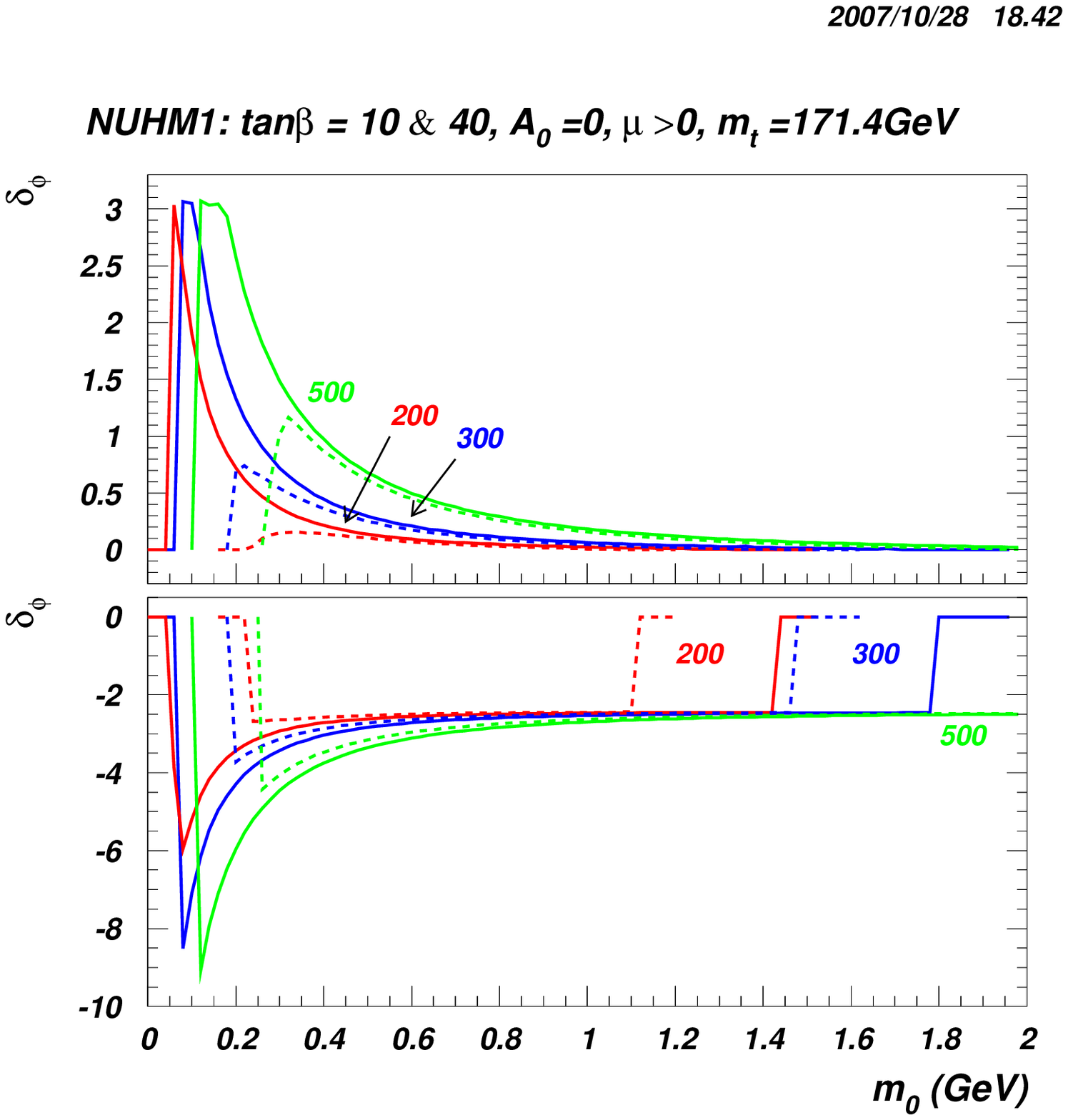,width=10cm,angle=0} 
\caption{\label{fig:nuhm1} The values of $\delta_\phi$  needed to
bring various mSUGRA points into accord with the measured relic density
versus $m_0$ for $A_0=0$ and $\tan\beta =10$ (solid lines) and 40
(dashed lines) in the NUHM1 model.  The curves correspond to
$m_{1/2}$ values of 200, 300 and 500~GeV.}}

{\it Implications for collider searches:}

In the case with $m_\phi > m_0$ where we have MHDM (even though $m_0$
can be much lower than its typical HB/FP value in mSUGRA), the
low value of $|\mu |$ implies that {\it all} the charginos and
neutralinos will be quite light. Thus, they are more likely to be seen
either via direct -ino pair production at the LHC, or to be produced at
large rates in gluino and squark cascade decays.  In general, for small
$|\mu|$, $\tg$ and $\tq$ cascade decay patterns become much more complex
because many squark and gluino decay chains that are normally suppressed
in the mSUGRA case become relevant. In addition, for small $|\mu |$,
there is a smaller $\tz_2-\tz_1$ mass gap, so $\tz_2$ 2-body ``spoiler
decays'' $\tz_2\to\tz_1 Z$ and $\tz_2 \to \tz_1 h$ are likely closed, and
leptonic 3-body decays $\tz_2\to\tz_1\ell\bar{\ell}$ occur at observable
levels. The dilepton mass edge from these three-body decays frequently
serves as the starting point for sparticle mass reconstruction in SUSY
cascade decays at the LHC~\cite{mlledge}. Since $|\mu|$ is small, it is
also possible that $\tz_3$ also only decays via three-body channels. In
this case, the dilepton mass distribution will contain three mass edges
(though these may not all be observable), and its shape may provide further
information about the nature of the neutralinos.

In the case where $m_\phi <0$ so that $m_A\sim 2m_{\tz_1}$, 
$A$ as well as the other heavier Higgs
bosons $H$ and $H^\pm$ are much lighter than expected in mSUGRA, even at
low-to-moderate $\tan\beta$ values. In this case -- for a fixed value of
$\tan\beta$ -- direct detection of the heavier Higgs states $A,\ H$ and
$H^\pm$ at the LHC is more likely than in mSUGRA, and further, these
states may also be produced in the gluino and squark decay chains
via the decays of secondary chargino and neutralinos~\cite{bisset}.

{\it Implications for DM searches:}

For the case of $m_\phi \gg m_0$ with small $|\mu |$ and MHDM, the
enhanced higgsino component of the $\tz_1$ leads to {\it both} enhanced
direct and indirect DM detection rates compared to mSUGRA. This case has
excellent detection prospects in the next generation of detectors such
as XENON-100, LUX or mini-CLEAN.

For the case where $m_\phi <0$ with $m_A\sim 2m_{\tz_1}$, the $\tz_1$
remains nearly pure bino, so direct detection rates and indirect
detection via neutralino annihilation to neutrinos in the solar core
remain low, at values typical of mSUGRA models. However, indirect
$\tz_1$ detection via halo annihilations to gamma rays or anti-matter
are all enhanced relative to mSUGRA (but not always to observable
levels), since the halo neutralinos can still annihilate through the
$s$-channel pseudoscalar resonance~\cite{bo}.

\subsubsection{Non-universal Higgs mass: two extra parameters case}
\label{ssec:nuhm2}

{\it Motivation:} In SUSY GUT models based upon the gauge group $SU(5)$,
each of the MSSM Higgs superfields lives in {\it different}
representations of the gauge group: $H_u\in {\cal H}_5$, while $H_d\in
{\cal H}_{5^*}$.  The two-extra-parameters NUHM model
(NUHM2)~\cite{nuhm2} assumes independent Higgs field soft masses
$m_{H_u}^2$ and $m_{H_d}^2$. The simplest assumption for the matter
scalars is that they all acquire a common GUT scale mass $m_0$, although
they also would exist in separate $5^*$ and $10$ dimensional
representations under $SU(5)$.

{\it Parameter space:} One form of parameter space for the NUHM2 model
is 
\be 
m_0,\ m_{H_u}^2,\ m_{H_d}^2,\ m_{1/2},\ A_0,\ \tan\beta ,\
sign(\mu )\ \ \ {\rm (NUHM2')} .  
\ee 
However, the EWSB minimization
conditions allow the new GUT scale parameters $m_{H_u}^2$ and
$m_{H_d}^2$ to be traded for the weak scale parameters $|\mu |$ and
$m_A$ which are frequently easier to work with for phenomenological
analyses: 
\be 
m_0,\ \mu ,\ m_A,\ m_{1/2},\ A_0,\ \tan\beta ,\ sign(\mu
)\ \ \ {\rm (NUHM2)} .  
\ee 
Both forms of parameter space are allowed in
Isajet spectra and event generation.  In addition, Blazek {\it et
al.}~\cite{bdr} adopt a Higgs SSB parameterization wherein the Higgs
soft masses are split evenly about a common mass which most generally is
an independent parameter, $m_{10}$. 
Their parameterization is given by: 
\be
m_{H_{u,d}}^2=m_{10}^2\ (1\mp \delta_H), 
\ee 
where $\delta_H$ is
dimensionless and can take either positive or negative values. If we
choose $m_{10}=m_0$, we obtain a one-parameter extension that we refer
to as the Higgs-splitting (HS) model.

In Fig.~\ref{fig:hs}, we illustrate the values of $\delta_H$ needed to
move the mSUGRA relic density prediction into accord with
(\ref{eq:relic}) by lowering $\delta_H$ to negative values. This case
gives rise to models with low $\mu$ {\it and} low $m_A$ in the HS model.
We show curves of $\delta_H<0$ needed to pull the mSUGRA relic density
into accord with Eq.~\ref{eq:relic} versus $m_0$ for $m_{1/2}=200,\ 300$
and 500~GeV, and for $\tan\beta =10$ and 40, with $A_0=0$ and $\mu >0$.
Here the situation is similar to the NUHM1 case in that less dialing is
required from larger $m_0$ due to the increasing top Yukawa coupling
effect. As in the previous figure, less dialing is needed for the larger
value of $\tan\beta$. We also mention that
if instead $\delta_H$ is raised to large positive values
(not shown in the figure), then instead one enters a WMAP-allowed region
via $\tell_L/\tnu$ or $\tu_R/\tc_R$ co-annihilation as discussed in
Ref.\cite{nuhm2}.
\FIGURE[tbh]{
\epsfig{file=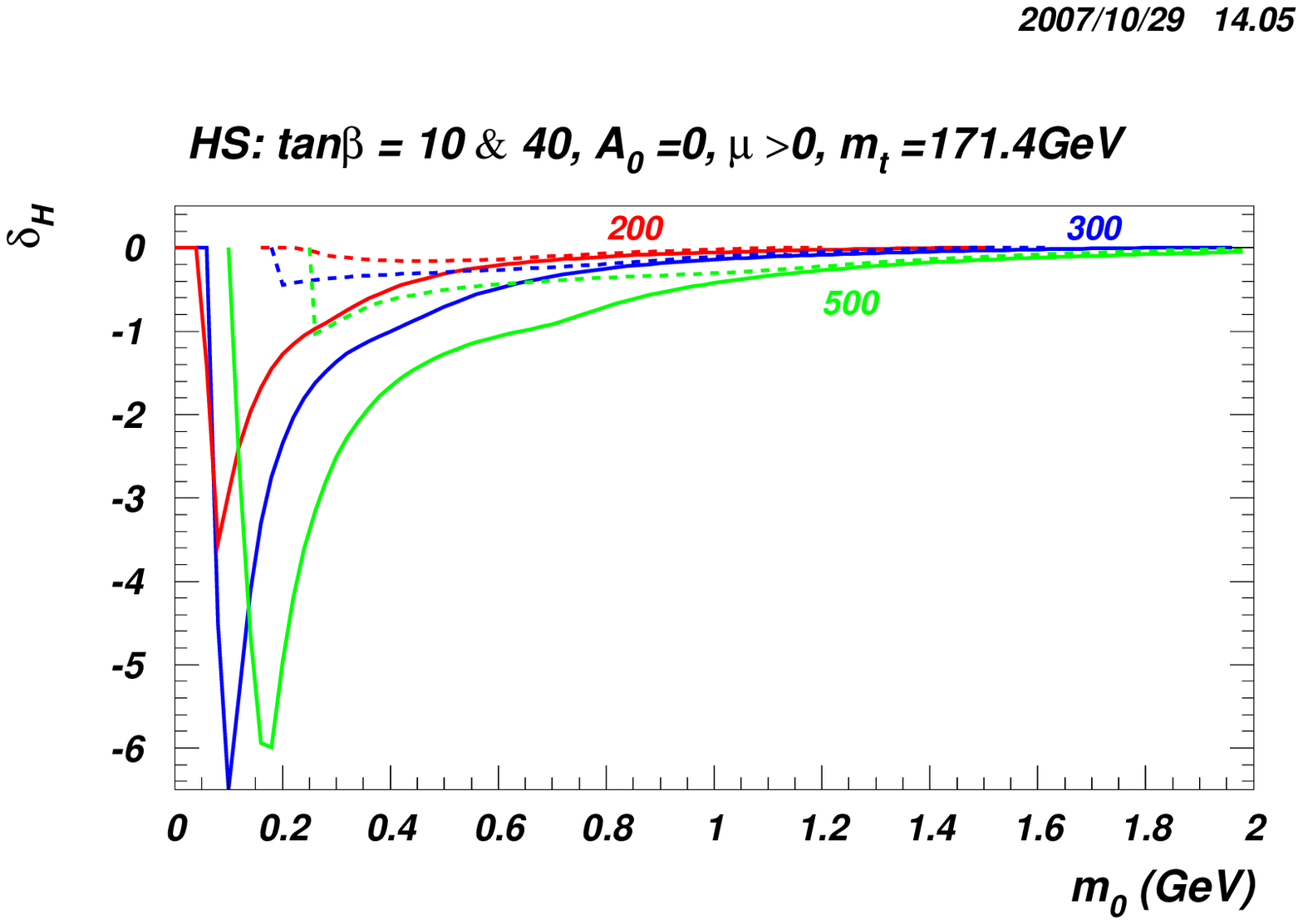,width=10cm,angle=0} 
\caption{\label{fig:hs} Values of $\delta_H$  needed to bring
various mSUGRA points into accord with the measured relic density versus
$m_0$ for $A_0=0$ and $\tan\beta =10$ (solid lines) and 40 (dashed
lines) in the NUHM2 HS scenario.  The curves correspond to $m_{1/2}$
values of 200, 300 and 500~GeV.  }}

{\it Implications for collider searches:}

In the NUHM2 model, since $\mu$ and $m_A$ are free parameters, one is
free to choose {\it both} $\mu$ and $m_A$ to be small, so one can have
MHDM and $A$-funnel annihilation contributions simultaneously.  This
type of model leads to the possibility of having light -inos {\it and}
light $A,\ H$ and $H^\pm$ at the same time. This would lead to a very
complex and rich pattern of gluino and squark cascade decays at the LHC.

A new distinct possibility also arises in the NUHM2 model. In the MSSM
scalar mass RGEs (see {\it e.g.} Ch. 9 of Ref.~\cite{wss}), the
right-hand-side includes a term $$S=m_{H_u}^2-m_{H_d}^2+Tr\left[{\bf
m}_Q^2-{\bf m}_L^2-2{\bf m}_U^2+{\bf m}_D^2+{\bf m}_E^2\right]$$ which
vanishes in the mSUGRA case, but is obviously non-zero if the Higgs soft
masses are split.  If the Higgs mass splitting is large, then the
$S$-term helps push the $m_L^2$ and $m_U^2$ soft terms to small values,
leading to cases with left-slepton neutralino co-annihilation, or to
cases with very light $\tu_R$ and $\tc_R$ squark masses, with
squark-neutralino co-annihilation acting to reduce the relic density.
The presence of very light left-sleptons or light $\tu_R$, $\tc_R$
squarks, with other sfermions at the TeV scale, 
might be an indication of the HS scenario with large, positive
$\delta_H>0$. In this case, contrary to the prediction of many models,
the lighter stau is dominantly $\ttau_L$.

{\it Implications for DM searches:}

For the NUHM2 model, all direct and indirect DM detection rates are
strongly enhanced if $\mu$ is small, while only halo annihilation rates
are enhanced if $2m_{\tz_1}\sim m_A$. If instead parameters are in the
region where agreement with the observed relic density occurs because
the left-type sleptons are relatively light, we do not expect increases
in either direct or indirect detection rates.  Finally, in the light
$\tu_R$, $\tc_R$ region, rates for direct detection of DM, as well as
for its indirect detection via observation of high energy $\tnu_{\mu}$'s
at IceCube will be  enhanced since these depend mainly on
neutralino-nucleon scattering via squark exchange, and $\tu_R$ squarks
are light. Halo annihilation rates are also somewhat enhanced, since
neutralinos can more easily annihilate into $u\bar{u}$ and $c\bar{c}$
pairs, giving rise to gamma ray and anti-matter signals~\cite{nuhm2}.

\subsection{Models with non-universal gaugino masses}
\label{ssec:nugm}

{\it Motivation:} In mSUGRA, it is assumed that the gaugino mass
parameters $M_1$, $M_2$ and $M_3$ unify to $m_{1/2}$ at $Q=M_{\rm
GUT}$. This holds true in supergravity models if the gauge kinetic
function $f_{AB}\sim \delta_{AB}f(h_M)$, where $A,B$ are gauge indices,
and $f(h_M)$ is an arbitrary function of hidden sector fields $h_M$, but
common to all the gauge groups.  More generally, the gauge kinetic
function need only transform at the symmetric product of two
adjoints. In this more general case, if the auxiliary field that breaks
supersymmetry also breaks the grand unification gauge symmetry, GUT
scale gaugino mass parameters need not unify~\cite{anderson}. 
Non-unified masses also
occur in models of gaugino-mediated SUSY breaking~\cite{dermisek} and in
various string-motivated models~\cite{ibanez}. In models with mixed
moduli-anomaly mediated SUSY breaking (MMAMSB)~\cite{mmamsb}, the
gaugino masses are again split at $M_{\rm GUT}$, with the splitting
proportional to the gauge group $\beta$-functions.  Motivated by these
considerations, in the phenomenological models that we consider below,
we will allow independent gaugino mass parameters at $Q=M_{\rm GUT}$.
To isolate the effect,
we will assume that just one of these mass parameters
deviates from it unified value, and tune it to reproduce the measured
value of the DM relic density leaving the other two at $m_{1/2}$.

\subsubsection{Mixed wino dark matter}
\label{ssec:mwdm}

{\it Parameter space:} In mSUGRA, at the GUT scale one assumes
$M_1=M_2\equiv m_{1/2}$ which leads to $M_1\sim {M_2\over 2}$ at the
weak scale due to RG running. The $\tz_1$ is usually a nearly pure bino
state with too large a relic density. In anomaly-mediated SUSY breaking
(AMSB) models~\cite{rs}, the $\tz_1$ is nearly pure wino-like, with too
low a relic density. There exists, therefore, an intermediate
situation with $M_1\sim M_2$ at the weak scale which gives the observed
relic density, so that the $\tz_1$ is
{\it mixed wino dark matter} (MWDM)~\cite{mwdm}. Starting from the mSUGRA model
boundary conditions at $M_{\rm GUT}$, one can either increase $M_1$ so
that $M_1> M_2=M_3=m_{1/2}$: 
\be 
m_0,\ M_1,\ m_{1/2},\ A_0,\ \tan\beta
,\ sign(\mu )\ \ \ ({\rm the\ MWDM1\ case}) 
\ee 
or lower $M_2$ such that $M_2< M_1=M_3=m_{1/2}$: 
\be 
m_0,\ M_2,\ m_{1/2},\ A_0,\ \tan\beta,\
sign(\mu )\ \ \ ({\rm the\ MWDM2\ case}) .  
\ee 
In the MWDM1 case, since
$M_1$, $M_2$ and $\mu$ will all be much closer together, the $\tz_1$ is
actually a mixed bino-wino-higgsino state, while in MWDM2, since
$M_1\sim M_2\ll \mu$, the $\tz_1$ is more a pure bino-wino mixed state.

In Fig. \ref{fig:mwdm}, we show curves which illustrate the GUT scale
ratio of $r_1=M_1/m_{1/2}$ needed to move the mSUGRA relic density
prediction into accord with (\ref{eq:relic}). We show curves versus
$m_0$ for $m_{1/2}=200,\ 300$ and 500~GeV, and for $\tan\beta =10$ and
40. We take $A_0=0$ and $\mu >0$.  At the lowest $m_0$ values $r_1=1$
since we are in the stau co-annihilation (or for $m_{1/2}=200$~GeV in
the bulk) region. For $m_0$ values beyond this, we have to dial $M_1$ so
as to obtain a mixed bino-wino-higgsino $\tz_1$ to be in concordance with
the observed relic density measurement.  For very large values of $m_0$
approaching the HB/FP region, the required value of $r_1$ once again
begins to reduce as long as $m_{\tz_1}>M_W$ so that $\tz_1\tz_1 \to
W^+W^-$ is accessible in the early universe . The up-turn in the
$m_{1/2}=200$~GeV curves at large values of $m_0$ occurs when this
reaction becomes kinematically suppressed: in this case, a larger value
of $r_1$ once again allows this reaction without which the relic density
tends to be too large. However, for $m_{1/2}=200$~GeV and $\tan\beta=10$,
and $m_0 \agt 1.45$~TeV, we are deep enough into the HB/FP region so
that even for mSUGRA the annihilation cross section via $Z^*$ exchange
is large enough to get agreement with (\ref{eq:relic}) even with
$m_{\tz_1}<M_W$.\footnote{This portion of the curve is excluded by the
LEP constraint on $m_{\tw_1}$.}  In the case of the curves for
$m_{1/2}=300$ and 500~GeV, $m_{\tz_1}$ always remains above $M_W$ so
that the HB/FP region of mSUGRA is smoothly reached. The dashed curves
exhibit analogous behaviour. 

\FIGURE[tbh]{
\epsfig{file=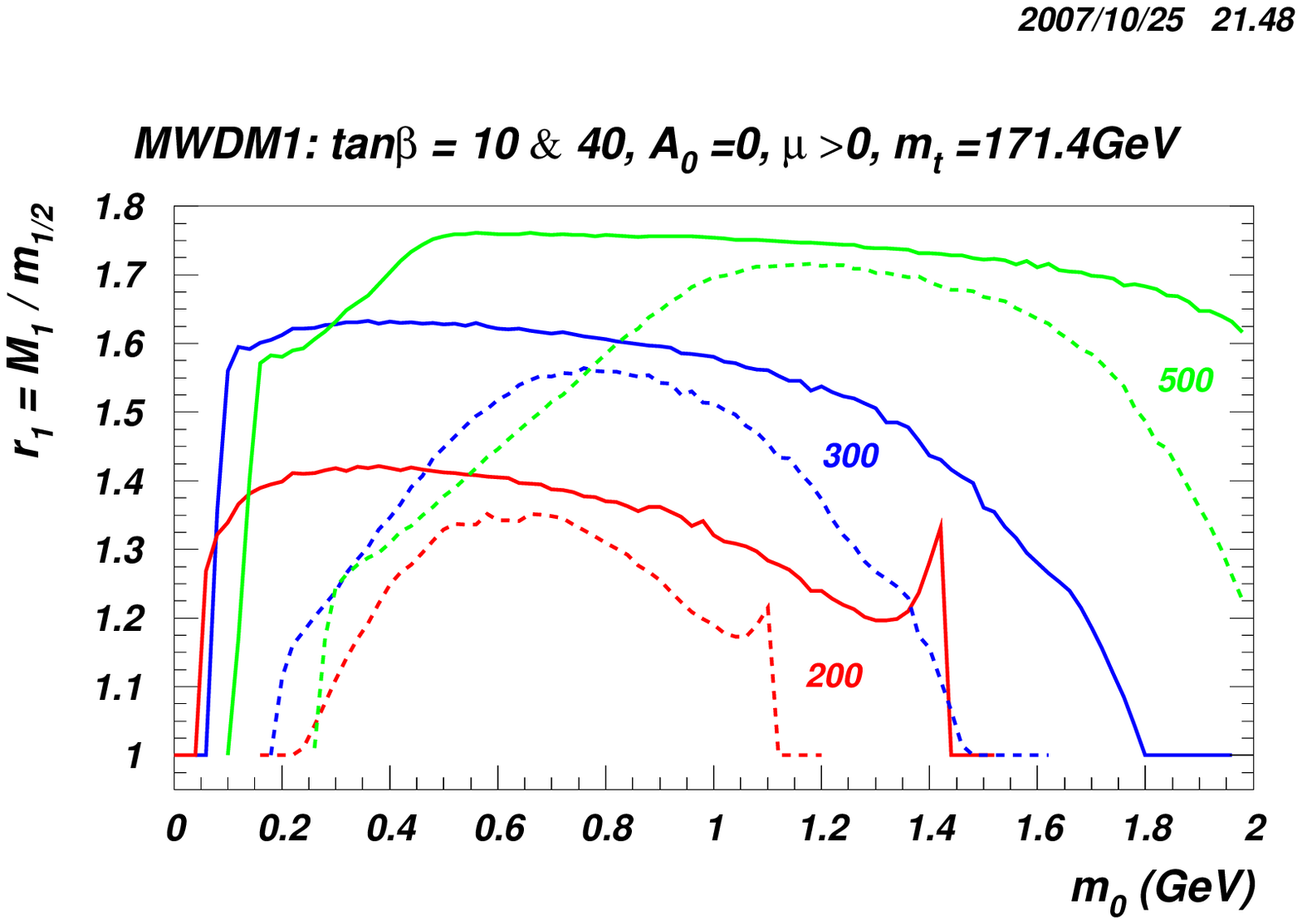,width=10cm,angle=0} 
\caption{\label{fig:mwdm} Values of $r_1=M_1/m_{1/2}$ needed to bring
various mSUGRA points into accord with the measured relic density versus
$m_0$ for $A_0=0$ and $\tan\beta =10$ (solid lines) and 40 (dashed
lines) in the MWDM1 scenario.  The curves correspond to $m_{1/2}$ values
of 200, 300 and 500~GeV. For the 200 (300)~GeV solid curves, the regions
with $m_0 \agt 1.4$ (1.94)~~TeV are excluded because the chargino is too
light.}}

{\it Implications for collider searches:}

In MWDM models with $M_1\sim M_2$ such that the relic density constraint
(\ref{eq:relic}) is fulfilled, the mass gap $m_{\tz_2}-m_{\tz_1}$
generally tends to be $\sim 20-30$~GeV for MWDM2 models (larger for
MWDM1, where the neutralino also has a higgsino component), somewhat
smaller than the mass gap of $\sim 50$~GeV found in models with MHDM
(such as the HB/FP region of mSUGRA). This means that at collider
experiments, the same-flavor/opposite-sign isolated dilepton mass
spectrum should have a single visible edge around 20-30~GeV arising from
$\tz_2\to\tz_1\ell\bar{\ell}$ decays. Moreover, the shape of the
dilepton spectrum should correspond to one where the neutralino
eigenvalues have the same sign. This should be distinct from MHDM models
which tend to have the higher mass edge and possibly a shape
corresponding to opposite signs of the neutralino mass eigenvalues~\cite{kitano}.  
Also since MHDM models have a small $\mu$ parameter,
edges due to $\tz_3$ decay may also be visible. The small
$\tz_2-\tz_1$ mass gap suppresses 3-body decays of $\tz_2$ more than
2-body decays: then the branching fraction for the radiative decay
$\tz_2\to\tz_1\gamma$, which is normally very suppressed in
mSUGRA, can reach the 10\% level in the MWDM models~\cite{bwca,z2z1g}. 
Finally, when $M_1$ is raised, it feeds into raising
masses of especially right-type sleptons relative to their mSUGRA
predictions. Likewise, when $M_2$ is reduced, left- squark and slepton
masses get reduced relative to mSUGRA.

{\it Implications for DM searches:}

In the MWDM1 scenario, the $\tz_1$ becomes a mixed bino-wino-higgsino
state, and as a result all direct and indirect DM detection rates are
boosted relative to mSUGRA -- sometimes by an order of magnitude or more
due to the enhanced higgsino component.  In the MWDM2 scenario, where
the $\tz_1$ develops a smaller higgsino component, direct detection and
$\nu_\mu$ indirect detection rates are only slightly enhanced, while
indirect DM detection rates from halo annihilations can again be boosted
by an order of magnitude or more, but not necessarily to observable
levels.

\subsubsection{Bino-wino co-annihilation (BWCA)}
\label{ssec:bwca}

{\it Parameter space:} In the BWCA scenario~\cite{bwca}, where $M_1 \sim
-M_2$ at the weak scale, there is very little mixing between the bino
and neutral wino states even when these are very close in mass. If
$|M_1|$ is just slightly smaller than $|M_2|$, and $|\mu|$ is relatively
large, the lightest neutralino remains bino-like, but because
$m_{\tz_1}\simeq m_{\tz_2,\tw_1}$, bino-wino co-annihilation processes
in the early universe reduce the relic density to the observed level. In
the BWCA case, the parameter space is the same as in the MWDM case,
except that $M_1$ and $M_2$ are now opposite in sign. As with the MWDM
case, one can either raise $|M_1|$ by about a factor 2, or lower $|M_2|$
by about the same factor to attain $-M_1\sim M_2$ at the weak scale.

In Fig. \ref{fig:bwca}, we show curves which illustrate the GUT scale
ratio $r_2=M_2/m_{1/2}$ needed to move the mSUGRA relic density
prediction into accord with (\ref{eq:relic}). We show results versus
$m_0$, for $m_{1/2}=200,\ 300$ and 500~GeV, and for $\tan\beta =10$ and
40. We take $A_0=0$ and $\mu >0$ (upper frame) and $\mu < 0$, the sign
favored by the 
E821 $(g-2)_{\mu}$ experiment (lower frame), at least for lower ranges
of $m_0$ and $m_{1/2}$.  That the ratio $r_2$ remains close to $-1/2$ in the
upper frame, for all $m_0$ values in between the stau-coannihilation and
the HB/FP regions, reflects the fact that the evolution of gaugino
masses does not depend on sfermion masses at 1-loop. The solid lines for
$\tan\beta=10$ in the lower frame show very similar behaviour, except
that $\mu$ is now negative. In contrast, there is a large flat region at
low values of $m_0$ in the $\tan\beta=40$ case shown by the dashed
lines. We have traced this to the fact that the Higgs funnel region has
already opened up even for $\tan\beta=40$, and that the funnel region is
contiguous to the stau co-annihilation (and for $m_{1/2}=200$~GeV, also the
bulk) region.

\FIGURE[tbh]{
\epsfig{file=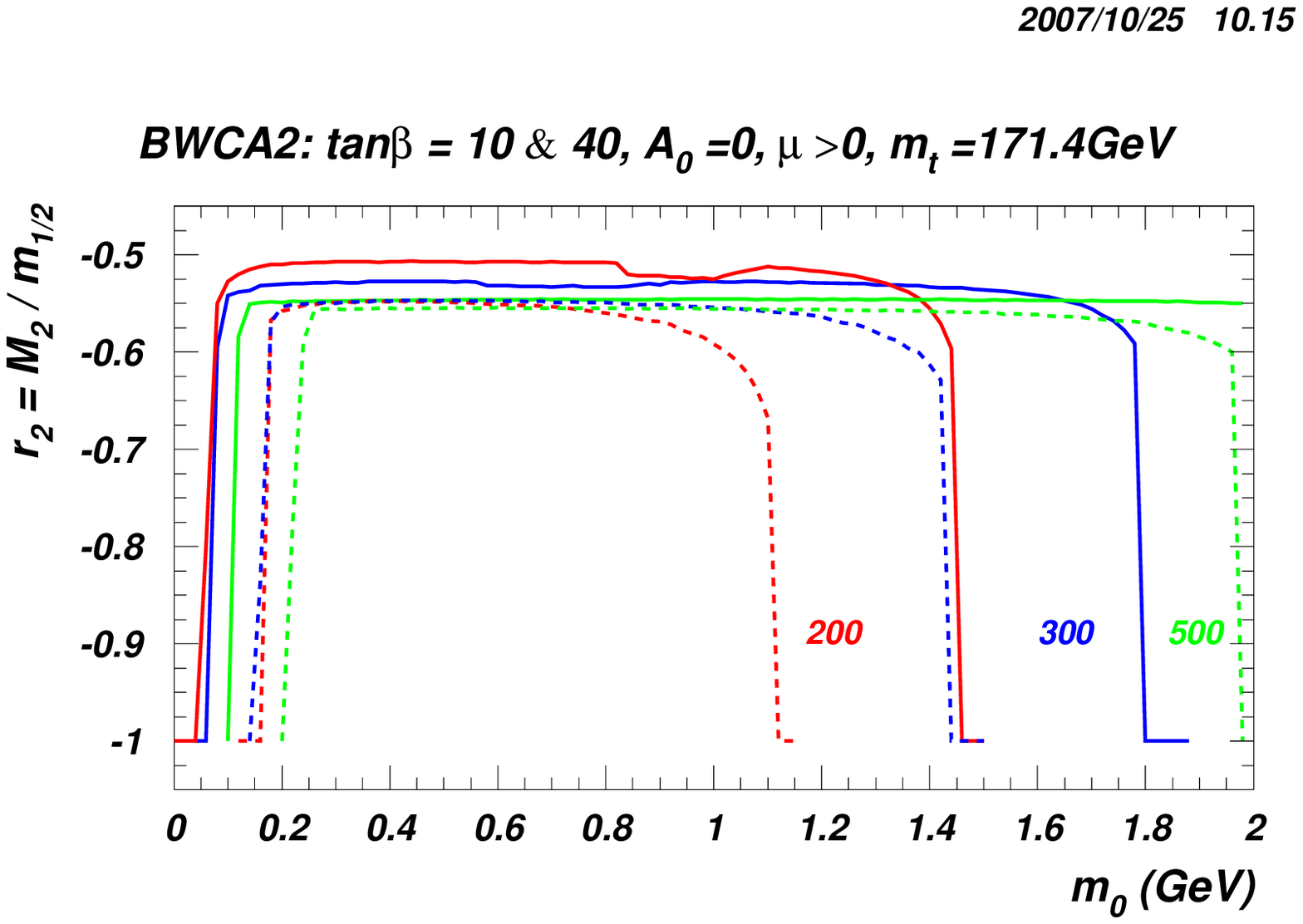,width=10cm,angle=0} 
\epsfig{file=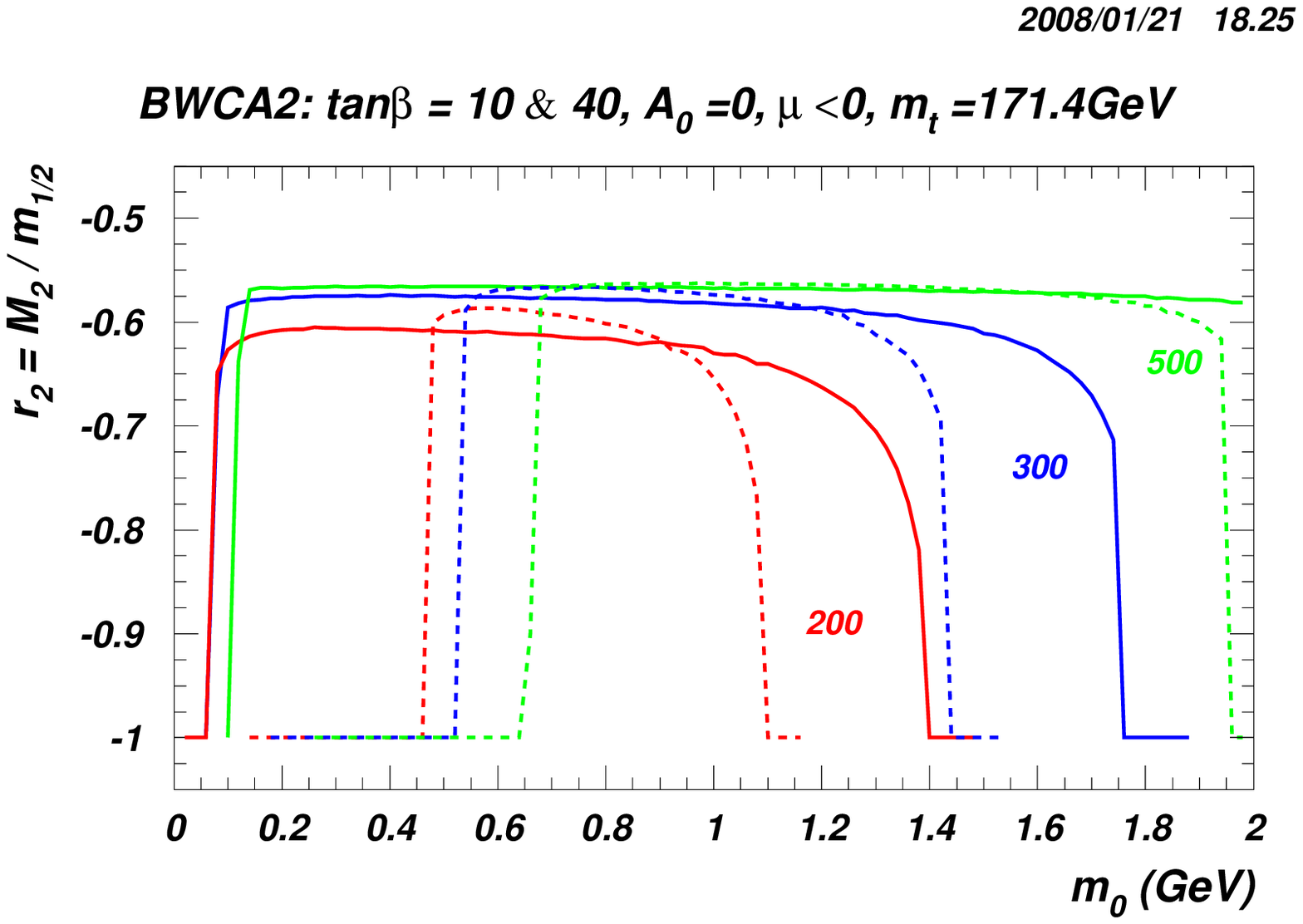,width=10cm,angle=0}
\caption{\label{fig:bwca} Values of $r_2=M_2/m_{1/2}$ needed to bring
various mSUGRA points into accord with the measured relic density versus
$m_0$ for $A_0=0$ and $\tan\beta =10$ (solid lines) and 40 (dashed
lines) in the BWCA scenario.  The upper frame is for $\mu > 0$ while in
the lower frame we take $\mu <0$. The curves correspond to $m_{1/2}$ values
of 200, 300 and 500~GeV.  }}

{\it Implications for collider searches:} In the BWCA scenario, the
$\tz_2-\tz_1$ mass gap becomes very small: of order 15-30~GeV
typically. Thus, as in the MWDM case, one expects a $m(\ell\bar{\ell})$
mass edge to be visible owing to $\tz_2\to\tz_1\ell\bar{\ell}$ decay,
since all $\tz_2$ two-body spoiler decay modes are kinematically
closed. The mass gap should be much smaller than that typically expected
from mSUGRA models. The $m(\ell^+\ell^-)$ distribution for opposite-sign/
same flavor dileptons 
should contain a single mass-edge, 
``one hump bump'', unlike the case of MHDM, which favors a
``two-hump-bump'' since $\tz_3$ would be light as well.\footnote{We
thank Dr. Theodore Geisel for coining related expressions.}

The  small mass gap (required for effective co-annihilation)
strongly favors two-body decays over three-body decays and results in
large branching fraction for the radiative decay $\tz_2\to\tz_1\gamma$:
reaching over 30\% in the scans presented in Ref.~\cite{bwca}. If
$\tz_2$ were at rest, the $\gamma$ from the radiative decay would be
mono-energetic but rather soft.  However, for fast moving $\tz_2$
secondaries from gluino and squark cascade decays, the photon energy
gets boosted, and in the BWCA case there should be an observable signal
also in the multi-jet plus isolated photon plus $\eslt$ channel at the
LHC: see Fig.~16 of Ref.~\cite{bwca}.

At $e^+e^-$ colliders, the relative minus sign between $M_1$ and
$M_2$  leads to
enhanced production of $\tz_1\tz_2$ pairs compared to mSUGRA
predictions, and also the predictions in the MWDM case.  
This provides a way of distinguishing between the BWCA and MWDM
frameworks which otherwise have a very similar mass spectrum. 
Moreover, operating just above $\tz_1\tz_2$ threshold, one
might make detailed studies of $\tz_2$ decay branching fractions,
including the radiative mode $\tz_2\to\tz_1\gamma$, which will result in
events with almost mono-energetic single photons recoiling against ``nothing''.

{\it Implications for DM searches:} In the BWCA case, since the $\tz_1$
remains purely bino-like, rates for indirect DM detection remain low,
similar to results from the corresponding mSUGRA case in many instances.
Rates for direct DM detection can be far below the sensitivity of any
proposed detector
if the sign of $M_1\mu$ is negative, {\it i.e.} for $\mu>0$
in the BWCA2 case, and for $\mu <0$ in the BWCA1 case, because of
cancellations in neutralino couplings that enter the direct
DM detection rate calculations~\cite{bwca}.

\subsubsection{Low $|M_3|$ dark matter: compressed SUSY}
\label{ssec:lm3dm}

{\it Parameter space:} The low $|M_3|$ dark matter (LM3DM) scenario 
arises by starting with
mSUGRA parameter space, but (for $m_0 \alt 1-2$~TeV) {\it lowering} the
GUT scale value of $|M_3|$ relative to
$M_1=M_2=m_{1/2}$~\cite{belanger,m3dm}.  Lowering $|M_3|$ results in
smaller gluino and, via RGE effects, also squark masses.  These effects
feed into the MSSM RGEs and affect the running of $m_{H_u}^2$ --
effectively diminishing the downward push from the top quark Yukawa
coupling -- resulting in lower $|\mu|$ values, and hence MHDM.  The MHDM
case can be easily compatible with the observed relic density
constraint since there is enhanced neutralino annihilation to $WW$,
$ZZ$ and $Zh$ states in the early universe.  Thus, the parameter space
is given by 
\be m_0,\ m_{1/2},\ M_3,\ A_0,\ \tan\beta ,\ sign(\mu )\ \ \
({\rm the\ LM3DM\ case}) .  
\ee 
Here, $M_3$ can be of either
sign. Although the first and second generation sfermion masses are
essentially unaffected by the sign flip, $m_{\tst_1}$, and through $\mu$,
also chargino and neutralino masses, do show clear dependence on the
relative sign between $M_3$ and $M_{1,2}$.

In Fig.~\ref{fig:lm3dm}, we show the GUT scale ratio of
$r_3=M_3/m_{1/2}$ needed to move the mSUGRA relic density prediction
into accord with (\ref{eq:relic}) versus $m_0$. As always, we show
results for $m_{1/2}=200,\ 300$ and 500~GeV, and for $\tan\beta =10$ and
40. We take $A_0=0$ and $\mu >0$.  For the solid curves, we see that
once we are away from the stau co-annihilation (and for the
$m_{1/2}=200$~GeV case, also bulk) region, we need to reduce $|M_3({\rm
GUT})|$ to obtain the correct relic density.  Since in this scenario we
are lowering $\mu$, the degree of dialing is generally smaller for
larger $m_0$ due to increasing top Yukawa coupling effects, just as in the
NUHM1 model with positive $\delta_\phi$.  The situation is more
complicated for the dashed curves where $\tan\beta =40$. For $M_3<0$,
the Higgs funnel already starts to appear for this relatively low value
of $\tan\beta$. For the red and blue dashed curves corresponding to
$m_{1/2}=200$ and 300~GeV, respectively, this funnel region is
contiguous with the bulk/stau co-annihilation region, so that $r_3$
remains at -1 for $m_0 \alt 600$~GeV. For yet larger values of $m_0$, 
$|r_3|$ needs to be dialed down though, because of the proximity of the
Higgs funnel, by not quite as much as for the corresponding 
$\tan\beta=10$ case,
until the HB/FP region is reached. For the $m_{1/2}=500$~GeV curve, the
Higgs-funnel region occurs for 450~GeV $\alt m_0 \alt
800$~GeV, and is well separated from the stau co-annihilation
region. Thus some tuning of $r_3$ (but again, not as much as in the
$\tan\beta=10$ case) is needed for $m_0$ values away from the very
narrow stau co-annihilation region, and again for large $m_0$ values
outside the Higgs-funnel region, until the HB/FP region is reached at
$m_0\sim 1.9$~TeV. For $M_3>0$, although the Higgs-funnel does not occur
for $\tan\beta=40$, $s$-channel $A/H$ exchange does significantly
enhance neutralino annihilation amplitudes for $m_0 \alt 2m_{1/2}$: as a
result, the value of $r_3$ needed varies more slowly at the low $m_0$
end for the dashed curves than for the solid curves.
\FIGURE[tbh]{
\epsfig{file=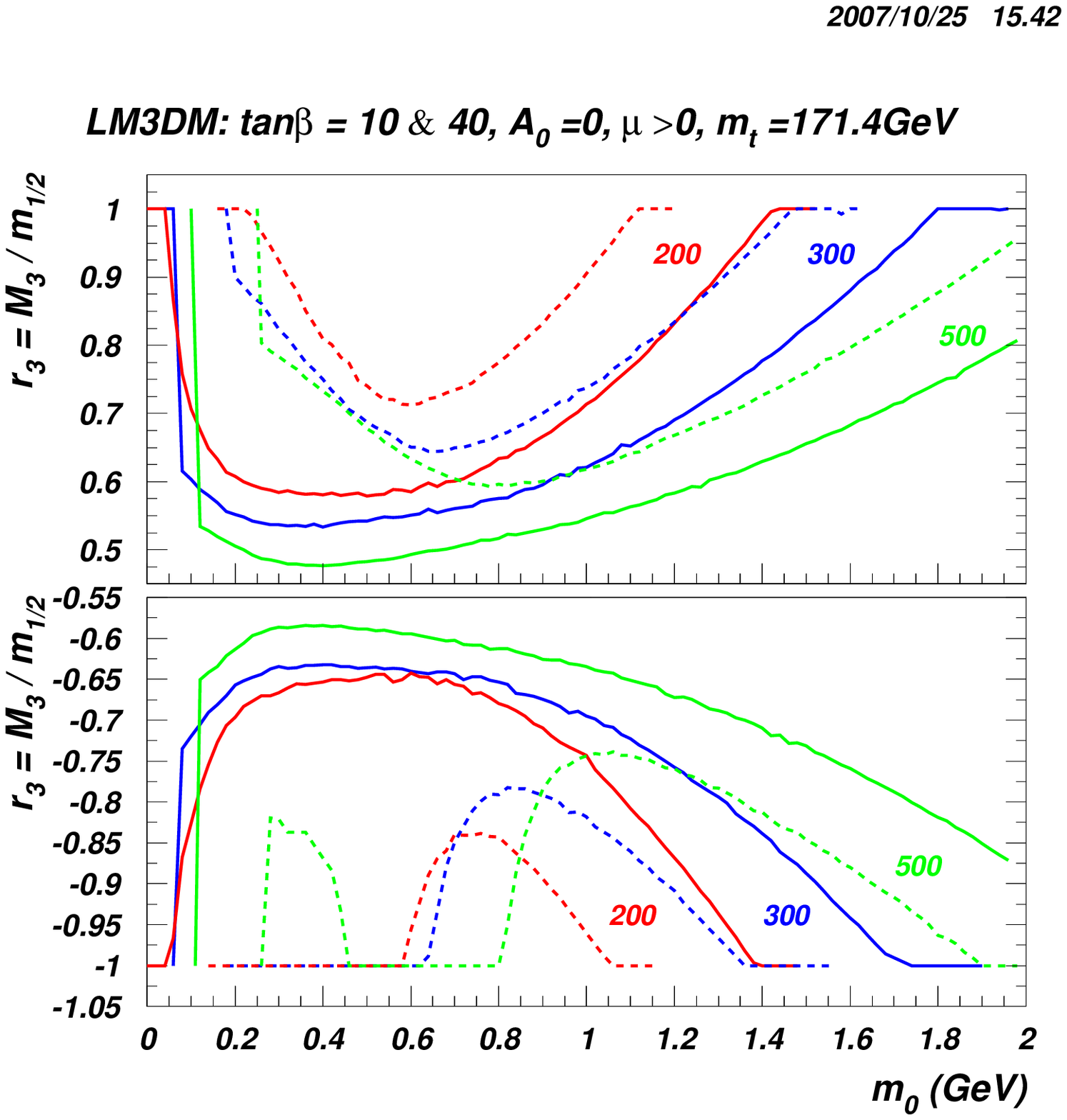,width=10cm,angle=0} 
\caption{\label{fig:lm3dm} Values of the GUT scale ratio 
$r_3=M_3/m_{1/2}$ needed to bring
various mSUGRA points into accord with the measured relic density versus
$m_0$ for $A_0=0$ and $\tan\beta =10$ (solid lines) and 40 (dashed
lines) in the LM3DM scenario.  The curves correspond to $m_{1/2}$ values
of 200, 300 and 500~GeV.  }}

We should mention that a related scenario, dubbed ``compressed SUSY'', has been
suggested by Martin~\cite{compsusy}.  In compressed SUSY, $M_3$ is
lowered, but also by choosing large negative values of the $A_0$
parameter, a rather light $\tst_1$ state can be generated. Then if
$m_{\tz_1}>m_t$, neutralino annihilation
via $\tz_1\tz_1\to t\bar{t}$, which does not suffer the
usual $P$-wave suppression because of the large top quark mass, can dominate in
the early universe, resulting in the observed relic density in a
different way. The phenomenology of compressed SUSY models has been examined
in Ref.~\cite{compsusy2}.

{\it Implications for collider searches:} Since $m_{\tg}$ and $m_{\tq}$
are lowered relative to $m_{\tw_1}$ values, the Tevatron and LHC reach
is enhanced compared to corresponding mSUGRA predictions. In mSUGRA, the
chargino mass limit $m_{\tw_1}>103.5$~GeV from LEP2 usually pre-empts
the limit from direct gluino searches at the Tevatron because in models
with gaugino mass unification, $m_{\tg}\sim 3.5 m_{\tw_1}$, and thus
$m_{\tg}\agt 350$~GeV, which is not far from the reach of a 2~TeV
$p\bar{p}$ collider. However, in LM3DM with non-unified gaugino masses,
the gluino can have a mass as low as $\sim 200$~GeV, while the chargino
remains in the LEP2 allowed region, $m_{\tw_1}>103.5$~GeV .  Thus, a significant
chunk of LM3DM parameter space is open to gluino pair and gluino-squark
searches at the Tevatron~\cite{m3dm_tev}.  Since the gluino mass is
lowered with respect to $m_{\tz_1}$, the radiative gluino decay
$\tg\to\tz_1 g$ may be the dominant decay mode of the $\tg$,
with a branching fraction~\cite{gltoz1g} as high as 85\% in regions of
parameter space where squarks are very heavy. In this case, the decay of
a gluino
leads to a single high $p_T$ jet, and gluino pair production will
look more like squark production.

Since the $\tz_1$ is MHDM in the LM3DM scenario, the $\tz_2-\tz_1$ mass
gap is again lowered compared to mSUGRA predictions, and the
$\tz_2\to\tz_1\ell\bar{\ell}$ decay should be visible at LHC
searches. In the LM3DM case, the mass gap is typically of order
30-80~GeV, and there should be $\tz_3$ contributions to the
$m(\ell\bar{\ell})$ distribution as well, thus allowing the MHDM
scenario to be distinguished from BWCA or MWDM. The large higgsino
content of the neutralino $\tz_1$ will also tend to lead to a large
branching ratio for gluino decays to third generation quarks so that the
reach of the LHC will be enhanced by tagging $b$-jets~\cite{mmt}.

For ILC searches, it is more likely that squark pair production
(especially $\tst_1\bar{\tst}_1$ production) will be accessible. In
addition, since the $\tz_3$, $\tz_4$ and $\tw_2$ states are lighter than
the corresponding mSUGRA parameter space points, it is more likely that
various -ino pair production reactions would be accessible to ILC
searches, allowing the reconstruction of chargino and neutralino mass
matrices.

{\it Implications for DM searches:} In the LM3DM scenario, since we
expect {\it both} light squarks and low $\mu$ (leading to MHDM), direct
DM search rates are enhanced relative to mSUGRA by up to {\it two orders
of magnitude}! Rates for $\nu_\mu$ events at IceCube are enhanced by up
to three orders of magnitude!  Similar enhancements are seen in gamma
ray and anti-matter search predictions arising from neutralino
annihilation in the galactic halo.  Thus, the LM3DM scenario seems a
boon for direct and indirect DM searches.

\subsubsection{High $|M_2|$ dark matter: left-right split SUSY}
\label{ssec:hm2dm}

{\it Parameter space:} In the high $|M_2|$ dark matter scenario (HM2DM),
the parameter set is the same as in the MWDM2 scenario, except that now
the $SU(2)$ gaugino mass parameter $M_2$ is dialed {\it up} in
magnitude~\cite{hm2dm}. This increased value of $|M_2|$ 
feeds into the MSSM
RGEs by first pushing $m_{H_u}^2$ to higher values (than in the universal
gaugino mass case) during its evolution  from $Q=M_{\rm
GUT}$. Then  as RG running continues, the top Yukawa
coupling terms take over, and  $m_{H_u}^2(Q)$ begins to be
reduced. Since its positive peak-value was higher than in the canonical
case with universal gaugino masses, $m_{H_u}^2$ attains a relatively
small negative value when the evolution is stopped at
the weak
scale. Finally, since $\mu^2\sim -m_{H_u}^2({\rm weak})$, 
we end up with a small weak scale $|\mu |$ parameter, and MHDM.

In the HM2DM scenario, the large value of $|M_2(M_{\rm GUT})|$, via RG
evolution, lifts the SSB masses of $SU(2)$ doublet matter scalars to
large values, so that left-sleptons, and to a smaller extent also
left-squarks, are much heavier than right- ones.  Thus, the HM2DM model
can be
regarded as left-right split SUSY.  In this model, all light third generation
matter sfermions ($m_{\tf_1}$) then tend to be predominantly right-
states, whereas in most models, the $\tb_1$ tends to be mainly $\tb_L$,
and $\tst_1$ tends to be a mixed left-right squark state.  Further,
since $|\mu|$ is small, the HM2DM model leads to a  spectrum with light
$\tz_1$, $\tz_2$, $\tz_3$ and $\tw_1$ states, which tend to be
higgsino-like or mixed bino-higgsino. The $\tw_2$ and $\tz_4$ are nearly
pure winos, and also very heavy.

In Fig.~\ref{fig:hm2dm}, we show values of the GUT scale
ratio of $r_2=M_2/m_{1/2}$ needed to bring the neutralino relic density
prediction into accord with (\ref{eq:relic}). We show curves versus
$m_0$ for $m_{1/2}=200,\ 300$ and 500~GeV, for $\tan\beta =10$ and
40,
with $A_0=0$. Since $\Delta a_{\mu}^{SUSY} \propto \mu M_2$ we take
$sign(\mu)=sign(M_2)$ so their product is positive in accordance with
the measured value of $(g-2)_{\mu}$~\cite{pdb}.  As in the LM3DM model,
flipping the sign of $M_2$ (without changing its magnitude) causes the
A-funnel to open up for $\tan\beta=40$. But now the funnel region
remains merged with the stau-coannihilation region even for
$m_{1/2}=500$~GeV, so that for the lower range of $m_0$ no additional
dialing is necessary and $r_2$ remains at $-1$.
\FIGURE[tbh]{
\epsfig{file=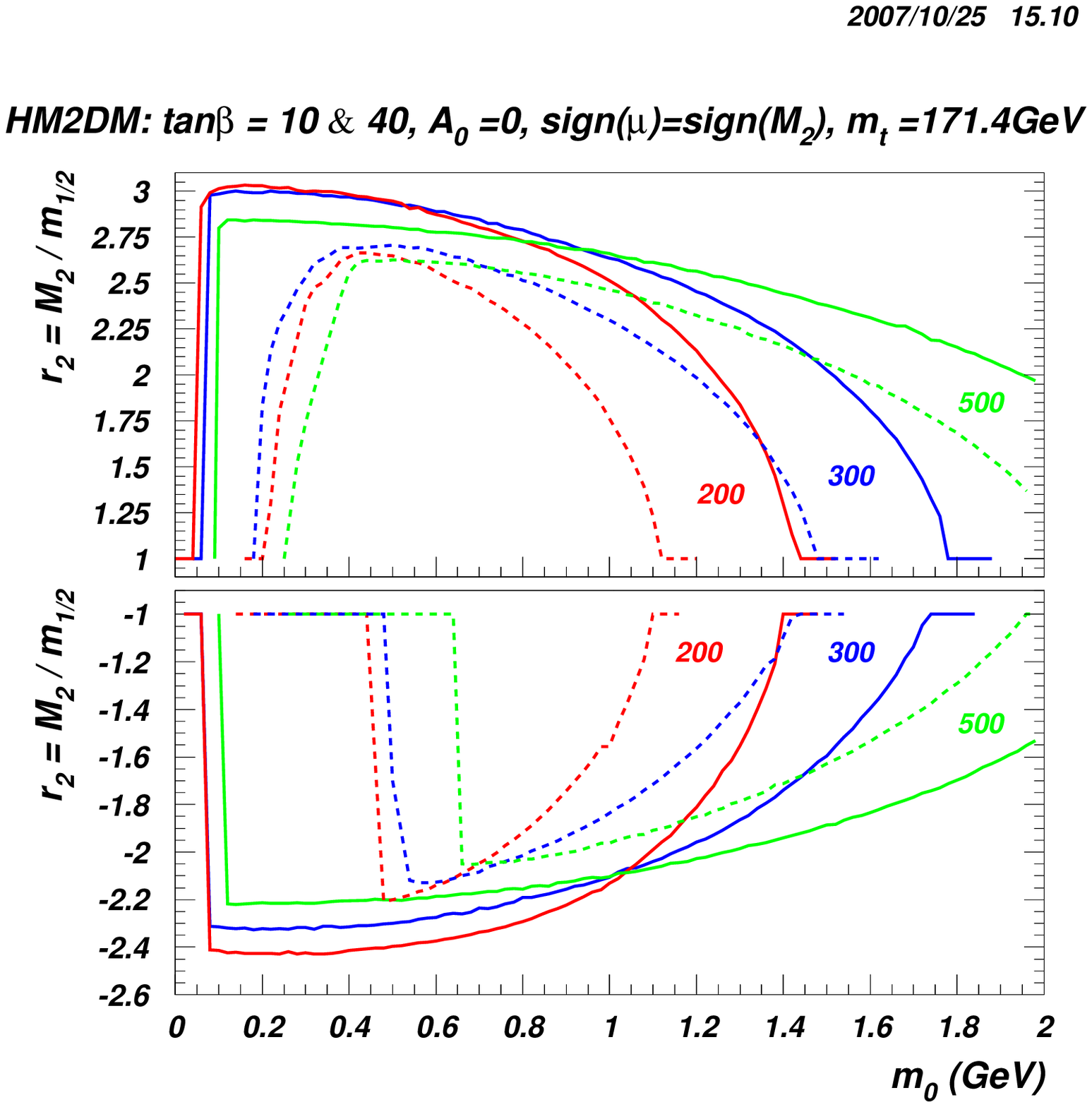,width=10cm,angle=0} 
\caption{\label{fig:hm2dm} Values of the GUT scale ratio
$r_2=M_2/m_{1/2}$ needed to bring various mSUGRA points into accord with
the measured relic density versus $m_0$ for $A_0=0$ and $\tan\beta =10$
(solid lines) and 40 (dashed lines) in the HM2DM scenario.  The curves
correspond to $m_{1/2}$ values of 200, 300 and 500~GeV.  }}

{\it Implications for collider searches:} In the HM2DM scenario, since
we again have MHDM, the $\tz_2-\tz_1$ mass gap tends to be in the range
30-80~GeV, so that the mass edge from $\tz_2\to\tz_1\ell\bar{\ell}$
decays, and possibly also from $\tz_3$ decays,  should be
seen in gluino and squark cascade decay events at the LHC. The shape of
the distributions may also make it possible to glean information about
the relative signs of the neutralino mass eigenvalues. Also, the
increased higgsino content of the lighter states should again lead to
increased $b$-jet multiplicity in SUSY events at the LHC.

At the ILC, the production of chargino and neutralino pairs would vary
in a contrasting way compared to mSUGRA because the low lying gaugino
states would be essentially devoid of wino components. If third
generation squarks and sleptons are accessible to ILC searches, then a
variation in beam polarization would reveal all these states to be
predominantly right-type states, and their pair production cross sections
would decrease with increasingly left-polarized beams.

{\it Implications for DM searches:} Since in the HM2DM scenario, the
$\tz_1$ is a mixed bino-higgsino state, its signal for spin-independent direct
detection should be observable at the next generation of detectors
(super-CDMS or 100~kg noble liquid detectors) over much of the parameter
space. 
The $\nu_\mu$ signal at IceCube or Antares would also be boosted by up
to two orders of magnitude compared to mSUGRA and may be observable over
a substantial portion of parameter space.  Rates for detection of gamma
rays and anti-matter from neutralino halo annihilation are also boosted
relative to mSUGRA by 1-2 orders of magnitude.

\section{Illustrative benchmark cases}
\label{sec:bm}

In this section, we list some benchmark cases of models with
universality and non-universality.  We start with the mSUGRA model, and
adopt a point in parameter space with $$m_0,\ m_{1/2},\ A_0,\ \tan\beta
,\ sign (\mu ) =300\ {\rm GeV},\ 300\ {\rm GeV},\ 0,\ 10,\ +1$$ with
$m_t=171.4$~GeV.  This point is listed in column 2 of
Table~\ref{tab:nusm}. We see that we get $\Omega_{\tz_1}h^2=1.1$ 
for this point which  is conclusively in conflict with
(\ref{eq:relic}), and so excluded assuming standard Big Bang cosmology
and thermal relic neutralinos.  For every other model in
Table~\ref{tab:nusm}, we relax the universality assumption and allow one
additional parameter that we tune to bring the model into the DM-allowed
range with $\Omega_{\tz_1}h^2\sim 0.1$.  The point here is to be able to
compare and contrast the spectra along with other features of each of
these DM-allowed models with the corresponding spectrum of the mSUGRA
model, and with one another.  We also list at the bottom the Isatools
output of $\Omega_{\tz_1}h^2$, $BF(b\to s\gamma )$, the SUSY
contribution to the muon anomalous magnetic moment $\Delta
a_\mu^{SUSY}$, the spin-independent cross section for the elastic
neutralino-proton scattering $\sigma_{SI}(\tz_1p)$, and the Higgsino
content of the neutralino $R_{\tH}=\sqrt{v_1^{(1)2}+v_2^{(1)2}}$ in
the notation of Ref.~\cite{wss}. We use
ISAJET~7.76 to generate this Table. 

In the first non-universal case, the NMH model,  we see that the
first/second generation SSB parameter $m_0(1,2)=54$~GeV in order to
obtain the observed relic density. Then, the $\te_R$ and $\te_L$ states
(and also the $\tmu_R$ and $\tmu_L$ states) have much reduced masses
than the corresponding mSUGRA case, while $m_{\ttau_{1,2}}$, and also
gluino, chargino and neutralino masses are essentially unchanged from
their mSUGRA values.  The low value of $m_{\te_R}=128.9$~GeV and
10.5~GeV mass gap between $\te_R$/$\mu_R$ and $\tz_1$ ensure a high rate
for neutralino annihilation and co-annihilation in the early universe.
In addition, while $BF(b\to s\gamma )$ remains near the measured and SM
value because third generation squarks and charged Higgs bosons are
heavy, the value of $\Delta a_\mu^{SUSY}$ is enhanced, thus reconciling
these two possibly disparate measurements.

The second non-universal case, labelled NUHM1$_\mu$, comes from the
NUHM1 model where $m_\phi$ is dialed up to $549$~GeV so that $\mu$
becomes small and we have mixed higgsino DM, even though $m_0$ is far
smaller than in the mSUGRA HB/FP region. In this case, $m_{\tw_1}$ has
been reduced so much that the point is actually LEP2 excluded. Note that
$R_{\tH}$ has risen to 0.84, signaling a higgsino-like $\tz_1$.  The
branching ratio for the decay $b\to s\gamma$ is slightly reduced. The
salient feature is the direct detection cross section, which is now in
the range of well-tempered neutralino models\cite{wtn,wtn_dd} with MHDM,
is 37 times higher than the corresponding mSUGRA value, and very close
to the current 90\% CL limit from Xenon-10 search~\cite{xenon10}.

The third non-universal case, labelled NUHM1$_A$, dials $m_\phi$ to
$-728\ {\rm GeV}$, which raises $\mu$ to large values but lowers
$m_A$ to be just above $2m_{\tz_1}$ so that $A$-funnel annihilation
reduces the relic abundance, even though $\tan\beta$ is not large. In
this case, sfermions and lighter -inos have essentially the same masses
as in mSUGRA, but now we have relatively light Higgs boson states $A$, $H$ and
$H^\pm$ accessible to LHC searches. Also, $BF(b\to s\gamma )$ is
somewhat enhanced due to the light $H^\pm$ entering the $tH^\pm$ loop
contribution to this decay.

The fourth non-universal case is the HS model and has the two Higgs soft
masses split about the common value $m_0$, and has both MHDM and some
$A$-funnel annihilation, with a light spectrum of Higgs bosons,
as well as charginos and neutralinos.
All the Higgs bosons and all the charginos and neutralinos should be
accessible at an electron-positron collider operating at a center of
mass energy just above 500~GeV. The higgsino component of the $\tz_1$,
which is larger than in mSUGRA, leads to considerable
enhancement of the direct detection cross section.

\TABLE{
\begin{tabular}{lccccc}
\hline
Model & mSUGRA & NMH & NUHM1$_\mu$ & NUHM1$_A$ & HS \\
\hline
parameter & --- & $m_0(1,2)$ & $m_{\phi}$ & $m_{\phi}$ & $\delta_H$ \\
special value       & ---   & 54    & 549   & -728  & -1.36  \\ \hline
$\mu$       & 385.1 & 386.5 & 105.8 & 748.5 & 269.3  \\
$m_{\tg}$   & 729.7 & 722.1 & 731.4 & 733.4 & 728.9  \\
$m_{\tu_L}$ & 720.8 & 658.4 & 724.3 & 720.5 & 720.1  \\
$m_{\tst_1}$& 523.4 & 526.5 & 484.1 & 624.5 & 505.8  \\
$m_{\tb_1}$ & 656.8 & 659.8 & 642.2 & 689.5 & 645.4  \\
$m_{\te_L}$ & 364.5 & 216.2 & 364.8 & 365.8 & 373.4  \\
$m_{\te_R}$ & 322.3 & 128.9 & 322.5 & 321.9 & 301.8  \\
$m_{\ttau_1}$& 317.1& 317.6 & 317.8 & 316.4 & 299.3  \\
$m_{\tw_2}$ & 411.7 & 412.7 & 264.7 & 754.8 & 321.1  \\
$m_{\tw_1}$ & 220.7 & 219.5 & 91.1  & 234.9 & 196.6  \\
$m_{\tz_4}$ & 412.5 & 413.5 & 268.1 & 754.6 & 322.9  \\
$m_{\tz_3}$ & 391.3 & 392.7 & 137.3 & 747.1 & 277.1  \\ 
$m_{\tz_2}$ & 220.6 & 219.4 & 117.4 & 234.5 & 198.1  \\ 
$m_{\tz_1}$ & 119.2 & 118.4 & 69.0  & 121.5 & 115.4  \\ 
$m_A$       & 520.3 & 521.9 & 584.5 & 268.5 & 279.0  \\
$m_{H^+}$   & 529.8 & 531.4 & 593.8 & 281.6 & 292.0  \\
$m_h$       & 110.1 & 110.1 & 109.8 & 110.5 & 109.8  \\ \hline
$\Omega_{\tz_1}h^2$& 1.1 & 0.10 & 0.11 & 0.11 & 0.10 \\
$BF(b\to s\gamma) \times 10^4$ & 3.0 & 3.1 & 2.5 & 4.3 & 3.4  \\
$\Delta a_\mu \times 10^{10}$ & 12.1 & 27.2 & 17.9 & 9.3 & 13.7 \\ 
$\sigma_{SI} (\tz_1p ) \times 10^9\ ({\rm pb})$ & 
2.1 & 2.1 & 78 & 1.2 & 27 \\
$R_{\tH}$ & 0.15 & 0.14 & 0.84 & 0.06 & 0.26 \\
\hline
\end{tabular}
\caption{A comparison of the characteristics
 of mSUGRA with corresponding characteristics in
models with scalar mass non-universality that lead to the observed relic
 abundance of DM.
Input parameters and resultant sparticle masses in GeV units,
together with the predicted neutralino relic density, $BF(b\to s\gamma)$ and $\Delta
a_{\mu}$, the SUSY contribution to the anomalous magnetic moment of the
muon, the direct detection cross section for the $\tz_1$, and finally, the
higgsino content of the $\tz_1$. In each case, we fix
$m_0=m_{1/2}=300$~GeV, $A_0=0$, $\tan\beta =10$ and $m_t=171.4$~GeV, and
for each non-universal model tune the parameter in the first row to its
 special value shown in row 2 to reproduce the observed relic abundance.
}
\label{tab:nusm}}

%%%%%%%%%%

In Table~\ref{tab:nugm}, we continue this comparison with the same
mSUGRA point in Table~\ref{tab:nusm}, but this time for models with
non-universal gaugino mass parameters. We show the results for this 
mSUGRA point once again for the convenience of the reader. In column~3,
we consider the MWDM1 model where we raise the 
GUT scale value of $M_1$ to 490~GeV, resulting in a $\tz_1$ state that is 
a mixed bino-wino-higgsino state. The heightened wino and higgsino
components of $\tz_1$ allow for enhanced $\tz_1\tz_1\to W^+W^-$ in the
early universe, thus putting the model into accord with the measured DM
abundance. The model has direct detection rates enhanced by a factor $\sim 7$
over mSUGRA, to the $10^{-8}$~pb range. 
Except for $\tz_1$ which now has a mass of 195~GeV compared to 119~GeV
 in mSUGRA, the sparticle spectra are almost the same in the two
 cases. The heavier $\tz_1$ state implies that
the $\tz_2-\tz_1$ mass gap is
about 29~GeV compared to $\sim 100$~GeV in mSUGRA, and that its
 higgsino-content is somewhat larger than in mSUGRA. 

In the next column we consider a BWCA scenario where we dial $M_1(M_{\rm
  GUT})$ to $-480$~GeV to obtain the observed relic density. Again, we see
  that except for $m_{\tz_1}$ which moves to 202~GeV, sparticle and
  Higgs boson masses as well as the higgsino content of the $\tz_1$ are
  essentially the same as for the corresponding mSUGRA model. The
  $\tw_1-\tz_1$ mass gap is now just 18~GeV, so bino-wino
  co-annihilation acts to reduce the relic density, even though the
  $\tz_1$ remains in a nearly pure bino-like state. The small
  $\tz_2-\tz_1$ mass gap compared to the value of $m_{\tg}$, which might
  be deduced at LHC, would signal a model with non-universal gaugino
  masses.  Notice that for reasons detailed in Ref.~\cite{wtn_dd}, the
  $\tz_1$ direct detection cross section is far lower than any proposed
  detector can probe.

In column 5, we attain the observed relic density by dialing $M_3({\rm
M_{GUT}})$ from 300~GeV to 160~GeV to obtain a viable LM3DM model.  We see
that we now have a much lighter spectrum of sparticles than in mSUGRA,
not only squarks and gluinos, but also charginos and neutralinos (but
not sleptons). In this case, we would expect
huge sparticle production cross
sections at LHC and complicated cascade decay chains. 
Since the reduction of $M_3$ hardly affects slepton masses, sleptons are
not much lighter than squarks even though the sfermion mass scale is
just 300-400~GeV.
The $\tz_1$ has a significant higgsino content leading to MHDM with a
$\tz_2-\tz_1$ mass gap of 58~GeV, while the $\tz_3-\tz_1$ mass gap is
only slightly below $M_Z$, so 3-body $\tz_3$ decays will also occur,
though the dilepton mass will be peaked close to $M_Z$.  The large
higgsino content also results in a correspondingly enhanced $\tz_1$
direct detection cross section -- $7\times 10^{-8}$~pb -- close to the
90\% CL limit from Xenon-10 DM searches.  We also see that the
relatively light top squarks and charginos significantly reduce $BF(b\to
s\gamma )$.

Finally, in the last column in Table~\ref{tab:nugm}, we show results for the
HM2DM model where we get agreement with (\ref{eq:relic}) by raising
$M_2(M_{\rm GUT})$ to 900~GeV. We see that this gives a very low value
of $\mu$ so that the $\tz_1$ becomes MHDM with $R_{\tH}=0.67$. We see
that the left-type squarks and especially the left-type sleptons are now
much heavier than in mSUGRA, while $\tq_R$ and $\tell_R$ are hardly
affected, leading to the left-right split spectrum referred to earlier.
The lighter chargino, and the three lightest neutralinos are all lighter
than in mSUGRA, while $\tz_4$ is very heavy.  The $\tz_2-\tz_1$ mass gap
is $\sim 47$~GeV, while the $\tz_3-\tz_1$ mass gap is just 64~GeV. The
branching ratio for $b\to s\gamma$ is reduced from its mSUGRA value
because $\tst_1$ and the higgsino-like chargino are relatively light,
while the DM direct detection cross-section is large, as is typical of
models with MHDM.
\TABLE{
\begin{tabular}{lccccc}
\hline
Model & mSUGRA & MWDM & BWCA & LM3DM & HM2DM \\
\hline
parameter & --- & $M_1(M_{\rm GUT})$ & $M_1(M_{\rm GUT})$ & $M_3(M_{\rm GUT})$ &
$M_2(M_{\rm GUT})$ \\
special value       & ---   & 490   & -480  & 160   & 900  \\ 
\hline
$\mu$       & 385.1 & 385.9 & 376.6 & 185.3 & 134.8 \\
$m_{\tg}$   & 729.7 & 729.9 & 731.7 & 420.2 & 736.4 \\
$m_{\tu_L}$ & 720.8 & 721.2 & 722.0 & 496.9 & 901.8 \\
$m_{\tu_R}$ & 702.7 & 708.9 & 709.9 & 467.0 & 696.3 \\
$m_{\tst_1}$& 523.4 & 526.5 & 536.3 & 312.2 & 394.3 \\
$m_{\tb_1}$ & 656.8 & 656.0 & 658.9 & 443.2 & 686.4 \\
$m_{\te_L}$ & 364.5 & 371.5 & 371.4 & 366.1 & 669.3 \\
$m_{\te_R}$ & 322.3 & 353.3 & 352.2 & 322.6 & 321.3 \\
$m_{\tw_2}$ & 411.7 & 412.4 & 404.5 & 282.9 & 719.7 \\
$m_{\tw_1}$ & 220.7 & 220.8 & 220.0 & 152.5 & 136.5 \\
$m_{\tz_4}$ & 412.5 & 414.5 & 403.3 & 285.2 & 723.1 \\
$m_{\tz_3}$ & 391.3 & 391.9 & 385.8 & 194.4 & 160.2 \\ 
$m_{\tz_2}$ & 220.6 & 223.2 & 219.2 & 163.6 & 142.3 \\ 
$m_{\tz_1}$ & 119.2 & 194.6 & 201.7 & 105.5 & 94.8 \\ 
$m_A$       & 520.3  & 525.9 & 518.6 & 398.3 & 670.7 \\
$m_{H^+}$   & 529.8  & 535.3 & 528.1 & 408.7 & 679.8 \\
$m_h$       & 110.1  & 110.2 & 109.8 & 106.0 & 111.9 \\ 
\hline
$\Omega_{\tz_1}h^2$& 1.1 & 0.10 & 0.10 & 0.10 & 0.10 \\
$BF(b\to s\gamma) \times 10^4$ & 3.0 & 3.0 & 3.1 & 2.0 & 2.3 \\
$\Delta a_\mu \times 10^{10}$ & 12.1 & 11.8 & 10.1 & 16.4 & 3.1 \\ 
$\sigma_{SI} (\tz_1p ) \times 10^9\ ({\rm pb})$ & 
2.1 & 15 & 0.031 & 72 & 34 \\
$R_{\tH}$ & 0.15 & 0.25 & 0.16 & 0.50 & 0.67 \\
\hline
\end{tabular}
\caption{A comparison of the characteristics of mSUGRA with
 corresponding characteristics in models with gaugino mass
 non-universality that lead to the observed relic abundance of DM.
 Input parameters and resultant sparticle masses in GeV units, together
 with the predicted neutralino relic density, $BF(b\to s\gamma)$ and
 $\Delta a_{\mu}$, the SUSY contribution to the anomalous magnetic
 moment of the muon, the direct detection cross section for the $\tz_1$, and
 finally, the higgsino content of the $\tz_1$. In each case, we fix
 $m_0=m_{1/2}=300$~GeV, $A_0=0$, $\tan\beta =10$ and $m_t=171.4$~GeV,
 and for each non-universal model tune the parameter in the first row to
 its special value shown in row 2 to reproduce the observed relic
 abundance.  }
\label{tab:nugm}}

\section{General characteristics of relic-density-consistent models}
\label{sec:scan}

In this section, we abstract general features of the various models that
we have introduced earlier by performing scans over model parameters,
where we keep only parameter points which lead to a relic density
$\Omega_{\tz_1}h^2\sim 0.11$. We also reject models with
$m_{\tw_1}<103.5$ GeV from LEP2 searches. In the case of the mSUGRA
model, we scan parameters over the range: $m_0:0-5$~TeV, $m_{1/2}:
0-2$~TeV, $A_0=0$, $\tan\beta =10,30,45,50,52,55$ and both signs of
$\mu$. Thus, our scans will include the stau co-annihilation region, the
HB/FP region at large $m_0$ and the $A$-funnel at large $\tan\beta$.
For models with non-universality, in order to have manageable parameter
space scans, we restrict ourselves to a scan over mSUGRA parameters
$m_0:0-2$~TeV, $m_{1/2}:0-1.5$~TeV, $A_0=0$, $\tan\beta =10$ and $\mu
>0$ (except for the HM2DM and BWCA2 models with negative $M_2$ where we
take $\mu <0$ since this is somewhat favored by the measurement of
$(g-2)_{\mu}$).  For models with non-universality, 
for every set of mSUGRA parameters in our scan,
we adjust the special additional  parameter listed in the first rows of 
Tables~\ref{tab:nusm} and \ref{tab:nugm} to bring the 
neutralino
relic density $\Omega_{\tz_1}h^2$  into accord with 
(\ref{eq:relic}). 
Our upper limit on $m_{1/2}$ is chosen somewhat arbitrarily to avoid too
much fine-tuning. We allow much larger values of $m_0$ just in the
mSUGRA region, since in this case it has been argued that fine-tuning in
the HB/FP region is not very large~\cite{hb_fp}.  We limit ourselves to
lower values of $m_0<2$~TeV in the models with non-universality since
MHDM characteristic of the low $|\mu|$ values in the HB/FP region 
can be attained for all values of $m_0$. 

\subsection{Implications for collider searches}

In Fig.~\ref{fig:d_sqgl} we show the value of $m_{\tg}\ vs.\ m_{\tu_R}$
(as a representative value of the approximately degenerate squark mass)
for both signs of mSUGRA, as well as for eight other models with
non-universal SSB terms indicated in the legend on the figure. Each dot
shows the gluino and up-squark mass for 
a model, with parameters chosen  so that
the neutralino relic density saturates (\ref{eq:relic}).
Notice that in this figure, as in several subsequent ones, not all the
colors are visible since some model points are overwritten by other model points. 
The diagonal dashed line for $m_{\tu_R}=m_{\tg}$ shows that when we require
that the observed relic density be obtained, with the exception of the
two branches in the HB/FP region of mSUGRA where $m_0$ (and hence the
squark mass) is very large, all models yield $m_{\tg}\sim m_{\tq}$.  The
two dotted lines denote the approximate reach of the CERN LHC with
100~fb$^{-1}$ of integrated luminosity if $m_{\tg}\simeq m_{\tq}$, as
adapted from Ref.~\cite{lhcreach}.  Most of the models scanned lie
within reach of the CERN LHC, which is partly an artifact from the upper
limit we take on $m_{1/2}$ in our parameter space scans. For the two
branches in the HB/FP region of mSUGRA shown by the red and blue dots at
large $m_0$ in the figure, experiments at the LHC will be sensitive to
models where $m_{\tg}\alt 1.8$~TeV.
\FIGURE[tbh]{
\epsfig{file=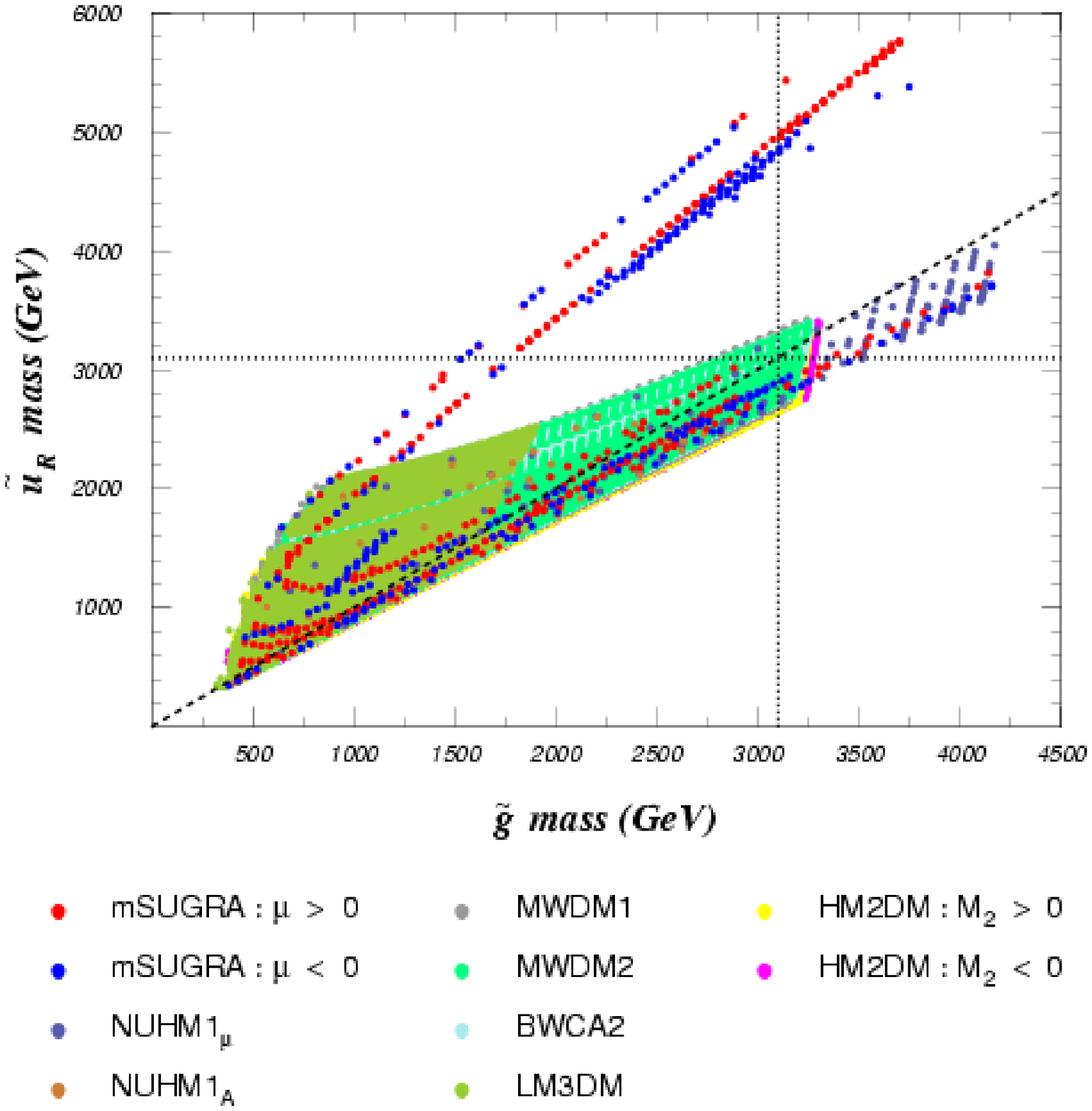,width=12cm,angle=0} 
\caption{\label{fig:d_sqgl} Predictions for $m_{\tg}\ vs.\ m_{\tu_R}$
from various models with $A_0=0$, $m_t=171.4$~GeV and
$\mu >0$ (except in the cases with the blue dots for the mSUGRA model, 
light blue dots for the BWCA2 model 
and magenta dots for the HM2DM model with $M_2<0$ for which we have $\mu
< 0$), but where the special parameter in the various non-universal mass
models has been dialed to yield $\Omega_{\tz_1}h^2\simeq 0.11$.  
We fix $\tan\beta=10$ except for the mSUGRA model where we allow 
$\tan\beta=10$, 30, 45, 50, 52 and 55.
The approximate 100~fb$^{-1}$ reach of CERN LHC is denoted by the dotted
lines, while a dashed line denotes where $m_{\tu_R}=m_{\tg}$. Here, and
in subsequent figures, dots for some of the models are covered up by
other dots, and so are not visible. }}

In Fig.~\ref{fig:d_ht1}, we show these DM-allowed models in the
$m_{\tst_1}\ vs.\ m_h$ plane. Here, we note a clear trend in all
models: heavier $\tst_1$ squarks are correlated with larger values of $m_h$,
largely because top-Yukawa radiative corrections to $m_h$
increase with the stop mass. 
For many models with $m_A \gg M_Z$, then $h \simeq H_{\rm SM}$ so that
the LEP2 lower bound of 114.1~GeV would be applicable.
We have not required this bound in our analysis for reasons discussed
earlier, 
and also to be able to show the trend of $m_h$ with other
observables. The lightest value of $m_{\tst_1}$ occurs in the LM3DM
model, which allows $m_{\tst_1}$ as low as $\sim 200$~GeV, although
even lighter values of $m_{\tst_1}$ 
(readily accessible even at the Tevatron~\cite{BST}) 
would be allowed 
if we also admitted  variation of  $A_0$.
\FIGURE[tbh]{
\epsfig{file=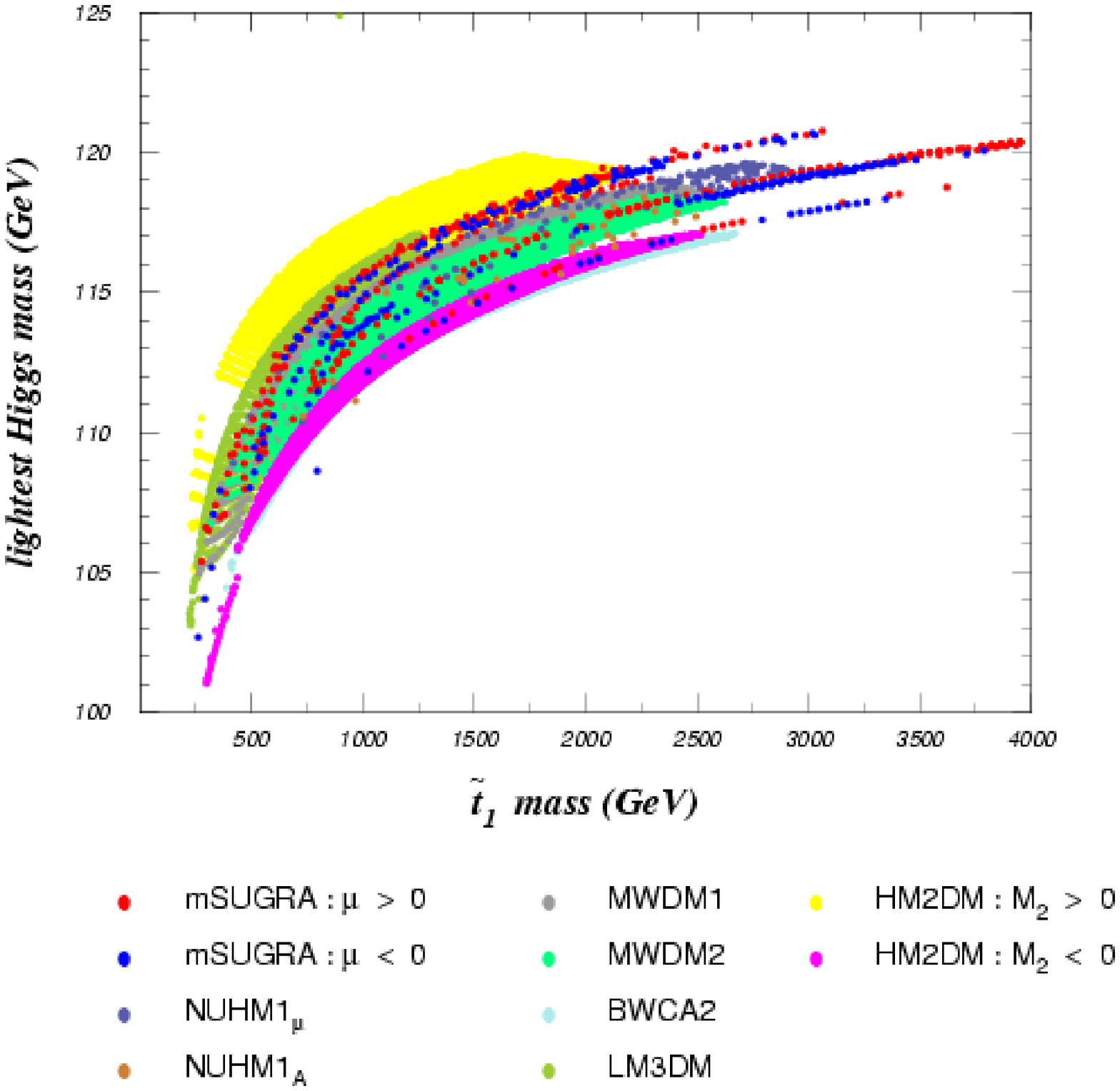,width=12cm,angle=0} 
\caption{\label{fig:d_ht1} Predictions for $m_h\ vs.\ m_{\tst_1}$ from
various models with $A_0=0$, $m_t=171.4$~GeV and the sign of $\mu$ as in
Fig.~\ref{fig:d_sqgl}, but where the special parameter of non-universal
mass models has been dialed to yield $\Omega_{\tz_1}h^2\simeq 0.11$. We
fix $\tan\beta=10$ except for the mSUGRA model where we allow
$\tan\beta=10$, 30, 45, 50, 52 and 55.  }}

In Fig.~\ref{fig:d_z2z1}, we show the lighter chargino mass $m_{\tw_1}\
vs.\ m_{\tz_2}-m_{\tz_1}$. We denote by the dashed line the region where
$m_{\tz_2}-m_{\tz_1}<M_Z$. Below this line, the spoiler decay modes
$\tz_2\to\tz_1 Z$ or $\tz_1h$ are kinematically closed, so that $\tz_2$
must decay via 3-body modes like $\tz_2\to\tz_1\ell\bar{\ell}$. The mass
edge in the invariant dilepton mass distribution from this decay~\cite{mlledge}, which can serve as the starting point for sparticle mass
reconstruction at the LHC as discussed earlier, will be visible as long
as its branching fraction is not strongly suppressed~\cite{btwino}.
From the  figure, we see that {\it most} of the
models that give rise to the correct relic density also predict that the 
spoiler modes are closed so that the
$m(\ell\bar{\ell})$ mass edge will likely be visible! Exceptions arise
from the stau-co-annihilation and $A$-funnel regions of mSUGRA, or at
vestiges of other models which already had the correct relic density
(because the mSUGRA parameters were in the stau co-annihilation region)
so that no special non-universality parameters needed to be adjusted to
get the correct relic density ({\it e.g.} the gray points from the MWDM1
model). Notice also that there are models where the mass gap is very
small. Even for these models it is likely that the $\tz_2$ will be
sufficiently boosted in its production via decays of much heavier
squarks/gluinos so that the daughter leptons have large enough
transverse momenta so as to be detectable. 
\FIGURE[tbh]{
\epsfig{file=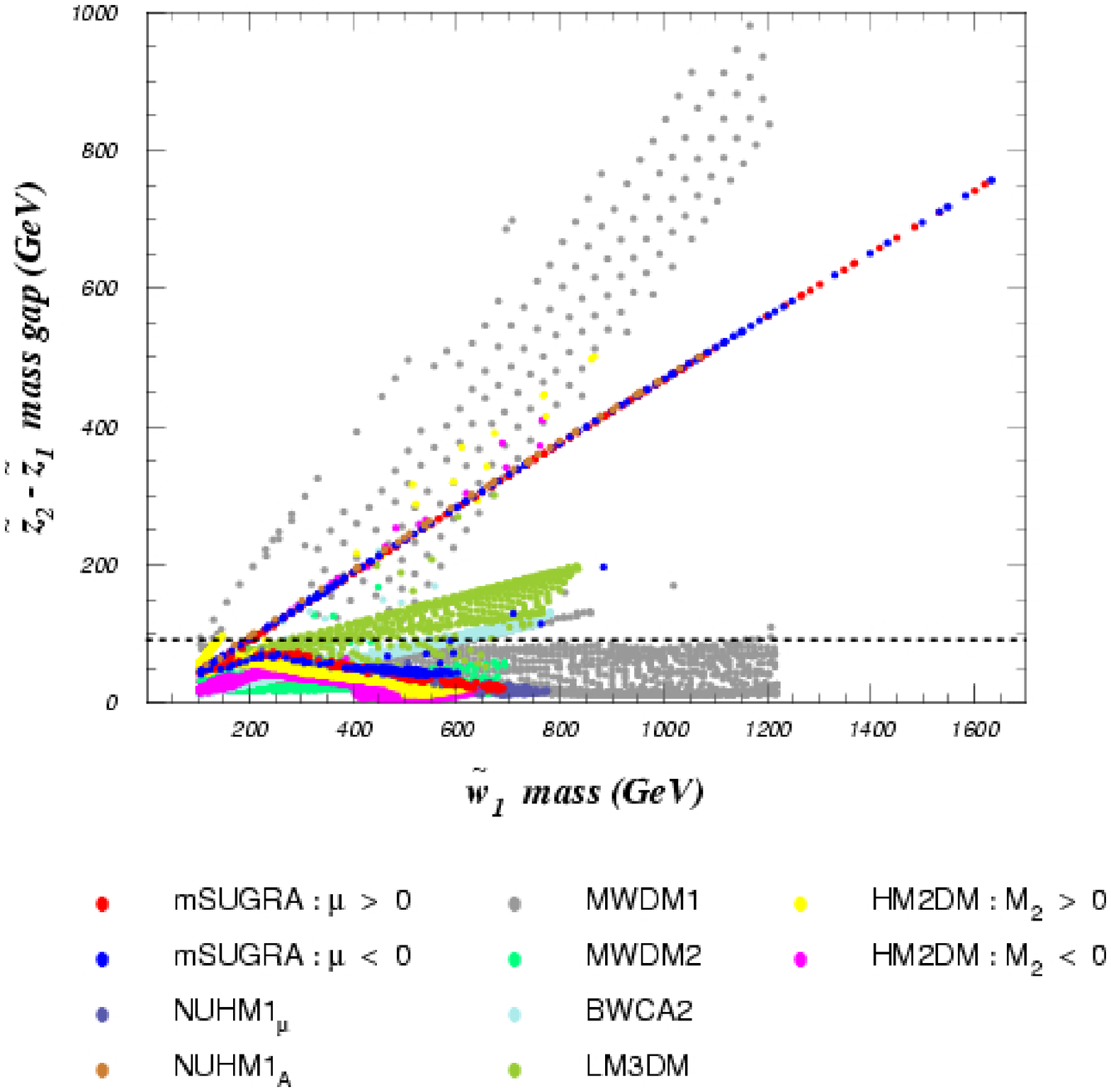,width=12cm,angle=0} 
\caption{\label{fig:d_z2z1} Predictions for $m_{\tw_1}\ vs.\
m_{\tz_2}-m_{\tz_1}$ from various models with $A_0=0$,
$m_t=171.4$~GeV and the
sign of $\mu$ as in Fig.~\ref{fig:d_sqgl}, but where the special parameter 
of non-universal mass models has been
dialed  to yield $\Omega_{\tz_1}h^2\simeq 0.11$.  
We fix $\tan\beta=10$ except for the mSUGRA model where we allow 
$\tan\beta=10$, 30, 45, 50, 52 and 55.
The dashed
line denotes the point where $m_{\tz_2}-m_{\tz_1}=M_Z$, where the
two-body spoiler decay mode $\tz_2\to \tz_1 Z$ turns on.  }}

In Fig.~\ref{fig:d_w1l1}, we show predictions for $m_{\tw_1}$ and
$m_{\ttau_1}$ for WMAP-allowed models. The approximate reach of the
ILC500 ($\sqrt{s}=500$~GeV) and ILC1000 (with $\sqrt{s}=1000$~GeV) are
shown by the dashed and dotted lines , respectively, which delineate the
kinematic limit for $\tw_1^+\tw_1^-$ or $\ttau_1^+\ttau_1^-$ pair
production.  Here, we see that
it is quite easy to evade the ILC reach
and still have a neutralino relic density consistent with
(\ref{eq:relic}).  This is in contrast to prejudices from studies in
the mid-1990s which favored the bulk annihilation region of mSUGRA,
which then implied sparticle mass ought to be quite light, and likely
accessible to LEP2 and ILC500 searches~\cite{bulk}.  The upper
bands of mSUGRA model parameter points correspond to the HB/FP region,
while the bands of points at low $m_{\ttau_1}$ but high
$m_{\tw_1}$ correspond to the stau co-annihilation region in mSUGRA or in
MWDM1 models.
\FIGURE[tbh]{
\epsfig{file=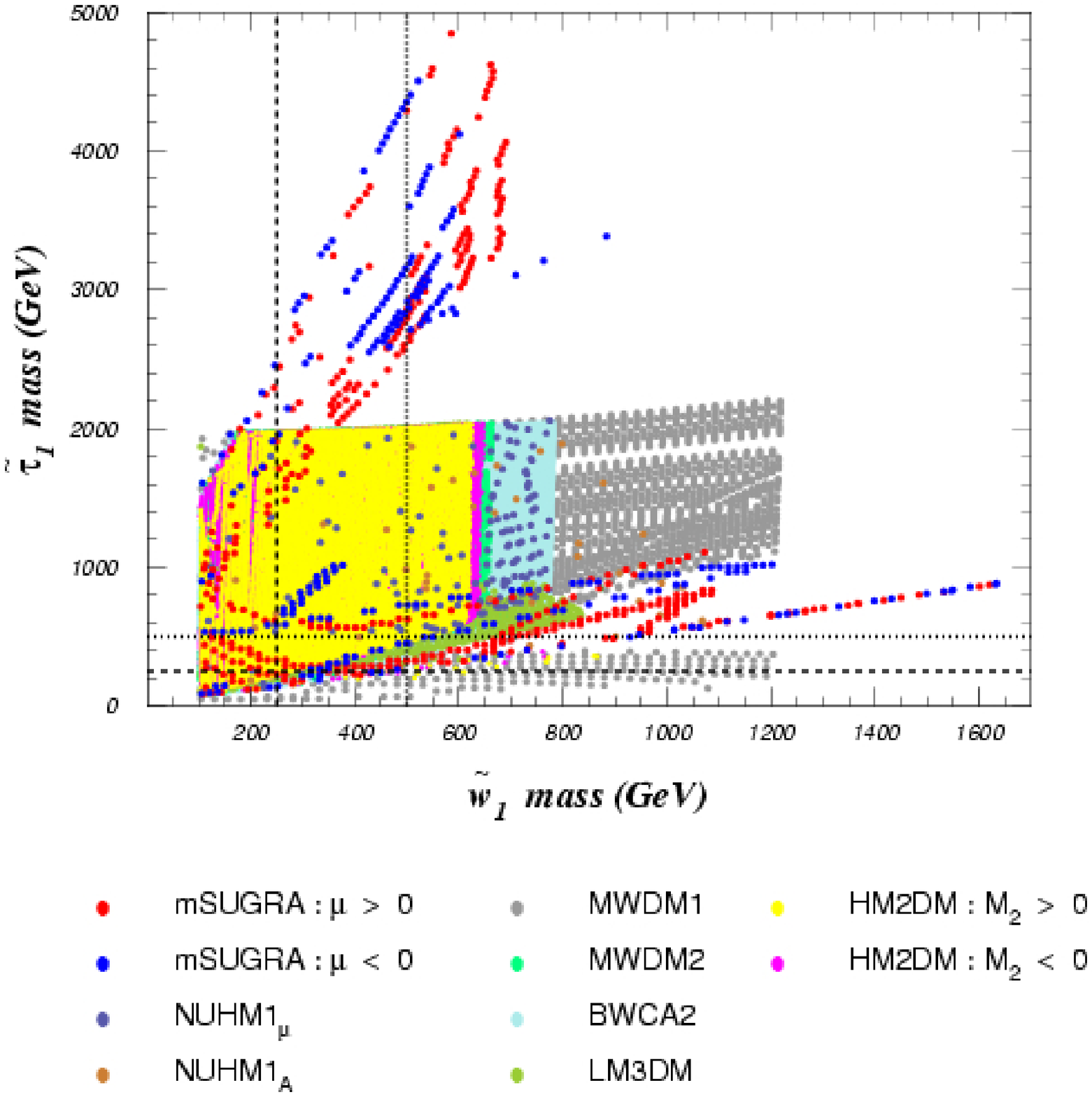,width=12cm,angle=0} 
\caption{\label{fig:d_w1l1} Predictions for $m_{\tw_1}\ vs.\
m_{\ttau_1}$ from various models with $A_0=0$, 
$m_t=171.4$~GeV and the
sign of $\mu$ as in Fig.~\ref{fig:d_sqgl}, but where the special parameter of non-universal mass
models has been dialed to yield $\Omega_{\tz_1}h^2\simeq 0.11$. 
We fix $\tan\beta=10$ except for the mSUGRA model where we allow 
$\tan\beta=10$, 30, 45, 50, 52 and 55.
 The
dashed lines denote the approximate reach of ILC500, while dotted lines
mark the approximate reach of ILC1000. }} 

\subsection{Implications for $(g-2)_\mu$ and $BF(b\to s\gamma )$}

The rare decay $b\to s\gamma$ has always been interesting for SUSY (as
well as other new physics) studies, because the SM and the new physics
contributions  both occur at the one-loop order, and so are likely to be
comparable if the particles in the new-physics loop have masses of about
the weak
scale. This is indeed the case for weak scale SUSY.  
The branching fraction $BF(b\to s\gamma )$ has been measured by the
CLEO, Belle and BABAR collaborations; a combined analysis \cite{bsg_ex}
finds the branching fraction to be $BF(b\to s\gamma )=(3.55\pm
0.26)\times 10^{-4}$, while a recent SM prediction~\cite{misiak} finds
$BF(b\to s\gamma )=(3.15\pm 0.23)\times 10^{-4}$. The theoretical error
in the SUSY case may be somewhat larger. In Figure~\ref{fig:d_bsg}, we
show predictions for $BF(b\to s\gamma )$ in SUSY models where
$\Omega_{\tz_1}h^2\simeq 0.11$, against the value of $m_{\tg}$. We see
that for models with low $m_{\tg}$, large deviations from the SM
prediction are likely, although cases in agreement can be readily
found. 
As $m_{\tg}$ increases, the SUSY loop contributions to $BF(b\to s\gamma )$  
are suppressed.
In the absence of an underlying theory of flavor, we should
be careful in drawing strong inferences from this figure since even a
small amount of flavor-violation in the textures of SSB parameters could 
significantly alter these predictions, with little impact on implications
for direct searches at the LHC.
That $BF(b\to s\gamma)$ has potentially larger SUSY contributions for
mSUGRA models than in non-universal mass models is, of course, an
artifact of our scans: for the mSUGRA model we scan large values of
$\tan\beta$ while we fix $\tan\beta=10$ for models with non-universal masses.
\FIGURE[tbh]{
\epsfig{file=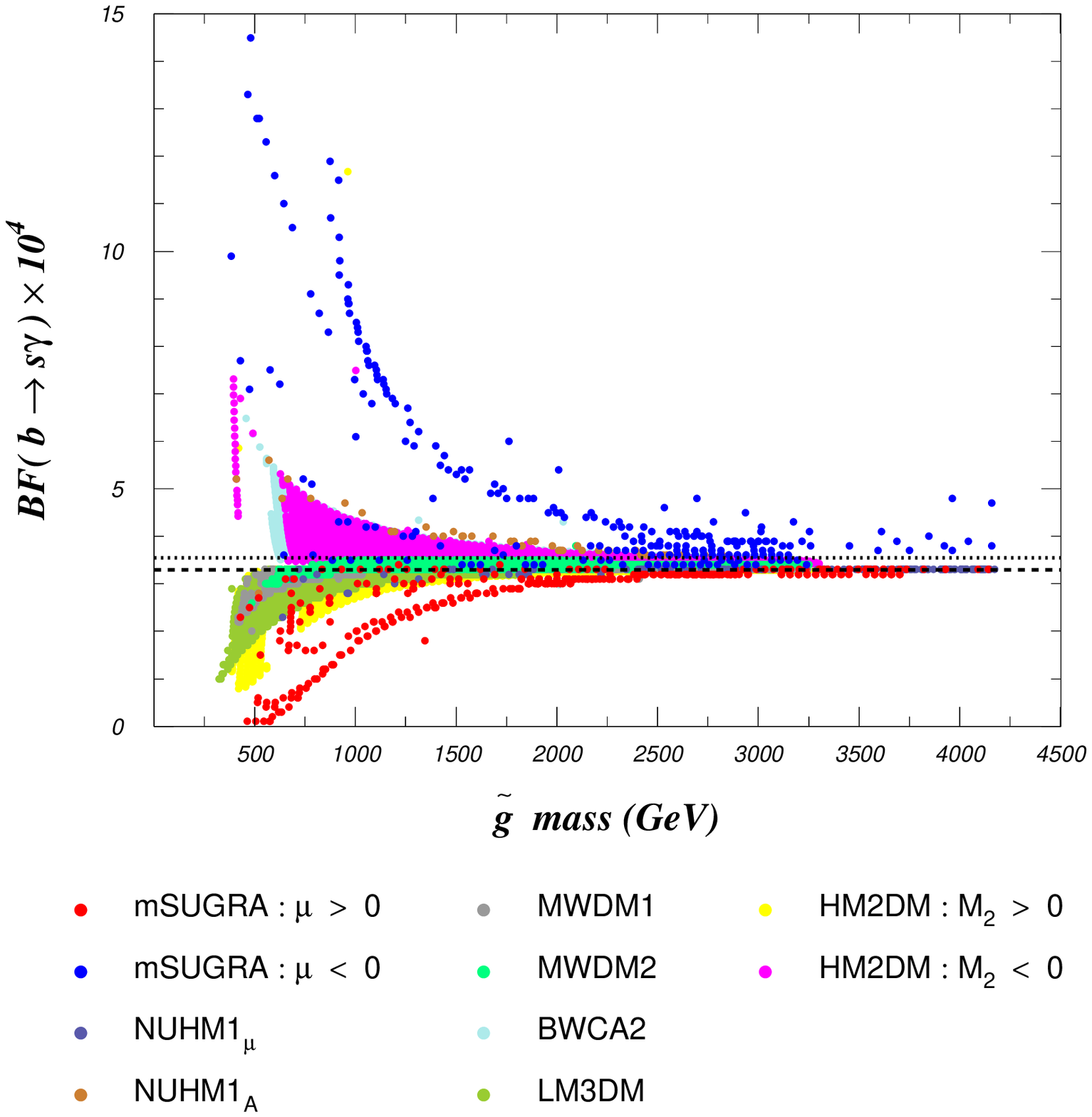,width=12cm,angle=0}
\caption{\label{fig:d_bsg} Predictions for $BF(b\to s\gamma )\ vs.\
m_{\tg}$ from various models with $A_0=0$, 
$m_t=171.4$~GeV and the sign of $\mu$ as in Fig.~\ref{fig:d_sqgl}, but
where the special parameter of non-universal mass models has been dialed
to yield $\Omega_{\tz_1}h^2\simeq 0.11$.  
We fix $\tan\beta=10$ except for the mSUGRA model where we allow 
$\tan\beta=10$, 30, 45, 50, 52 and 55.
The dotted line denotes the
central value of the combined experimental measurements, while the
dashed line denotes the corresponding SM prediction. The SUSY
contribution to the branching fraction is sensitive to $\tan\beta$ which
is varied for the scans of the mSUGRA model, but fixed at $\tan\beta=10$
for the scans in the case of non-universal mass models.}}  

Recent  measurements of the muon anomalous magnetic moment show an
apparent deviation from SM predictions. Combining QED, electroweak,
hadronic (using $e^+e^-\to {\rm hadrons}$ to evaluate hadronic loop
contributions) and light-by-light contributions, and comparing against
measurements from E821 at BNL, a {\it positive} deviation in
$a_\mu\equiv \frac{(g-2)_\mu}{2}$ of
\be
\Delta a_\mu =a_\mu^{exp} -a_\mu^{SM} =22(10)\times 10^{-10} 
\ee
is reported in the Particle Data Book~\cite{pdb}, {\it i.e.} a
$2.2\sigma$ effect.\footnote{More recent analyses\cite{davier} report a
larger discrepancy if only electron-positron data are used for the evaluation
of the hadronic vacuum polarization contribution; the significance of
the discrepancy is, however, reduced if tau decay data are used for this
purpose.} 

One-loop diagrams with $\tw_i-\tnu_\mu$ and $\tz_i-\tmu_{1,2}$ in the
loop would give supersymmetric contributions to $a_\mu$, perhaps
accounting for the (rather weak, yet persistent) discrepancy with the
SM. In Fig.~\ref{fig:d_amu}, we show $\Delta a_\mu^{\rm SUSY}$ versus
$m_{\tmu_L}$.  The dashed line indicates the central value of
the experiment/theory discrepancy as presented by the Particle Data
Group.  We see that a variety of models are able to account for the
discrepancy {\it as long as} $m_{\tmu_L}\alt 1-1.5$ TeV for mSUGRA, but
only about 500~GeV in the case of models with non-universal mass
parameters.  This is partly a consequence of the fact that in mSUGRA
our scans include large values of $\tan\beta$ in mSUGRA but are limited
to $\tan\beta=10$ for non-universal models.  Since $\Delta a_\mu^{\rm
SUSY} \propto \tan\beta$, had we allowed larger values of $\tan\beta$ in
our scans of models with non-universality, then 
the $\Delta a_\mu$ projections would increase beyond those plotted here,
and consistency with the present central value would be possible for
values of second generation slepton masses beyond the reach of a
1~TeV ILC. 
\FIGURE[tbh]{
\epsfig{file=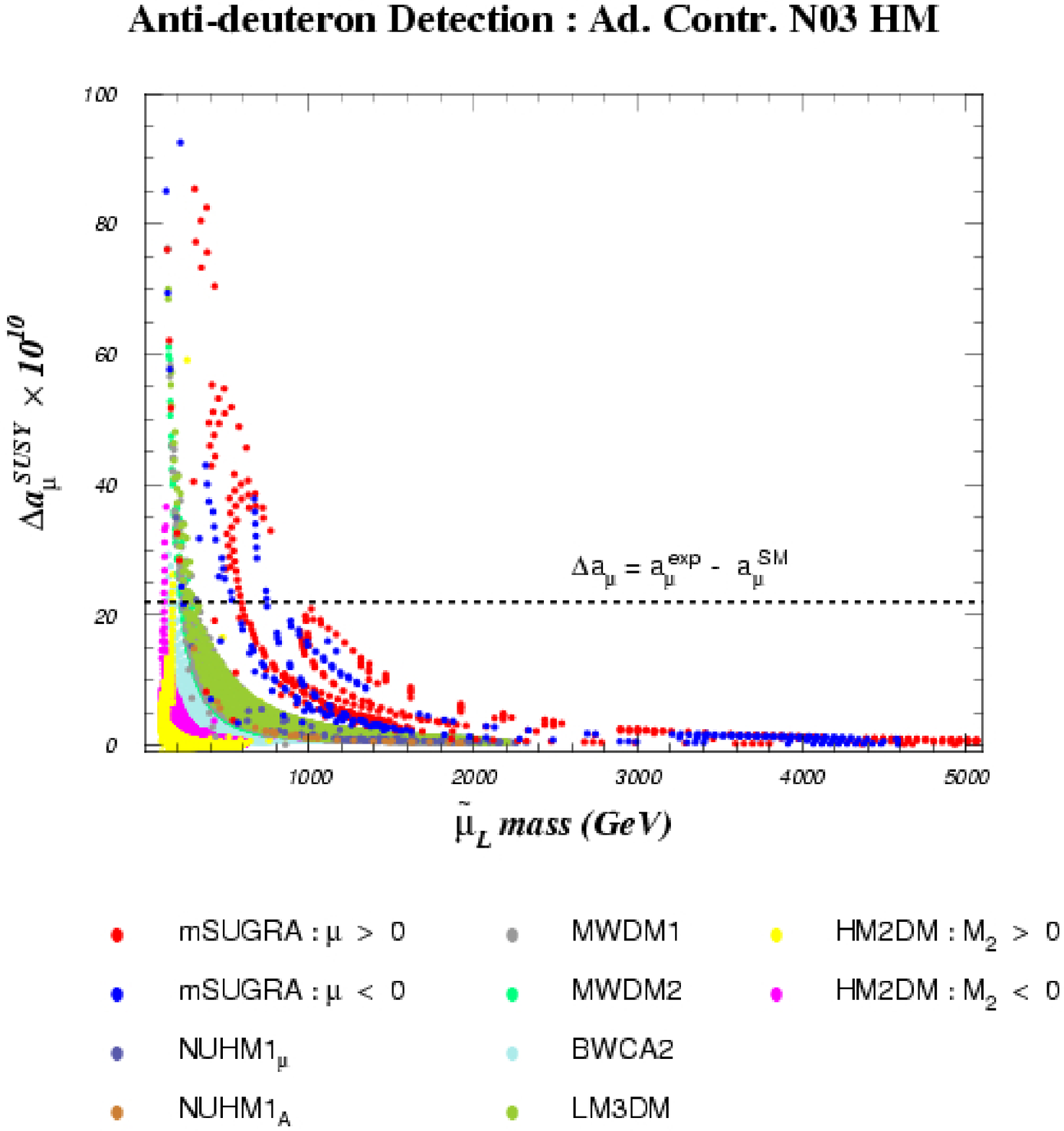,width=12cm,angle=0} 
\caption{\label{fig:d_amu} Predictions for $\Delta a_\mu^{SUSY}\ vs.\
m_{\tmu_L}$ from various models with $A_0=0$, 
$m_t=171.4$~GeV and the sign of $\mu$ as in Fig.~\ref{fig:d_sqgl}, but
where the special parameter of non-universal mass models has been dialed
to yield $\Omega_{\tz_1}h^2\simeq 0.11$.  
We fix $\tan\beta=10$ except for the mSUGRA model where we allow 
$\tan\beta=10$, 30, 45, 50, 52 and 55.
The dashed line denotes the
central value of the measured deviation from SM expectations as reported
by the Particle Data Group, though some recent analyses would infer an
even larger deviation as discussed in the text. Note that $\Delta a_\mu$
is sensitive to $\tan\beta$ which is varied for mSUGRA, but for scans of
the non-universal mass models, is fixed to be 10. }}

\subsection{Implications for direct detection of dark matter}

In Fig.~\ref{fig:dd}, we show the spin-independent neutralino-proton
scattering cross section, calculated with IsaReS program~\cite{isares}
from the IsaTools package, versus $m_{\tz_1}$.  A significant
uncertainty in the cross section comes from the value of the
pion-nucleon $\Sigma$-term~\cite{sigmaterm}.  In this plot we assumed
$\Sigma = 45$~MeV, but larger values can increase our predictions by
factor of about three.
This plot is an update
of similar results presented in Ref.~\cite{wtn_dd} in that it includes
additional models.  
We also show the current limit established by the
Xenon-10 collaboration~\cite{xenon10} (solid line), along with the
projected reaches for the SuperCDMS (25~kg)~\cite{supercdms} (dashed line), 
LUX 300~kg~\cite{lux}(dot-dashed line) and Xenon-1~ton~\cite{xenon} (dotted line) 
experiments. 
The reach contours have been generated assuming a standard local density and velocity profile.

We see two
distinct classes of models. In the first class, the neutralino-nucleon
cross section falls off with $m_{\tz_1}$ , while in the second class --
models with a well-tempered neutralino with significant higgsino component 
-- this
cross section asymptotes to about $10^{-8}$~pb, within the reach of the
next generation of detectors such as LUX-300~kg, Xenon-100 or super-CDMS. It is
important to realize that this second class includes {\it several} of
the specific models that we have considered where agreement with
(\ref{eq:relic}) is obtained via a significant higgsino component in the
$\tz_1$ so that we have either mixed higgsino DM or mixed wino-bino-higgsino
DM. This higgsino component then leads to a large cross section for
$\tz_1 p$ scattering via diagrams involving $h$ and/or $H$ exchanges,
where the Higgs bosons couple to the proton via both its quark and its
gluon content. The neutralino annihilation rate in the early universe
generally falls off with increasing  $m_{\tz_1}$, so that for heavier
neutralinos, a larger
higgsino content is necessary to maintain the
relic density at its observed value: it is precisely this increased
higgsino-content that maintains the direct-detection cross section
around 10$^{-8}$~pb even for large values of $m_{\tz_1}$ in the upper
branch of the figure. There are, however, many models where
accord with the observed CDM relic density is obtained by adjusting the
{\it masses} to get either stau co-annihilation or bino-wino
co-annihilation or Higgs funnel annihilation. In these cases of the
bino-like LSP, the direct detection cross section falls with $m_{\tz_1}$
to below the sensitivity of even 1t noble element detectors for
neutralino masses below about 400~GeV. There even are cases with
$m_{\tz_1}\alt 100$~GeV where -- due to interference between
various contributing amplitudes~\cite{ellis,wtn_dd} ({\it e.g.} mSUGRA
with $\mu < 0$) -- the neutralino-nucleon scattering cross section drops 
to well below $10^{-10}$~pb, which is below the projected sensitivity of all
proposed detectors to date. 
These same interference effects frequently
lead to a reduced cross section in the BWCA2 case where we also
take $\mu < 0$, the sign favored by the value of $\Delta a_\mu$. 

\FIGURE[tbh]{
\epsfig{file=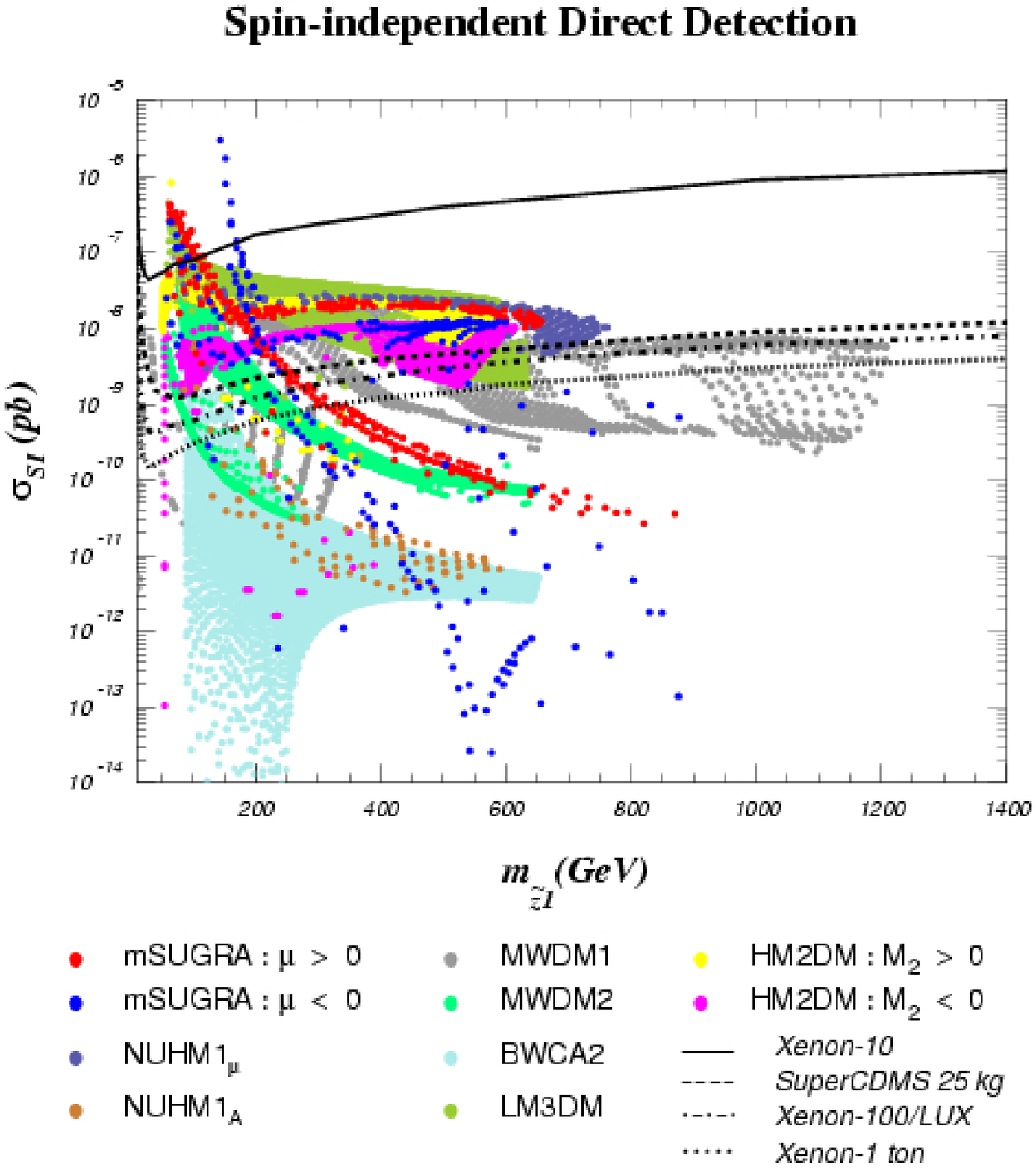,width=12cm,angle=0} 
\caption{\label{fig:dd} Predictions for $\sigma_{SI}(\tz_1 p)\ vs.\
m_{\tz_1}$, generally regarded as the figure of merit for direct
detection experiments, in various models with $A_0=0$, 
$m_t=171.4$~GeV and the
sign of $\mu$ as in Fig.~\ref{fig:d_sqgl}, but where the special parameter of
non-universal mass models has been dialed to yield
$\Omega_{\tz_1}h^2\simeq 0.11$.  
We fix $\tan\beta=10$ except for the mSUGRA model where we allow 
$\tan\beta=10$, 30, 45, 50, 52 and 55.
We also show the projected reach of
selected direct detection experiments.  }}

The neutralino may also scatter inside a detector via its spin-dependent
coupling to the nucleon due to its couplings to the $Z$ or to
squarks. In Fig.~\ref{fig:sisd} we show how this spin-dependent cross
section is expected to scale with the corresponding spin-independent
cross section in the various models that we have considered. It is
striking to see that while the spin-dependent cross section in
relic-density-consistent models
may be as
low as $10^{-8}$~pb, in well-tempered neutralino models where agreement
with the relic density is obtained by adjusting the higgsino content of
$\tz_1$ this cross section is always larger than $10^{-5}$~pb well above
the projected sensitivity, $\sigma_{\rm SD}(\tz_1 p) \agt 4\times
10^{-7}$~pb, of the proposed COUPP experiment with a target mass of 1t
\cite{coupp}. This is of course, because higgsinos have a large coupling
to the $Z$-boson. We note that there may be an observable signal via
spin-dependent couplings even for cases where the prospects for
direct detection via the spin-independent neutralino interaction appear to be
hopeless. We also remark that the 50 kg prototype of the COUPP
detector is projected to have a sensitivity $\sigma_{SD}(\tz_1 p)\sim
4\times 10^{-4}$~pb.

\FIGURE[tbh]{
\epsfig{file=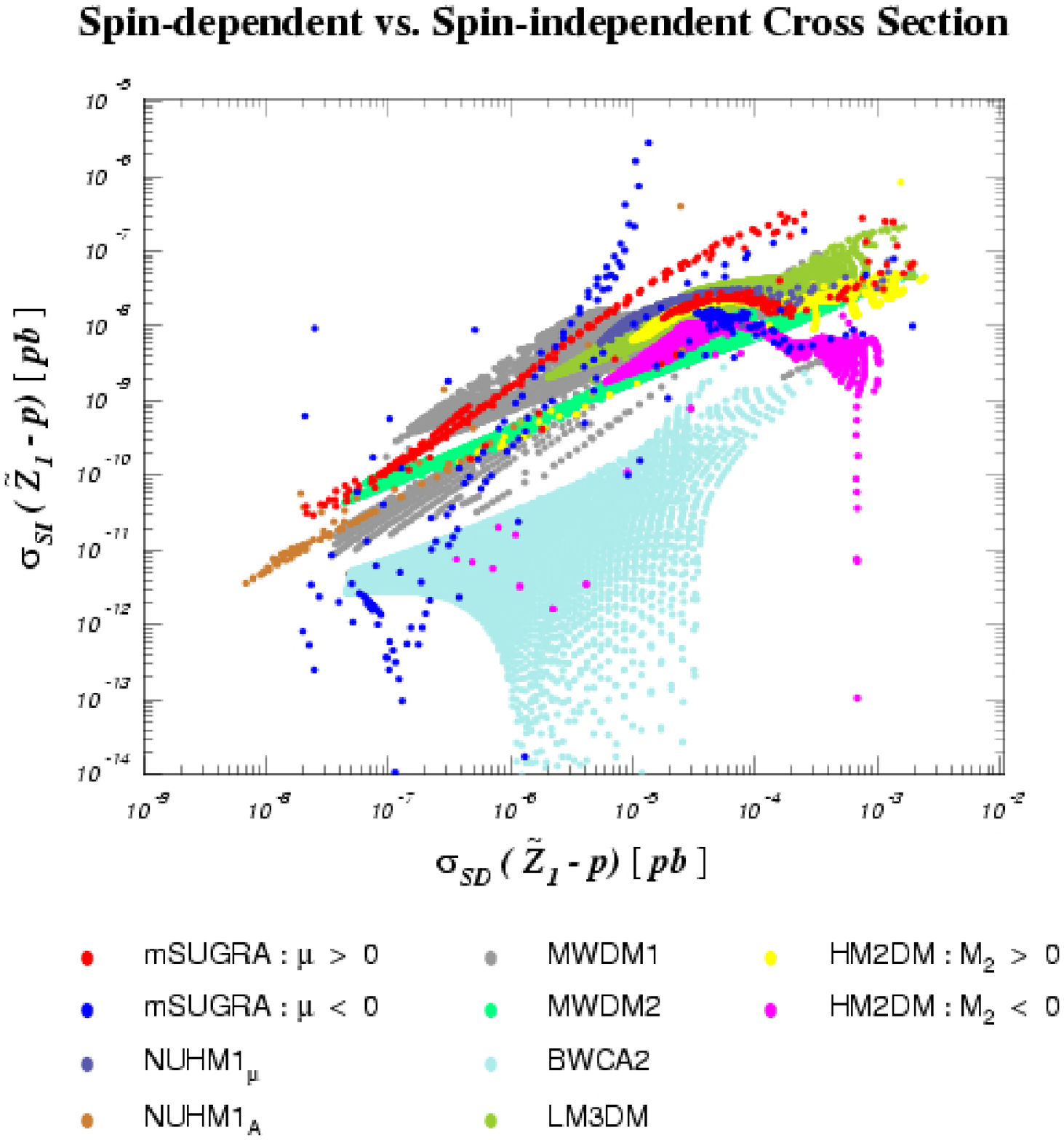,width=12cm,angle=0} 
\caption{\label{fig:sisd} Predictions for $\sigma_{SD}(\tz_1 p)\ vs.\
\sigma_{SI}(\tz_1 p)$ in various models with $A_0=0$, 
$m_t=171.4$~GeV and the
sign of $\mu$ as in Fig.~\ref{fig:d_sqgl}, but where the special parameter of
non-universal mass models has been dialed to yield
$\Omega_{\tz_1}h^2\simeq 0.11$.  
We fix $\tan\beta=10$ except for the mSUGRA model where we allow 
$\tan\beta=10$, 30, 45, 50, 52 and 55.}}

\subsection{Implications for indirect detection of dark matter: 
neutrino telescopes}

In Fig.~\ref{fig:idd_mu}, we show the flux of muons with $E_\mu >50$~GeV
which is expected from neutralino capture by the sun, with subsequent
neutralino annihilation in the solar core to $\nu_\mu$ states (
for some recent work, see Ref. \cite{gabe}). To
calculate these and subsequent indirect dark matter detection rates, we
use the DarkSUSY~\cite{darksusy} - Isajet interface. The flux of high
energy $\nu_\mu$ from neutralino annihilation in the solar core depends
on both the neutralino capture cross section as well as on the
neutralino annihilation cross section. The capture rate mainly depends
on the {\it spin-dependent} neutralino-nucleon cross section, the main
contribution to which comes from $Z$-exchange processes that are
enhanced if the neutralino has a significant higgsino content. Note that
it is easily possible to have a large spin-independent
neutralino-nucleon scattering cross section, and yet a small
spin-dependent cross section, leading to undetectable rates in the
IceCube experiment
This is exemplified in
Fig.~\ref{fig:idd_mu}, we {\it again} find that the well-tempered
neutralino models aggregate around an asymptotic regime of $\sim
10-100$~events/km$^2$/yr.  The approximate reach of IceCube for
$E_\mu>50$~GeV is around the 40~events/km$^2$/yr~\cite{icecube}, so again many of the
models with mixed gaugino/higgsino dark matter stand a good chance of
indirect detection via neutrino telescopes, whereas models where
(\ref{eq:relic}) is satisfied in other ways fall below the detectable
level. We remark though that if $m_{\tz_1} \sim 100$~GeV, even the signal from
BWCA2 may be in the detectable range. 
We mention that projections for IceCube, unlike those for
indirect detection from neutralino annihilation to anti-matter or gamma
rays (discussed next), are only slightly sensitive to the DM halo profile. 

\FIGURE[tbh]{
\epsfig{file=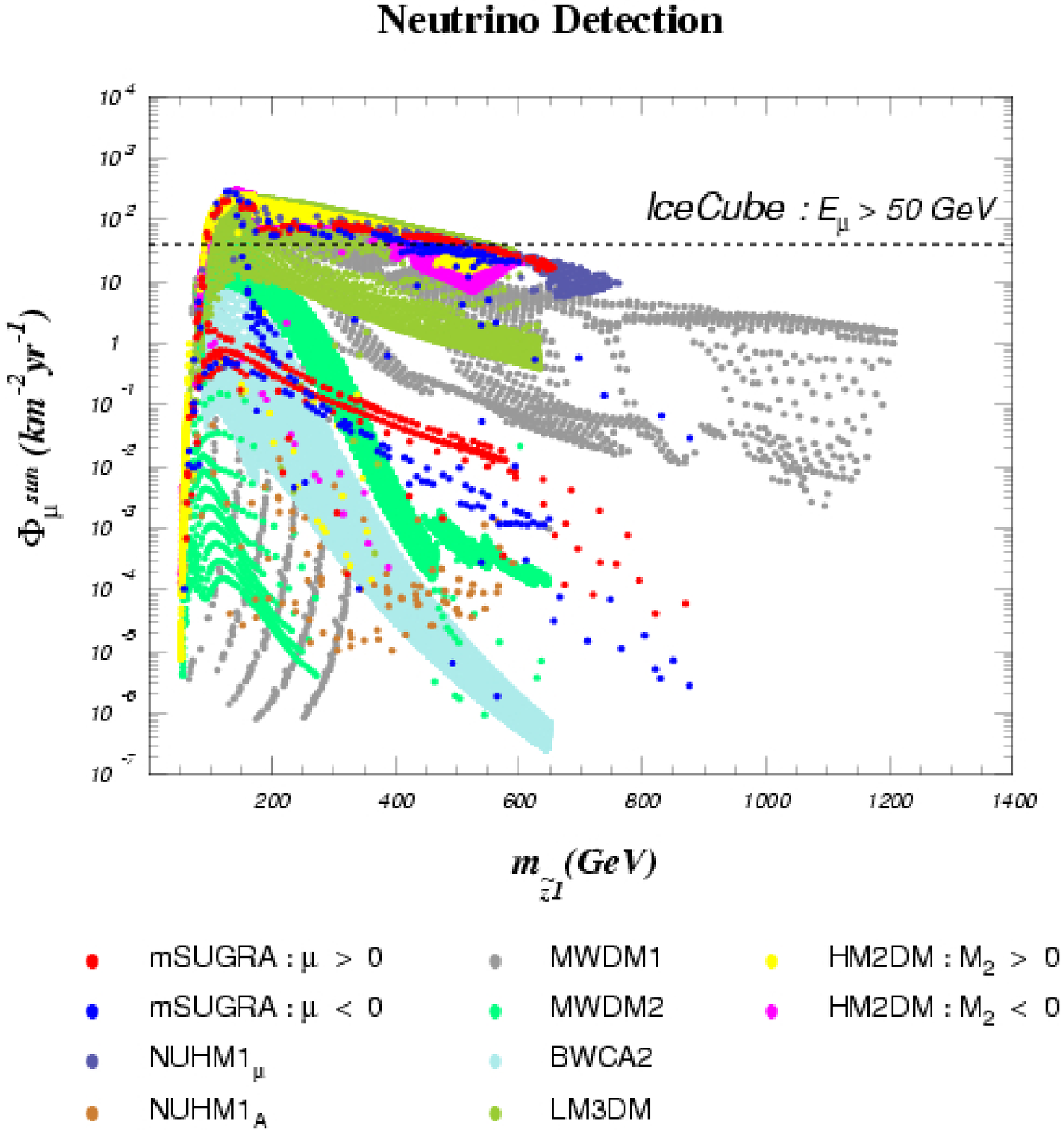,width=12cm,angle=0} 
\caption{\label{fig:idd_mu} Predictions for muon flux with $E_\mu
  >50$~GeV $\Phi_\mu\ vs.\ m_{\tz_1}$ from various models with $A_0=0$,
   $m_t=171.4$~GeV and the
sign of $\mu$ as in Fig.~\ref{fig:d_sqgl}, but where the special
  parameter of non-universal mass models has been dialed to yield
  $\Omega_{\tz_1}h^2\simeq 0.11$.  
We fix $\tan\beta=10$ except for the mSUGRA model where we allow 
$\tan\beta=10$, 30, 45, 50, 52 and 55.
The region above the dashed line
  denotes the approximate reach of the IceCube neutrino telescope.}}

\subsection{Implications for indirect detection of dark matter from halo 
annihilations}

An alternative method for indirect detection of dark matter is to search
for the debris resulting from dark matter annihilation in the galactic
halo. One promising method is to search for GeV-scale $\gamma$-rays,
which could come directly from $\tz_1\tz_1\to\gamma\gamma$ via a box and
triangle diagrams, or via $\tz_1\tz_1\to q\bar{q}$, where hadronization
and decay lead to gamma rays via the $q\to\pi^0\to \gamma$ chain. The
direct process occurs at low rates, but would have a characteristic
signal at the source with $E_\gamma\simeq m_{\tz_1}$, whereas gamma rays
coming from quark hadronization should be more abundant, but will yield
a continuum distribution in $E_\gamma$ with a cut-off at $m_{\tz_1}$.

Since $\gamma$-rays from neutralino annihilation should propagate
undeflected through the galaxy, a good place to look is the galactic
center, where the DM density is expected to be high.  In
Fig.~\ref{fig:idd_gam}, we show the flux of $\gamma$-rays coming from
the direction of the galactic center with $E_\gamma >1$~GeV, in units of
events/cm$^2$/s. The result is very sensitive to the choice of the galactic
dark matter density profile, as well as to (unknown) details of
how clumpy the halo distribution is.  We show results for the
Adiabatically Contracted N03 halo profile~\cite{n03}, 
where the deepening of gravitational potential wells caused by baryon
in-fall leads to a higher DM concentration in the center of the Milky
Way, and a concomitantly larger gamma ray flux. 
Other halo distributions, such as
the Burkert profile~\cite{burkert}, where the central DM halo cusp is
smoothed out by significant re-heating,
predict gamma ray fluxes that may be {\it four orders of
magnitude smaller}! The reach of the GLAST satellite experiment is indicated
by the dashed line~\cite{glast}. The important point is that models with large
$s$-wave neutralino annihilation cross sections cluster around an
asymptotic level of $\sim 10^{-7}/$cm$^2$/s, while models which rely on
co-annihilation such as BWCA or MWDM2 predict much lower
gamma ray fluxes.  A notable difference between signals from halo
annihilation versus signals from direct and neutrino detection is that
the halo annihilation signals can be enhanced by moving $2m_{\tz_1}$
onto the $A$-resonance~\cite{bo}: if neutralinos have enhanced
annihilation through the $A$-funnel in the early universe, then they can
also readily annihilate through the $A$-funnel in the galactic halo
(this does not hold true for the $h$ and $H$ resonances, which are
dominantly $p$-wave, or the $Z$ pole, which is not resonance
enhanced~\cite{griest}).  In the case of $\gamma$-ray signals, we see the
orange dots from NUHM1$_A$ model now populate higher rate levels than
the BWCA and MWDM2 cases, whereas for direct and $\nu_\mu$ signals in
Figs.~\ref{fig:dd} and \ref{fig:idd_mu}, the NUHM1$_A$ signal was
comparable to or even lower than the BWCA and MWDM2 models.
\FIGURE[tbh]{
\epsfig{file=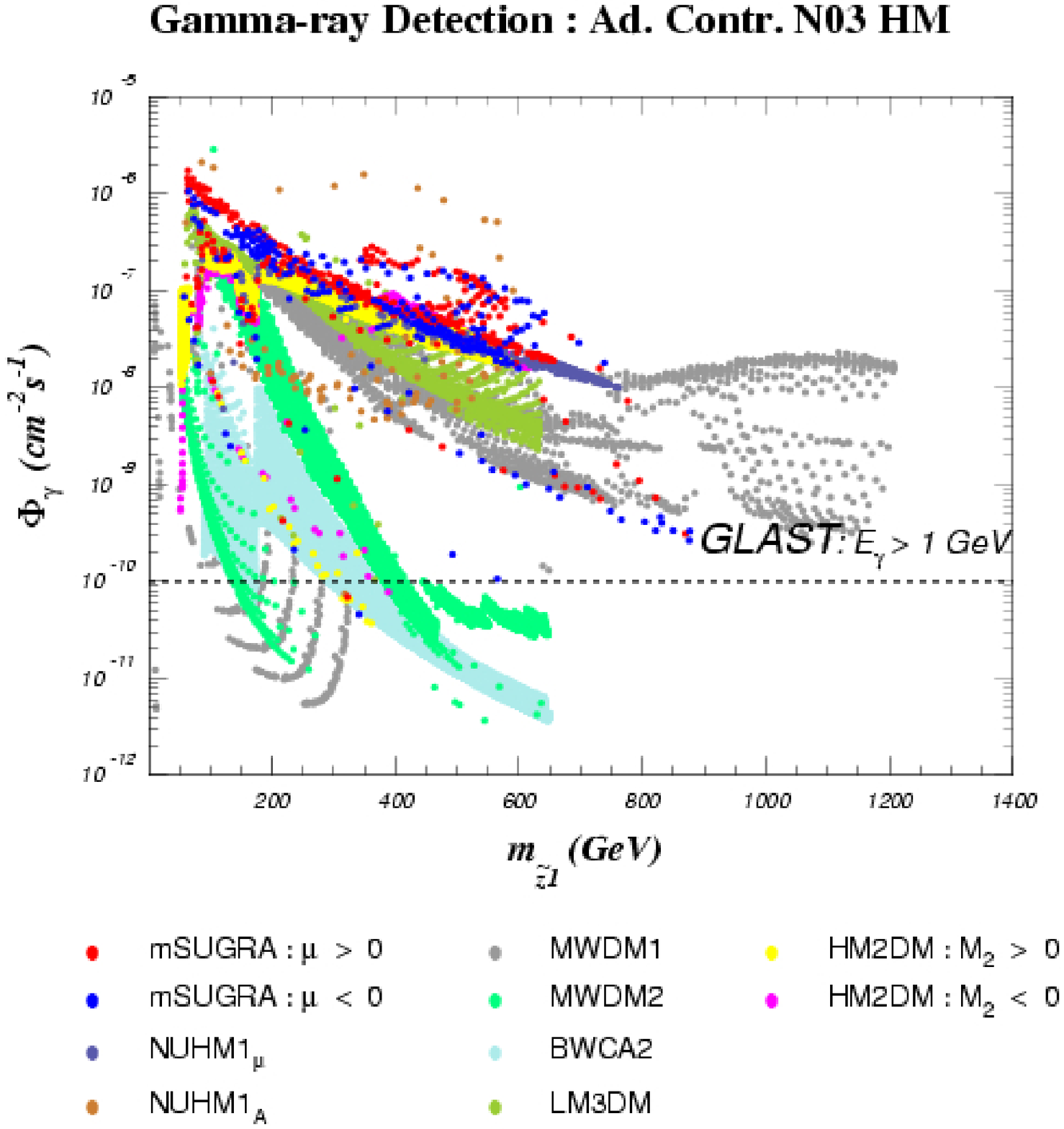,width=12cm,angle=0} 
\caption{\label{fig:idd_gam} Predictions for gamma ray flux with
$E_\gamma >1$~GeV from various models with $A_0=0$,
$m_t=171.4$~GeV and the sign of $\mu$ as in Fig.~\ref{fig:d_sqgl}, but
where the special parameter of non-universal mass models has been dialed
to yield $\Omega_{\tz_1}h^2\simeq 0.11$.  
We fix $\tan\beta=10$ except for the mSUGRA model where we allow 
$\tan\beta=10$, 30, 45, 50, 52 and 55.
We adopt the Adiabatically
Contracted N03 DM halo profile.  The region above the dashed line
denotes the approximate reach of the GLAST experiment. The flux
predicted from less cusped halo profiles may be down by as much as four
orders of magnitude.}}  

Another characteristic signature of DM halo annihilations is the
detection of large fluxes of anti-particles such as $\pbar$s, $e^+$s or
anti-deuterons $\bar{D}$.  For positrons and antiprotons, we evaluate
the averaged differential antiparticle flux in a projected energy bin
centered at a kinetic energy of 20~GeV, where we expect an optimal
statistics and signal-to-background ratio at space-borne antiparticle
detectors~\cite{antimatter,statistical}.  We take the experimental
sensitivity to be that of the Pamela experiment after three years of
data-taking as our benchmark.  For $\bar{D}$s, we evaluate the average
differential anti-deuteron flux in the $0.1<T_{\bar D}<0.25$~GeV range,
where $T_{\bar D}$ stands for the anti-deuteron kinetic energy per
nucleon, and compare it to the estimated GAPS sensitivity for an
ultra-long duration balloon-borne experiment~\cite{gaps} (see
Ref.~\cite{baerprofumo} for an updated discussion of the role of
antideuteron searches in DM indirect detection).

In Fig.~\ref{fig:idd_pbar} we show the flux of $\pbar$s assuming the
Adiabatically Contracted N03 halo profile; results from using the
Burkert profile yield results typically a factor of 10-20 below
these. The models with mixed higgsino dark matter cluster at high levels
of around $\sim 10^{-8}$ events/GeV/cm$^2$/s/sr while the $A$-funnel
annihilation case of NUHM1$_A$ populates the $10^{-9}-10^{-7}$
events/GeV/cm$^2$/s/sr range. The co-annihilation cases BWCA and MWDM2
and stau-co-annihilation in mSUGRA lie at much lower levels.
\FIGURE[tbh]{
\epsfig{file=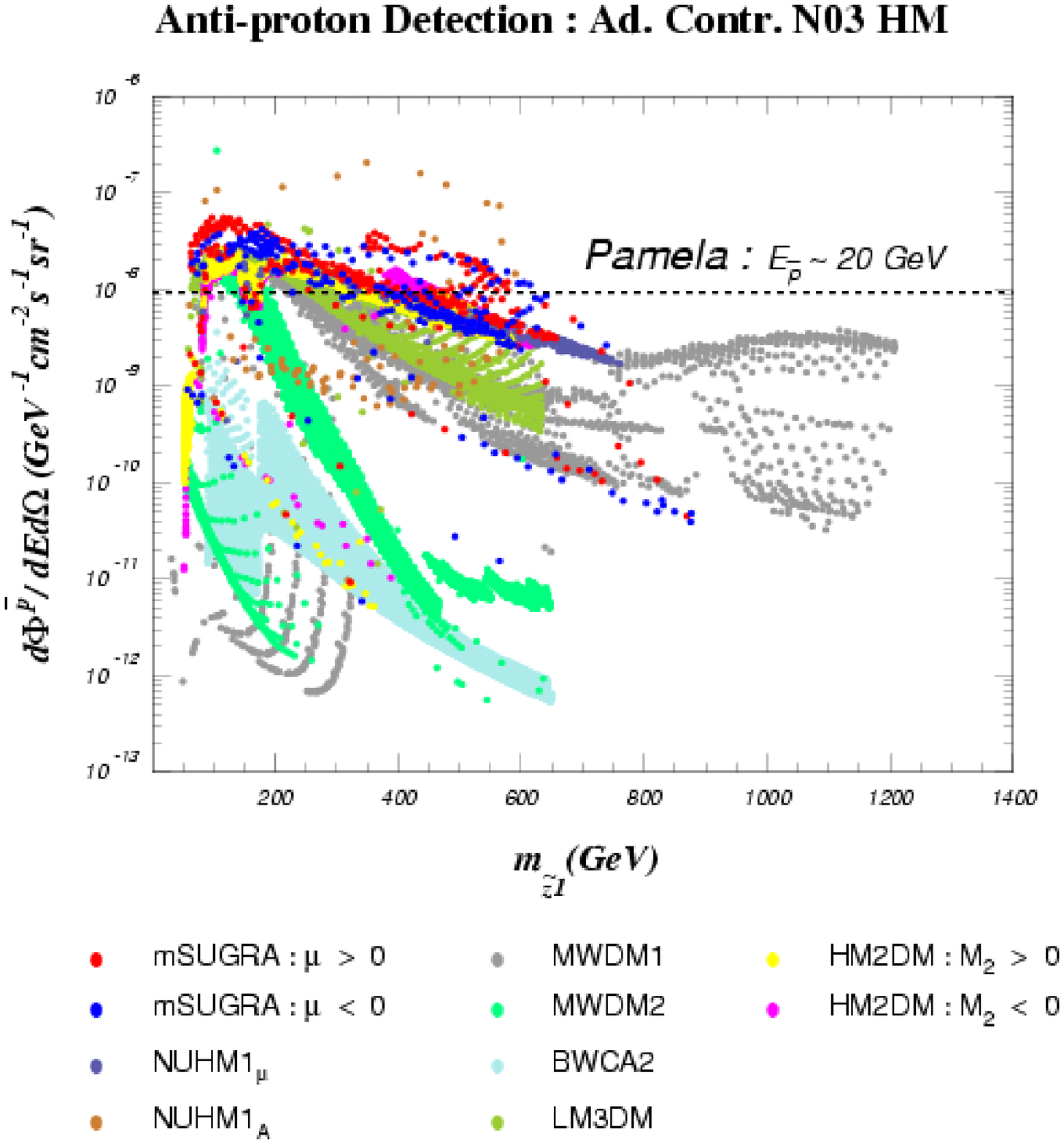,width=12cm,angle=0} 
\caption{\label{fig:idd_pbar} 
Predictions for anti-proton flux from various models with
$A_0=0$, $m_t=171.4$~GeV and the
sign of $\mu$ as in Fig.~\ref{fig:d_sqgl}, but where 
the special parameter of non-universal mass models 
has been dialed to yield $\Omega_{\tz_1}h^2\simeq 0.11$.
We fix $\tan\beta=10$ except for the mSUGRA model where we allow 
$\tan\beta=10$, 30, 45, 50, 52 and 55.
We adopt the Adiabatically Contracted N03 DM halo profile.
The region above the dashed line denotes the approximate reach of the 
PAMELA experiment.}} 

In Fig. \ref{fig:idd_ep} we show the flux of $e^+$s using the
Adiabatically Contracted N03 halo profile; results from using the
Burkert profile yield results about a factor of 3-5 lower\footnote{To
reach earth before losing too much energy and annihilating, the
positrons must originate from annihilation much closer to earth than for
$\bar{p}$s or $\gamma$s; thus, predictions for their flux are less
sensitive than those for anti-protons and gamma rays to the choice of
halo profile. Different halo distributions mainly differ on the DM
density near the galactic center, but agree on the local DM
density.}. Our projections are not optimistic. The models with mixed
higgsino dark matter cluster at the $\sim 10^{-9}$ events/GeV/cm$^2$/s/sr
level, which may be just below the Pamela reach~\cite{pamela}.  The $A$-funnel
annihilation case of NUHM1$_A$ populates the $10^{-10}-10^{-8}$
events/GeV/cm$^2$/s/sr range. The co-annihilation cases BWCA and MWDM2
and stau-coannihilation in mSUGRA are again at much lower levels.
\FIGURE[tbh]{
\epsfig{file=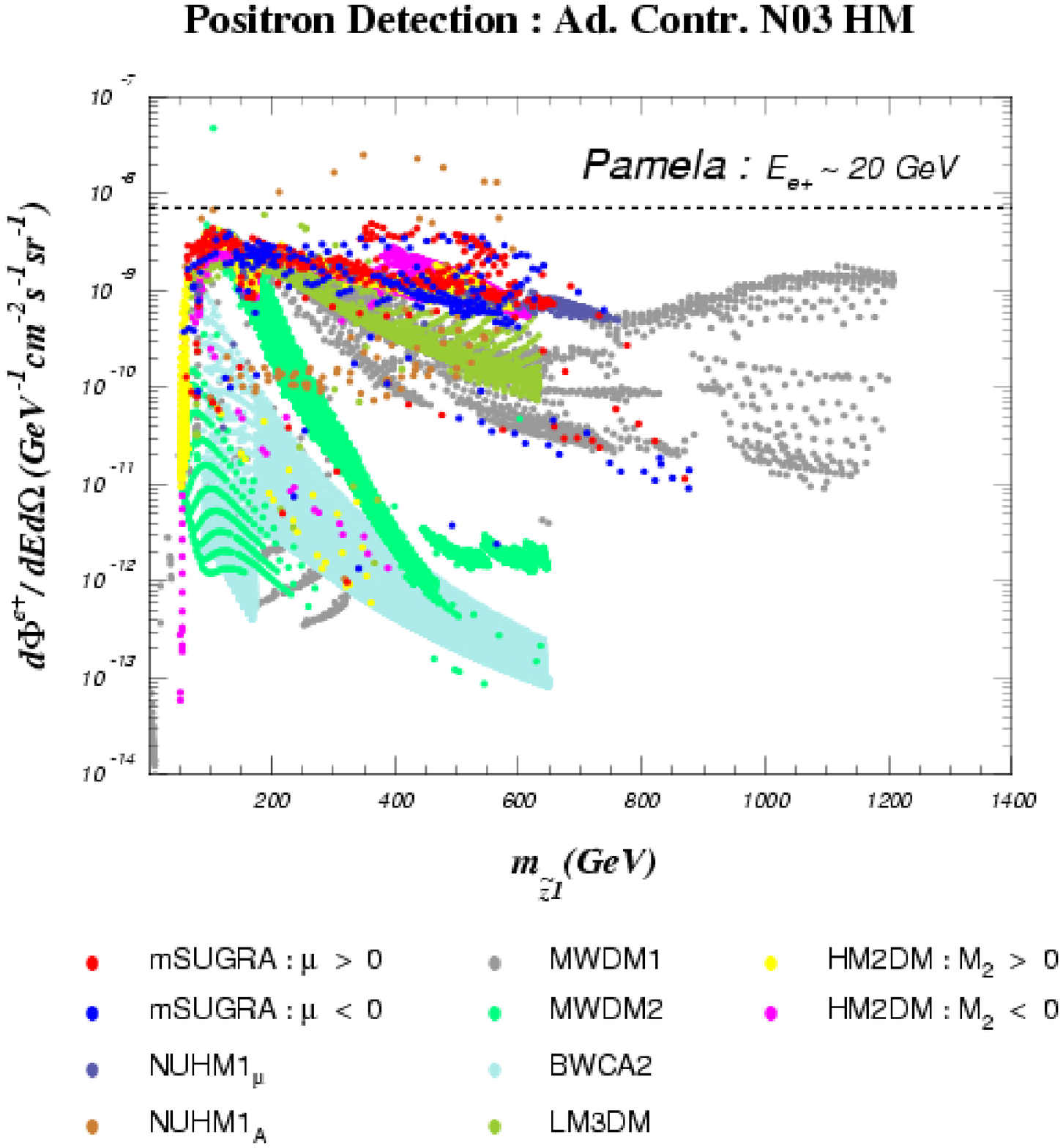,width=12cm,angle=0} 
\caption{\label{fig:idd_ep} Predictions for positron flux from various
models with $A_0=0$, $m_t=171.4$~GeV and the
sign of $\mu$ as in Fig.~\ref{fig:d_sqgl}, but
where the special parameter of non-universal mass models
has been dialed to yield
$\Omega_{\tz_1}h^2\simeq 0.11$.  
We fix $\tan\beta=10$ except for the mSUGRA model where we allow 
$\tan\beta=10$, 30, 45, 50, 52 and 55.
We adopt the Adiabatically Contracted
N03 DM halo profile.  The region above the dashed line denotes the
approximate reach of the PAMELA experiment.}}  

In Fig.~\ref{fig:idd_dbar}, we show the predicted flux of anti-deuterons
expected in a kinetic energy range $T_{\bar{D}}=0.1-0.25$~GeV using the
Adiabatically Contracted N03 halo profile, suitable for detection by the
proposed GAPS experiment.  Results using the Burkert profile tend to be
a factor of 10-20 lower, about the same as for anti-protons.  Models
with mixed higgsino dark matter populate the
$10^{-11}$~events/GeV/cm$^2$/s/sr level, and should be accessible to
GAPS via the long duration balloon flight. The NUHM1$_A$ model populates
points just below to just above the GAPS sensitivity level, while the co-annihilation
models give results which are generally beyond reach of any foreseeable
experiment.
\FIGURE[tbh]{
\epsfig{file=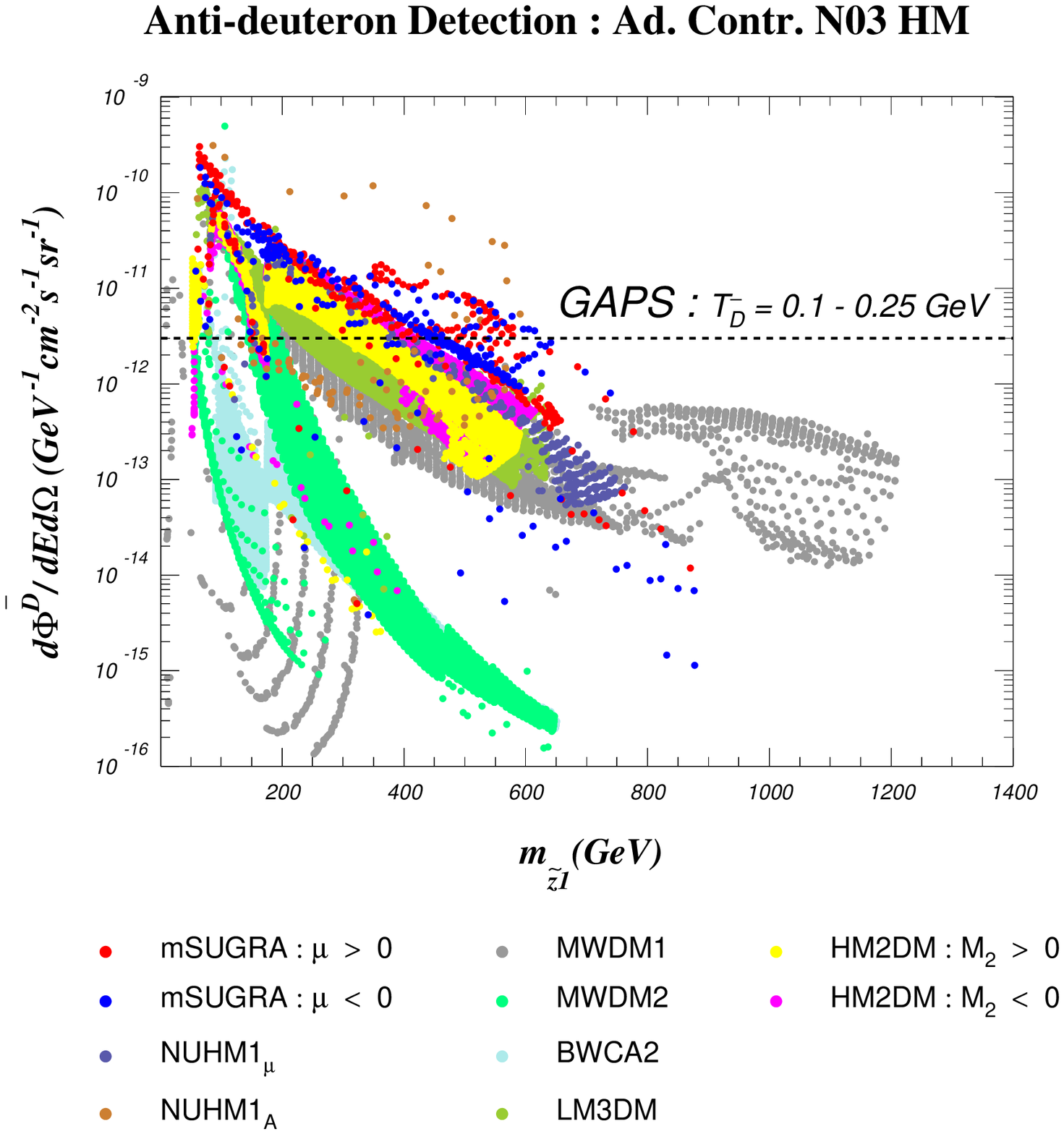,width=12cm,angle=0} 
\caption{\label{fig:idd_dbar} Predictions for anti-deuteron flux from
various models with $A_0=0$, $m_t=171.4$~GeV and the
sign of $\mu$ as in Fig.~\ref{fig:d_sqgl}, but where the special
parameter of non-universal models has been dialed to yield
$\Omega_{\tz_1}h^2\simeq 0.11$. 
We fix $\tan\beta=10$ except for the mSUGRA model where we allow 
$\tan\beta=10$, 30, 45, 50, 52 and 55.
 We adopt the Adiabatically Contracted
N03 DM halo profile.  The region above the dashed line denotes the
approximate reach of the GAPS experiment.}} 

In drawing inferences for prospects for indirect detection from the
scans of non-universal mass models discussed in this section, we should
keep in mind that we have shown results for just the Adiabatically
Contracted N03 halo profile (for detailed comparison of halo profiles,
see {\it e.g.} Ref.~\cite{nuhm2} and \cite{bo}) and fixed
$\tan\beta=10$. Typically, we have found that the indirect searches are
most sensitive to the higgsino component in the $\tz_1$. 
It is important to note that direct detection and indirect detection via
halo annihilation both grow as $\tan\beta$ is increased.

\section{Summary and concluding remarks}\label{sec:conclude}

If the observed cold dark matter \cite{wmap} is interpreted as thermal
relic neutralinos of $R$-parity conserving supersymmetric models, then
the determination of the CDM relic density (\ref{eq:relic}) provides a
very strong constraint, effectively reducing the dimension of model
parameter space by one unit. It is then reasonable to ask how this relic
density measurement constrains on what other experiments searching for
SUSY might or might not observe.

Indeed many groups have analyzed the implications of the measured value
of the CDM relic density for SUSY signals at the LHC within the mSUGRA
framework. Toward the end of Sec.~\ref{sec:intro}, we enumerated several
broad conclusions that were drawn from these studies.
In order to test the robustness of these conclusions,
it is necessary to
examine how these are affected if we relax the {\it untested}
universality assumptions that are the hallmark of the mSUGRA
framework. Motivated by this, as well as by the fact that most of the
relic-density-allowed  range of parameters lies on the
periphery of mSUGRA parameter space, we have examined a variety of models
where universality of scalar or gaugino SSB mass parameters
is relaxed via the introduction of just one
additional parameter that is then adjusted so that the
thermal relic density of neutralinos  matches
(\ref{eq:relic}),
by adjusting either the neutralino composition 
or its mass. 
In Sec.~\ref{sec:bm}, we show explicit examples of these various
models that lead to broadly similar sparticle 
mass spectra, and compare and contrast the
features of the different models with the paradigm mSUGRA framework, and
with one another.

Prior to the  analyses of non-universal models, there were several
prejudices inferred from studies based on mSUGRA, and frequently held to
be true, including:
\begin{enumerate}
\item The relic-density-consistent bulk region implies a variety of
  light sparticles accessible at the LHC, and possibly the ILC;
\item The Higgs-funnel region only occurs at large $\tan\beta$,
  where down-type Yukawa couplings are necessarily large, so that
  sparticle decay cascades are modified, with concomitant effects on
  collider signatures;
\item The higgsino-content of the neutralino LSP can only be large
  enough to get agreement with (\ref{eq:relic}) only if scalars are 
  essentially decoupled at the LHC;
\item The lighter $\tb$-squark is dominantly $\tb_L$ while the lighter
  stau is dominantly $\ttau_R$. 
\end{enumerate}
We have seen that even in relatively innocuous one-parameter extensions
of mSUGRA each of these conclusions is false. For instance, the HS model
allows rapid neutralino annihilation via light $\tu_R/\tc_R$ with other
sparticles much heavier, or via $\ttau_1$ which is dominantly
$\ttau_L$, the Higgs-funnel occurs for any value of $\tan\beta$ in the
NUHM1 model, and we can have MHDM for rather small scalar masses, also
in the NUHM1 model. While it is definitely worthwhile to correlate the
implications of one observation with what might and might not be seen in
other experiments, our analysis highlights the fact that such inferences
are frequently dependent on underlying assumptions. In particular, we
caution against drawing broad conclusions about what is or is not likely
at the LHC based upon studies of just the mSUGRA model.

In Sec.~\ref{sec:scan} we have performed scans over the parameter space
of the mSUGRA as well as over eight of its one-parameter extensions to
abstract features common to relic-density-consistent models. 
We end by summarizing our broad conclusions based on these scans.

\begin{itemize}

\item In mSUGRA, a well-tempered neutralino LSP can only be obtained in
the HB/FP region, where squark and slepton masses are far heavier than
the lightest charginos, neutralinos and gluino. In non-universal models,
we can easily have a well-tempered neutralino with $m_{\tq}\sim
m_{\tg}$. Indeed except for the HB/FP region of mSUGRA, squark and
gluino masses are typically comparable in relic-density-consistent
models. In a similar vein, we also note that while Higgs-funnel
enhancement is possible only for very large values of $\tan\beta$ in the
mSUGRA framework, if we allow for non-universality of Higgs SSB
parameters, we can have the Higgs funnel for any value of $\tan\beta$. 

\item In many relic-density-consistent models, the
$\tz_2 -\tz_1$ mass gap is usually less than $M_Z$, so that two-body
spoiler decays modes of $\tz_2$ are kinematically closed.  This means
that at least one dilepton mass edge (and perhaps more) is likely to 
be detectable at LHC. The location of the dilepton mass edge(s) 
is a rather clean signature of supersymmetric models, and 
often serves as the starting
point for sparticle mass reconstruction.

\item Most relic-density-consistent models should lead to observable
  signals at the LHC. In contrast, while models where accord with the
  observed relic density is obtained by tempering the higgsino-content
  of the neutralino will likely be accessible at a 1 TeV
  electron-positron collider, in other scenarios sparticles may
  simply be too heavy to be accessible. 

\item In well-tempered neutralino models, the mechanism that enhances
annihilation in the early universe also tends to enhance the direct DM
detection rate. In particular, models tempered via the higgsino content
of the LSP typically have $\sigma_{SI}(\tz_1 p)\sim 10^{-8}$ pb, which
ought to be accessible to the next set of direct detection experiments,
including LUX, Xenon-100, WARP, mini-CLEAN and SuperCDMS: see
Fig.~\ref{fig:dd}.  These experiments may also provide a measure of the
mass of the halo DM particle(s), assuming that it is not very heavy
compared to the target nucleus~\cite{mass}.\footnote{Direct detection
experiments with different target nuclei ranging over a wide range of
masses may thus provide clear evidence for multiple WIMP components in
the galactic halo.} If a signal is found in these direct detection
experiments, it can be directly compared to expectations based on SUSY
model parameters extracted in experiments at the LHC and especially the
ILC to test whether thermally produced neutralinos indeed saturate the
measured value of the cold DM density~\cite{peskin}, or whether DM, like
visible matter, turns out to have more than one component.

\item Likewise, these models have elevated rates for
indirect DM detection via neutrino telescopes. In this case, the flux of
muon neutrinos tends to be above $\Phi_\mu\sim 10$ events/km$^2$/year
for $E_\nu>50$ GeV. In many such models, the signal should be accessible
at the IceCube detector, as can be seen from Fig.~\ref{fig:idd_mu}. 

\item Finally, well-tempered neutralino models also have elevated rates
for indirect DM searches via neutralino annihilation in the galactic
halo into gamma rays and antimatter, especially if the higgsino
component is enhanced: see Fig.~\ref{fig:idd_gam}-\ref{fig:idd_dbar}.  These
rates have large uncertainties associated with the presently unknown
galactic dark matter density profile. But if a signal is found, it can
be compared to expectations using model parameters extracted from LHC and
ILC measurements, and the measured halo annihilation rate can be used to
determine the DM halo profile \cite{peskin}.

\end{itemize}

\acknowledgments

This research was supported in part by grants from the U.S. Department
of Energy. EKP was supported by the EU FP6 Marie Curie Research and Training Network
{\it UniverseNet} (MRTN-CT-2006-035863). EKP would like to thank the European Network of Theoretical
Astroparticle Physics ILIAS/ENTApP under contract number 
RII3-CT-2004-506222 for financial support.
        
% ---- Bibliography ----
%

\end{document}